\documentclass[preprint,aps,prc,showpacs,nofootinbib]{revtex4-1}\usepackage{amsmath}
\usepackage{amssymb}
\usepackage{graphics}
\usepackage{epsfig}

\usepackage{subfigure}

\newcommand\nn{\nonumber}
\newcommand\ba{\begin{eqnarray}}
\newcommand\ea{\end{eqnarray}}
\newcommand\be{\begin{equation}}
\newcommand\ee{\end{equation}}
\newcommand\ed{\end{document}}
\begin{document}

%%%%%%%%%%%%%%%%%%%%%%%%%%%%%%%%%%%%
\title{Dark photon manifestation in the tripletlike QED
processes $\gamma + \ell_i \to \ell^+_j \ell^-_j + \ell_i$,
$i\neq j,$ $i=e, \mu,$ $j=e, \mu,\tau$ }

%%%%%%%%%%%%%%%%%%%%%%%%%%%%%%%%%%%%

\author{G. I. Gakh}
\email{gakh@kipt.kharkov.ua}
\affiliation{\it National Science Centre, Kharkov Institute of
Physics and Technology, Akademicheskaya 1, and V. N. Karazin
Kharkov National University, Department of
Physics and Technology, 31 Kurchatov, 61108 Kharkov, Ukraine}
\author{M.I. Konchatnij}
 \email{konchatnij@kipt.kharkov.ua}
%% \cortext[cor2]{}
\affiliation{\it National Science Centre, Kharkov Institute of
Physics and Technology, Akademicheskaya 1, and V. N. Karazin
Kharkov National University, Dept. of
Physics and Technology, 31 Kurchatov, 61108 Kharkov, Ukraine}
\author{N.P. Merenkov}
\email{merenkov@@kipt.kharkov.ua}
\affiliation{\it National Science Centre, Kharkov Institute of
Physics and Technology, Akademicheskaya 1, and V. N. Karazin
Kharkov
National University, Dept. of
Physics and Technology, 31 Kurchatov, 61108 Kharkov, Ukraine}

\author{E. Tomasi--Gustafsson}
\email{egle.tomasi@cea.fr}
\affiliation{\it IRFU, CEA, Universit\'e Paris-Saclay, 91191
Gif-sur-Yvette, France}

\begin{abstract}
The tripletlike QED processes $\gamma + \ell_i \to \ell^+_j
\ell^-_j + \ell_i$ with $i\neq j$, and $i=e, \mu,$ $j=e,
\mu,\tau$ has been investigated as the reactions
where a dark photon, $A'$, is produced as a virtual state with
subsequent decay
into a $\ell^+_j \ell^-_j$ pair.
This effect arises due to the so-called kinetic mixing and is
characterized by a small parameter
$\epsilon$ describing the coupling
strength relative to the electric charge $e$.
In these processes, the dark photon appears as a resonance with Breit-Wigner
propagator in the $\ell^+_j \ell^-_j$ system. The distributions over the invariant mass of the
produced $\ell^+_j \ell^-_j$ pair is calculated and the value of the parameter $\epsilon$, as a function of the dark
photon mass, is estimated  for a given
number of  measured events in the specific experimental conditions. Assuming a standard deviation of $\sigma=2$ (corresponding to $\approx$ 95 $\%$
confidence limit) and a number of measured events equal to  $10^4$, we obtain the following limits for the parameter space of the dark photon:  $\epsilon^2\leq 10^{-7}$ ($\gamma+\mu^-\to
e^++e^-+\mu^-$) and $\epsilon^2\leq 10^{-7}\div 6\times 10^{-8}$  ($\gamma+e^-\to \mu^++\mu^-+e^-$) in 
 the $A'$ mass region 520 $< m_{A'} < $980 MeV. In  the  mass region 30 $< M_{A'} < $400 MeV we obtain 
$\epsilon^2\leq 10^{-7}$ ($\gamma+\mu^-\to e^++e^-+\mu^-$).
\end{abstract}
\maketitle

%%%%%%%%%%%%%%%%%%%%%%%%%%%%%%%%
\section{Introduction}
\label{Section:Introduction}
%%%%%%%%%%%%%%%%%%%%%%%%%%%%%%%%

Cosmological and astrophysical measurements give evidence of the existence of dark
matter. Its nature and interaction are unknown today. The recent experimental discoveries,
such as  neutrino oscillations (that imply massive neutrinos \cite{Fet98}), and the discrepancy
between the Standard Model (SM) prediction and the measured value of the anomalous magnetic moment of
the muon \cite{JN09},  as well as recent measurements of flavor-changing processes in B meson decays (showing 
tensions with the SM predictions \cite{W19}), lead one to consider physics
beyond the SM (see the reviews \cite{Rev12,Rev13,Rev15,Rev16}). The extension of  the SM includes various models predicting a number of new particles. The so-called dark photon (DP), $A'$, is one of the possible new particles. It is a massive vector boson that can mix with the ordinary photon
via "kinetic mixing" \cite{Holdom:1986eq}. Its mass and interaction strength are not predicted unambiguously
by the theory,  since  DPs can arise via different mechanisms. Various theoretically
motivated regions of the DP mass are given in Ref. \cite{Rev12}. A DP with mass larger
than 1 MeV  can  be produced in electron (proton) fixed-target experiments or at hadron
or electron-positron colliders (see the references in the review \cite{Rev12}). The properties of  a 
DP with a mass in the range eV - keV can be determined using astrophysical data.

A number of experiments searching for  DPs are carried out or planned in various laboratories. Using
APEX test run data obtained at the Jefferson Laboratory \cite{Essig:2010xa, Abrahamyan:2011gv},  the DP was searched in
electron-nucleus fixed-target scattering in the mass range 175$\div$250 MeV. The DarkLight Collaboration
\cite{Freytsis:2009bh} (JLab) studied the prospects for detecting a light boson with mass $m_X \leq $ 100
MeV in electron-proton scattering. The produced $A'$ decays to an  $e^+e^-$ pair leading to the $e^-p e^+e^-$
final state. An exclusion limit for the electromagnetic production of a light A$'$ decaying to $e^+e^-$
was determined by the A1 Collaboration at the Mainz Microtron MAMI ( \cite{Merkel:2011ze} using electron scattering
from a heavy nucleus. The Heavy Photon Search (HPS) experiment \cite{B19} searches for an electroproduced DP using an electron beam provided by the Continuous Electron Beam Accelerator Facility accelerator at JLab. HPS looks for DPs through two distinct methods - the search of a resonance in the $e^+e^-$ invariant mass distribution above the large QED background
(large DP-SM particles coupling region) - and  the search of a displaced vertex for long-lived DPs (small coupling region).

The manifestation of DPs is also searched for in the decay of known particles. The authors of
Ref. \cite{Frlez:2003pe} have studied the radiative pion decay, $\pi^ + \rightarrow e^+\nu\gamma$. The
measurements were performed at the $\pi$E1 channel of the Paul Scherrer Institute, Switzerland. The DP was also searched for in the decay of  $\pi^0$-mesons  ($\pi^0\rightarrow \gamma A'\rightarrow \gamma
e^+e^-$)  produced in proton-nuclei collisions at the Heavy Ion Accelerator Facility (China) \cite{Wang:2013bza}. The decay of
$\pi^0$-meson was the tool to search for the DP  also in the experiment at WASA-at-COSY(J\"{u}lich, Germany)
($\pi^0$-mesons were produced in the reaction $pp\rightarrow pp\pi^0$)  \cite{Moskal:2014dsa} and at
CERN, where $\pi^0$-mesons were produced in the decay of $K$-mesons, $K^{\pm}\rightarrow \pi^{\pm}\pi^0$ \cite{Goudzovski:2014rwa}.

The  search for a DP signal in inclusive dielectron spectra in proton-induced reactions on either a liquid
hydrogen target or on a nuclear target was performed at the GSI in Darmstadt \cite{Aet14}. An upper limit on the DP mixing
parameter in the mass range $m(A')=0.02 - 0.6$ GeV/$c^2$ was established. The constraints on the DP parameters from the data collected in  experiments with electron beams were summarized in Ref. \cite{Andreas:2012rm}. At JLab, it  was demonstrated that electron-beam fixed-target experiments would have a powerful discovery
potential for DPs in the MeV-GeV mass range \cite{IKST14}. Reference  \cite{Inguglia:2019gub} describes some of the main dark
sector searches performed by the Belle II experiment at the SuperKEKB energy-asymmetric $e^+e^-$ collider
(a substantial upgrade of the B factory facility at the Japanese KEK laboratory). The design luminosity
of the machine is $8 \times 10^{35}$ cm$^{-2}$s$^{-1}$. At  Belle II, the DP is searched for in the reaction
$e^+e^- \to \gamma_{ISR} A'$, with subsequent decays of the DP to dark matter $A' \to \chi \bar\chi$ (ISR stays for  initial state radiation). The main backgrounds come from QED processes such as
$e^+e^- \to e^+e^-\gamma (\gamma)$ and $e^+e^- \to \gamma\gamma (\gamma )$. Preliminary studies have been
performed and the sensitivity to the kinetic mixing parameter strength is given. With a small integrated
luminosity a very competitive measurement is possible, especially in the region for $M_{A'} > $ few times
$10$ MeV/$c^2$ where the BABAR experiment starts dominating in terms of sensitivity  \cite{Let17}. An
experiment to search for $A'$ was proposed at VEPP-3 (electron-positron collider) (Russia) \cite{Wojtsekhowski:2009vz}.
The search method is based on the missing mass spectrum in the reaction $e^+e^-\to\gamma A'$. 
At the future CEPC experiment (China) running at $\sqrt{s} = 91.2$ GeV it is possible for CEPC to perform a decisive measurement on a DP in the mass region 20 GeV $< m_{A'} <$ 60 GeV,  in about 3 months of operation \cite{JJ19}. $\sqrt{s} = $ 240 GeV is also achievable.

DP formation in various reactions was investigated theoretically in a number of papers. Bjorken {\it et al.}
\cite{Bjorken:2009mm} considered several possible experimental setups for experiments aimed at the search for $A'$
in the most probable range of masses from a few MeV to several GeV and confirmed that the experiments
at a fixed target perfectly suit for the discovery of DPs in this interval of  masses. DP production in the process of electron scattering on a proton or on heavy nuclei has been investigated in
Ref. \cite{BMV13} (Ref. \cite{Beranek:2013nqa}) for the experimental conditions of the MAMI (JLab) experiment
\cite{Merkel:2011ze} (Ref. \cite{Abrahamyan:2011gv}). The authors of Ref. \cite{Chiang:2016cyf} proposed to use rare leptonic decays of kaons
and pions $K^+(\pi^+)\rightarrow \mu^+\nu_{\mu}e^+e^-$ to study the light DP (with a mass of about 10 MeV). The
constraints on  DPs in the 0.01-100 keV mass range are derived in Ref. \cite{An:2014twa} (the indirect
constraints following from $A'\rightarrow 3\gamma $ decay are also revisited). The proposal to search for
light DPs using the Compton-like process, $\gamma e\rightarrow A'e$, in a nuclear reactor was suggested in
Ref. \cite{P17}. This suggestion was developed in Ref. \cite{Ge:2017mcq}, where the constraints
on some DP parameters were determined using the experimental data obtained at the Taiwan EXperiment On NeutrinO reactor (TEXONO). Some
results on the phenomenology of the DP in the mass range of a few MeV to GeV have been presented in Ref. 
\cite{Pospelov:2008zw}, where $g$ - 2 of muons and electrons together with other precision QED data, as well as radiative
decays of strange particles were analyzed.  

The process of the triplet photoproduction on a free electron,
$\gamma +e^-\to e^++e^-+e^-$, in which $A'$ can be formed as an intermediate state with subsequent decay
into an $e^+e^-$ pair was investigated in Ref. \cite{Gakh:2018ldx}. The advantage of this process is that the
background is a purely QED process $\gamma +e^-\to e^++e^-+e^-$, which can be calculated with the 
required accuracy. The analysis was done by taking into account the identity of the final electrons.
The constraints on parameter $\epsilon$ depending on the DP mass and the statistics (number) of events were obtained with a special method of gathering events where the invariant mass of one $e^+e^-$ pair remains fixed while the other pair is scanned. 

The search for  DPs produced at $e^+e^-$ colliders in the forward region was considered in Ref. \cite{Chen:2020bok}. An additional  detector set around BESIII can probe the DP coupling
parameter $\epsilon$ down to $2 \times 10^{-4}$, whereas at Belle-II, it would have a higher sensitivity down to $2 \times 10^{-5}$.

It should be mentioned that the constraints on the DP parameters can be obtained using  astrophysical data and that  the astrophysical constraints on the DP with masses in the range $M_{A'}\le$ (eV-keV) are stronger
than laboratory constraints (see, for example, Refs. \cite{Rev10, Raf96}). The calculation of the cooling rates for
the Sun gives the following constraint on the DP parameter space ($m_{A'} < 0.1$ eV) \cite{An:2013yfc}:
$$\epsilon \times \frac{m_{A'}}{eV} < 1.4 \times 10^{-11}. $$
The magnetometer data from $Voyager$ \cite{PCG15} probes DPs in the $10^{-24}$ to
$10^{-19}$ eV mass range. Values of the coupling parameter $\epsilon$ as low as $3 \times 10^{-5}$ for
the highest masses were excluded. The proposed search for  DPs using the Alpha Magnetic Spectrometer \cite{FST17a} has a potential discovery for  mass $m_{A'} \sim {\cal{O}}$(100) MeV and kinetic mixing parameters $10^{-10} \leq \epsilon
\leq 10^{-8}$. More information about DP (and dark matter) searches in astrophysical experiments
can be found in Ref. \cite{Battaglieri:2017aum}.

In this work, we suggest a possible way to detect the DP signal through the
reaction $\gamma + \ell_i \to \ell^+_j\, \ell^-_j + \ell_i$ ($\ell$ is a lepton), where a few tens MeV photon collides
with a high-energy electron or muon beam. The advantage of this reaction, as
compared with the reaction $\gamma +e^-\to e^++e^-+e^-$ considered in Ref.  \cite{Gakh:2018ldx},
is that there are no identical particles in the final state. So, the analysis
of the experimental results is simpler. We calculate the distributions over
the invariant mass of the produced $\ell^+_j \ell^-_j$- pair and search for the
kinematical region where the Compton-type diagram contribution is not suppressed
with respect to the Borsellino ones. The distributions for the proposed reaction
are more interesting  (corresponding to a larger number of events) due to the
increased phase space, since it is not necessary to fix the parameters of the second pair.
We estimated the value of the parameter $\epsilon$ as a function of the DP
mass,  for a given number of measured events.

This work is organized in the following way: Section II contains the formalism for
calculating the distribution over the invariant mass of the produced lepton pair.
The kinematics of the proposed reaction is briefly considered in Sec. IIA. The
calculation of the distribution over the invariant mass of the produced lepton pair,
which is caused by the QED mechanism, is given in Sec. IIB.
Section III is devoted
to the analysis of the DP effects in the proposed reaction. Here we estimate
the coupling constant $\epsilon$ as a function of the DP mass and the
number of the measured events. The analytic form for the double differential cross
section is given in the Appendix. Finally,  in Sec. IV we summarize  our conclusions.

%%%%%%%%%%%%%%%%%%%%%%%%%%%%%%%%%%%%%%%%%%%

\section{Formalism}
\label{Section:Formalism}

%%%%%%%%%%%%%%%%%%%%%%%%%
We assume  that the DP can manifest itself as
some intermediate state with the photon quantum numbers that
decays into a $\ell^+_j \ell^-_j$ pair. In such a case, the
process
\be
\gamma(k)+\ell^-_i(p)\rightarrow
\ell^+_j(p_3)+\ell^-_j(p_1)+\ell^-_i(p_2)\,,
\label{Eq:eq1}
\ee
in which a high-energy electron or muon interacts with a few
tens MeV photon, can be used, in principle, to probe the $A'$
signal in a wide range of $A'$ masses, from a few MeV up to 10
GeV.
%As concerns the "DP" state which interacts predominantly to the second and third
%lepton generation, the only channel $(\ell^-_i=\mu^-, \ell^{\pm}_j=\tau^{\pm} )$ is
%suitable to look for its signal.

\begin{figure}
%\captionstyle{flushleft}
\includegraphics[width=0.44\textwidth]{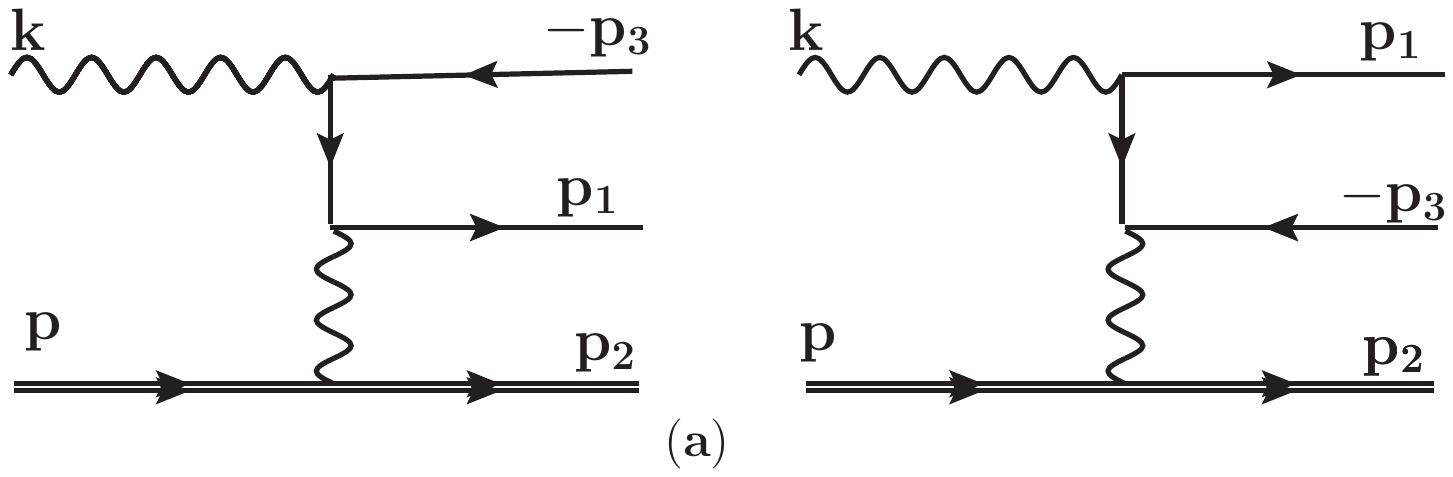}
\hspace{0.4cm}
\includegraphics[width=0.44\textwidth]{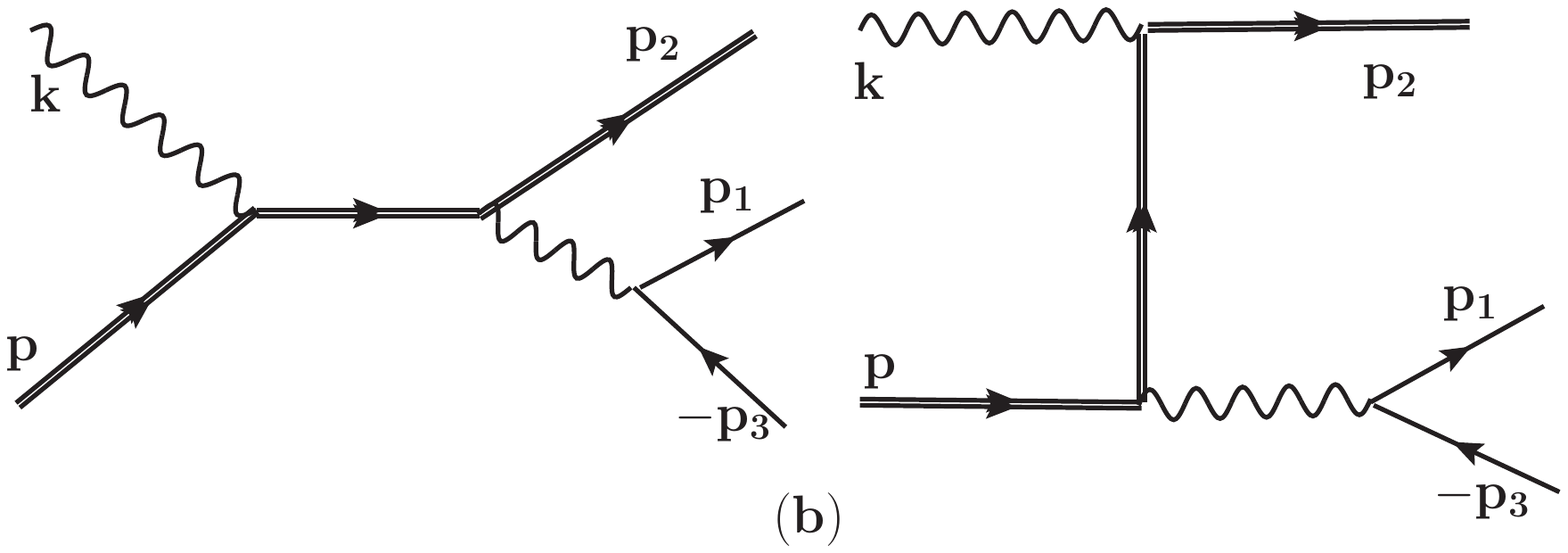}
\caption{The Feynman diagrams for the process
(\ref{Eq:eq1}). Diagrams ({\it a}) are called the Borsellino
diagrams, and
diagrams ({\it b}) are the Compton-type diagrams.}
\label{Fig:fig1}
\end{figure}

We suggest to scan the differential cross section
over the $\ell_j^+ \ell_j^-$ invariant mass, $s_1=(p_3+p_1)^2 $.
The background, due to a pure QED mechanism,
exceeds essentially the DP effect and
has to be calculated with high accuracy. In the lowest 
approximation, the
QED amplitude of the process (\ref{Eq:eq1}) is given by the four
diagrams shown in Fig.~\ref{Fig:fig1}. The DP effect is calculated  as a modification of the photon
propagator in the single-photon (the Compton type) diagrams Fig.~\ref{Fig:fig1}({\bf b}).

The applied method is the following: We calculate first the double differential cross
section over the invariant variables $s_1$ and $u=(k-p_2)^2$.
Next, we find the kinematical region, in terms of these variables, 
where the single-photon amplitudes
give a comparable or larger contribution with respect to the
double-photon amplitudes of the Borsellino diagrams in Fig.~\ref{Fig:fig1}({\bf a}). This region
can be delimited by excluding the large $|u|$ values with
appropriate cuts. Then, we perform an integration over the variable
$u$ in this restricted region and include the effect due to the
DP  contribution.

%%%%%%%%%%%%%%%%%%%%%%%%%%%%%%%%%%%%%%%%%%%%%%%
\subsection{Kinematics}
%%%%%%%%%%%%%%%
To describe the process (\ref{Eq:eq1}) we introduce the following set of invariant
variables \cite{Bykling:1972}:
\ba
&s=&(k+p)^2=(p_1+p_2+p_3)^2,\ s_1=(p_1+p_3)^2=(k+p-p_2)^2,\nn \\&s_2=&(p_2+p_3)^2=(k+p-p_1)^2, \label{Eq:eq2} \\
&t_1=&(k-p_1)^2=(p_2+p_3-p)^2, \ t_2=(p-p_2)^2=(p_1+p_3-k)^2.
\nn
\ea
In terms of these variables, we have:
\ba
&2(k\,p_2)=&s-s_1+t_2-M^2, 2(k\,p)=s-M^2, \ 2(k\,p_1)=m^2-t_1,
\nn\\
&2(k\,p_3)=&s_1+t_1-t_2-m^2, \ u=2M^2-s+s_1-t_2,\nn\\
&2(p\,p_1)= &s-s_2+t_1, \ 2(p\,p_2)=2M^2-t_2, \
2(p\,p_3)=s_2-t_1+t_2-M^2,\nn\\
&2(p_1\,p_3)=&s_1-2m^2, \ 2(p_2\,p_3)=s_2-M^2-m^2, \
2(p_1\,p_2)=s-s_1-s_2+m^2,
\label{Eq:eq3}
\ea
where $M~(m)$ is the mass of the initial (created) lepton.

\begin{figure}
%\captionstyle{flushleft}
\center
\includegraphics[width=0.4\textwidth]{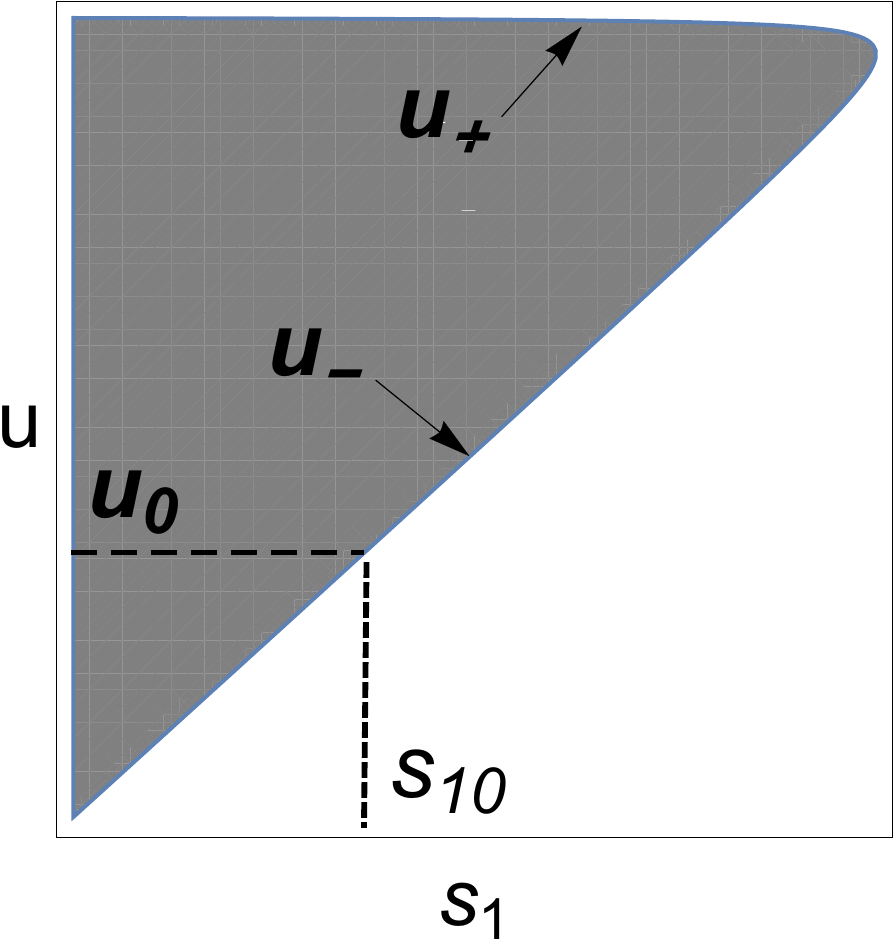}
\caption{Kinematical region for $(s_1,u)$. The quantity $s_{10}$
is the solution of the equation $u_0=u_-$, where $u_0$ is the negative cut parameter.}
\label{Fig:fig2}
\end{figure}

For  back-to-back events  azimuthal symmetry applies and
the phase space of the final particles can be written as
 \cite{Bykling:1972}:
\be
\label{Eq:eq4}
d\,R_3=\frac{d^3p_1}{2\,E_1}\,\frac{d^3p_2}{2\,E_2}\,\frac{d^3p_3}{2\,E_3}\,\delta(k+p-p_1-p_2-p_3)=
\frac{\pi}{16(s-M^2)}
\frac{dt_1\,dt_2\,ds_1\,ds_2}{\sqrt{-\Delta}}\,,
\ee
where $\Delta$ is the Gramian determinant. The limits of integration are defined by the condition of
the positiveness of $(-\Delta).$ In Ref.\,\cite{Gakh:2018ldx} these limits have been obtained for the triplet
production process where the masses of all leptons are equal. Similarly, we obtain for the considered reactions
\be
\label{Eq:eq5}
t_{1-}<t_1<t_{1+}\,, \ \ \ t_{1\pm}=\frac{A\pm 2\sqrt{B}}{(s-s_1)^2-2(s+s_1)M^2+M^4}\,,
\ee
where
\ba
A&=& \left\{-M^4 s_1 + s_1 s_2 (s_1 - t_2) + s^2 t_2 +
M^2 \left [ s_1^2 + s_1 s_2 + s (s_1 - t_2) + s_2 t_2\right ] - \right .  \nn \\
&& \left .s \left [ s_2 t_2 + s_1 (s_2 + t_2) \right ] + m^2
\left [ M^4 + s (s - s_1) + (s + s_1) t_2 - M^2 (2 s + 3 s_1 +
t_2) \right ] \right\}\,,\nn \\
B&=&\left [ st_2(s-s_1+t_2)+M^2(s_1^2-2st_2-s_1t_2)+
M^4t_2\right ]\times \left [m^2 M^4 + m^4 s_1 -m^2 M^2 (2 s +
s_1) + \right. \nn\\
&&  \left . \ \ \ m^2 (s^2 - s s_1 - 2 s_1 s_2) +
 M^2 s_1 (s - s_2) + s_1 s_2 (-s + s_1 + s_2)\right]\,.
 \nn
 \ea
At fixed $s_1$ and $t_2$, we have $s_{2-}<s_2<s_{2+}$ where 
\be
\label{Eq:eq6}
s_{2\pm}=
\displaystyle\frac{2 m^2 + M^2 + s - s_1\pm \lambda_1
\sqrt{1-\displaystyle\frac{4 m^2}{s_1} }}{2}\,,
 \lambda_1=\sqrt{(s-s_1)^2-2M^2(s+s_1)+M^4}\,.
 \ee
The corresponding range for $t_2$, $t_{2-}<t_2<t_{2+}$ and for
$s_1$, $4m^2<s_1< (\sqrt{s}-M)^2$, 
is limited by the curves
\be
\label{Eq:eq7}
t_{2\pm}=\frac{1}{2s}\left[C\pm \lambda_1(M^2 - s)\,\right ]\,,\
C=s_1(s+M^2)-(s-M^2)^2\,.
\ee
Furthermore, it is preferred to use the variable $u$ instead of $t_2$ because
one of the Compton-type diagrams has a pole behavior precisely in
the $u$ channel.
The kinematical region ($s_1,u$) is shown in Fig. \ref{Fig:fig2}, where:
\be
\label{Eq:upm}
u_{\pm}=\frac{1}{2s}\left[\bar C\pm \lambda_1(s-M^2)\,\right
]\,,\ \bar C=M^4+M^2(2s-s_1)-s(s-s_1)\,.
\ee

%%%%%%%%%%%%%%%%%%%%%%%%%%
\subsection{Calculation of the QED cross section}
%%%%%%%%%%%%%%%%%%%%%%%%%%

In the case of unpolarized particles, we have to average over (to
sum over) the polarization states of the initial (final)
particles. The differential cross section can be written in the
form
\be
\label{Eq:eq8}
d\sigma=\frac{1}{4}\,\frac{e^6}{4(k\,p)\,(2\pi)^5}\,\sum_{pol}|M|^2\,dR_3\,,
\ee
where $M$ is the matrix element of the process (\ref{Eq:eq1}):
\be
\label{Eq:eq9}
\sum_{pol}|M|^2=|M_b|^2+|M_c|^2+2{\cal Re}\big(M_bM^*_c\big),\
\ee
and the index $b\,\,(c)$ corresponds to the Borsellino
(the Compton-type) diagrams.

The double differential cross section, as a function of the
variables $s_1$ and $t_2$ [or the $(s_1,t_2)$ distribution],
is
\be
\label{Eq:eq10}
\frac{d\,\sigma}{d\,s_1\,d\,t_2}=\frac{\alpha^3}{64\,\pi(s-M^2)^2}\int\int\frac{d\,s_2\,d\,t_1}{\sqrt{-\Delta}}\sum_{pol}|M|^2\,.
\ee
The calculation of the matrix element squared (\ref{Eq:eq9}) gives:
\ba
|M_b|^2&=&\displaystyle \frac{8 }{t_2^2} \left\{ \displaystyle
\frac{8 d^2 m^2}{d_3^2}+t_2^2
\left[\left(\frac{1}{d_3^2}+\displaystyle\frac{1}{d_1^2}\right)
m^2-\displaystyle\frac{2 M^2}{d_1 d_3}-
\displaystyle \frac{2 }{d_3}\left(\displaystyle
\frac{d}{d_1}+1\right) \right] +
\right .    \nn\\
&& 8 (\text{pp}_3)^2
\left[\left(\frac{1}{d_3}+\frac{1}{d_1}\right){}^2
m^2-\frac{t_2}{d_1 d_3}\right ]+\nn\\
  && 4 (\text{pp}_3)
  \left [  t_2
      \left(
  -m^2\left( \frac{1}{d_3}+\frac{1}{d_1}\right)^2
   -\frac{1}{d_1}+\frac{2 d}{d_1 d_3}+\frac{1}{d_3}
   \right )
-\frac{4 d }{d_3}\left(\frac{1}{d_3}+\frac{1}{d_1}\right)
m^2+\frac{t_2^2}{d_1 d_3}
   \right ] +\nn\\
    &&t_2 \left [ 2
\left( \frac{1}{d_3}+\frac{1}{d_1}\right) ^2 m^4+2
\left(\frac{1}{d_3}+\frac{1}{d_1}\right)^2 m^2 M^2+4
\left(\frac{1}{d_3}+\frac{1}{d_1}\right)
    \left(\frac{d}{d_3}-1\right ) m^2- \right . \nn\\
  &&
  \left . \left . 4
\left(\frac{1}{d_3}+\frac{1}{d_1}\right) M^2-2 \left (
\frac{d_1}{d_3}+\frac{2 d
}{d_1}\left(\frac{d}{d_3}-1\right)+\frac{d_3}{d_1}\right) \right
]
   -4
\left( \frac{d_1}{d_3}+\frac{d_3}{d_1}\right)
M^2-\frac{t_2^3}{d_1 d_3}
    \right \} \,, \nn
    \ea
     \ba
|M_c|^2&=&|M_b|^2(p\leftrightarrow-p_3,p_1\leftrightarrow p_2,m
\leftrightarrow M)\,, \ d_i=(k p_i). \nn
   \ea
On the level of the full differential cross section, the interference of $M_c$ and $M_b$ contributes also, but
in the experimental setup considered in this paper, where only the scattered $\ell_i$ lepton is recorded, it vanishes due to the Furry theorem \cite{Furry:1937zz}.

Introducing the short notation:
\be
\displaystyle\frac{\pi}{64\,(s-M^2)}\int\limits_{s_{2-}}^{s_{2+}}d\,s_2
\int\limits_{t_{1-}}^{t_{1+}}d\,t_1\displaystyle\frac{W}{\sqrt{-\Delta}}
\equiv \overbrace{W},
\nn
\ee
and applying the relation (\ref{Eq:eq3}) between the variables $t_2$ and $u$, we have
\ba
\overbrace{|M_c|^2}&=& \displaystyle\frac{2\pi^2
\sqrt{1-\displaystyle\frac{4 m^2}{s_1}}(2 m^2 + s_1)}{3 (M^2 -
s)^2 s_1^3 (M^2 - u)}\times
\label{Eq:eq15} \\
&&
\left \{ -4 M^6 + (s - s_1 + u) (5 M^4 - 2 s_1^2 + s u) -
M^2 \left [s^2 - 4 s_1^2 + s_1 u + u^2 + s (s_1 + 6 u) \right ]
\right \} \,,
 \nn
 \ea
 %%%%%%%%%%%%%%%%%%%
\ba
&\overbrace{|M_b|^2}&=\frac{4 \pi ^2 }{\left(M^2-s\right)
\left(2 M^2-s+s_1-u\right)^2}\times
  \nn \\
&&
\left \{
\ln \left(\frac{\sqrt{s_1-4 m^2}+\sqrt{s_1}}{2 m}\right)
 \left [  \frac{4 s_1 [s_1^2-2 m^2 (2 m^2-3 s_1 )]  (M^2-s )^2}
 { (-2  M^2+s+u )^4}- \right . \right .
 \nn \\
&&
\frac{4 (M^2-s ) [2 m^2 \left(2 m^2 (M^2-s-s_1 )+s_1(-5 M^2+5
s+3 s_1 )\right)+s_1^2 \left(-2 M^2+2 s+s_1\right)] }
{\left(2   M^2-s-u\right)^3}-\nn \\
&&
\frac{2 [ -4 m^4+m^2 (-8 M^2+8 s+2 s_1)+(M^2-s )^2+s_1 (-M^2+3
s+2 s_1)] }{-2 M^2+s+u}+\nn \\
&&
\frac {1 }{(-2 M^2+s+u )^2} 2 [2 m^2 \left( 4 (M^2-s )^2+s_1
(10 s+s_1-8 M^2)-2 m^2 (2
   s+s_1)  \right )  +
  \nn \\
&&
s_1 \left (3 (M^2-s )^2+s_1 (4 s+s_1-2 M^2) \right ) ] + 2
M^2+s+3 s_1-u \bigg ] - \nn \\
 %  \ea
 %  \ba
&&
\sqrt{1-\frac{4 m^2}{s_1}} \bigg[ -\frac{4 (M^2-s ) [ m^2 s_1
(M^2-s-s_1 )+2 s_1^2 (2 M^2-2
   s-s_1 )] }{(-2 M^2+s+u )^3}+\nn \\
%\ea
%\ba
&&
\frac{4 s_1^2 (M^2-s )^2(m^2+2 s_1)}{(-2 M^2+s+u
)^4}+
\nn \\
&&
\frac{s_1 [2 m^2 (2 s+s_1)+9 (M^2-s)^2+2 s_1 (-6 M^2+8 s+s_1)]
}{(-2 M^2+s+u)^2}-\nn \\
&&
    \frac{s_1 (2 m^2-5 M^2+9 s+4 s_1)+ (M^2-s)^2}{-2
   M^2+s+u}+\frac{1}{2} (2 M^2+s+5 s_1-u)\bigg ] \bigg\}\,.
\label{Eq:eq16}
\ea
Following Eq.~(\ref{Eq:eq10}) and the definition of the quantity
$\overbrace{W}$, the double differential cross section can be
written as
\be\label{Eq:sud}
\frac{d
\sigma}{d\,s_1\,d\,u}=\frac{\alpha^3}{\pi^2(s-M^2)}\overbrace{\sum_{pol}|M|^2}\,,
\ee
where $\sum_{pol}^{ }|M|^2$ is defined by Eq.~(\ref{Eq:eq9}). To
measure the differential cross section $d\,\sigma/(ds_1du),$ it
is sufficient to detect the final muon 4-momenta.

The integration of the double differential cross section (\ref{Eq:sud}) with
respect to the variable $u$ in the limits $u_-< u < u_+$ gives
\ba
\frac{d \sigma_c}{ds_1}&=&\displaystyle\frac{\alpha ^3
\sqrt{1-\displaystyle\frac{4 m^2}{s_1}} (2 m^2+s_1)}{3 \pi s_1^2
(s-M^2)^3}
\left \{ \frac{\lambda_1 [ M^6-M^4(s+s_1 )+M^2 s (15 s+2 s_1
)+s^2 (s+7 s_1)] }{2 s^2}-\right .
\nn \\
&&
\left . [3 M^4+M^2 (6 s-2 s_1)-s^2+2(s-s_1) s_1] \ln
\left[\frac{(M^2+s-s_1+\lambda_1)^2}{4 M^2 s}\right] \right\}
\,,
   \label{Eq:eq18}
\ea
\ba
\frac{d \sigma_b}{d\,s_1}&=& \displaystyle\frac{2 \alpha ^3}{\pi    (s-M^2)}
    \left \{
\displaystyle\frac{2 \lambda _1 \sqrt{1-\displaystyle\frac{4
m^2}{s_1}}}
    {3 s_1^3 (M^2-s)^2} \right .
     \label{Eq:eq19}\\
   &&
   \left  [
m^2 [17 (M^2-s)^2+2 s_1 (4 M^2-2 s+s_1) ] + s_1 [7 (M^2-s)^2+s_1
(4 M^2-2 s+s_1) ] \bigg ] +
    \right .\nn\\
   &&
   \frac{1}{s_1^4 (s-M^2 )}
   \Bigg [ s_1 \sqrt{1-\displaystyle \frac{4 m^2}{s_1}}
    \bigl [  2 m^2 [ 2 (M^2-s )^2+   s_1 (6 M^2-2 s+s_1 ) ]
      \nn \\
   &&
   +s_1
   [(M^2-s )^2+s_1(5 M^2-s ) ] \bigr  ] +
   2 \bigl [ 4 m^4  [ 2 (M^2-s)^2+s_1 (6M^2-2 s+s_1)]  -
   \nn \\
    &&
    -2 m^2 s_1 [ 2 (M^2-s )^2+s_1 (6 M^2-2  s+s_1 )]-
   \nn \\
   &&
   s_1^2 [ (M^2-s )^2+s_1 (3 M^2-s )]  \bigr ]
\ln \left( \frac{\sqrt{s_1-4 m^2}+\sqrt{s_1} }{2 m} \right )
\Bigg ] \times
   \nn\\
   &&
    \ln
\left(\frac{\left(\lambda _1-M^2+s+s_1\right)^2}{4 s
s_1}\right)-
   \nn \\
%\ea
%\ba
   &&
\frac{2 \lambda _1 }{3 s_1^4 (M^2-s)^2}\bigl [ -4 m^4 [17 (M^2-s
)^2+2 s_1
   (4 M^2-2 s+s_1 )]+
    \nn\\
   &&
   6 m^2 s_1 (M^2-s ) (7 M^2-7 s+2 s_1)+
    \nn\\
   &&
   s_1^2 [8 (M^2-s )^2+ s_1 (5 M^2-s+2 s_1)] \bigr] \times
\ln \left(\frac{\sqrt{s_1-4 m^2}+\sqrt{s_1}}{2 m}\right)+\nn\\   &&
\frac{1}{2 s_1^4 (s-M^2)}\left [ s_1 \sqrt{1-\frac{4 m^2}{s_1}}    \bigl [ 4 m^2 [ 2 (M^2-s)^2+s_1 (6 M^2-2 s+s_1 )] +
    \right . \nn\\
   &&
   +s_1 [2 (M^2-s )^2+s_1 (10 M^2-2 s+s_1)] \bigr ]+
   \nn\\
   &&
2 (8 m^4-4 m^2 s_1-s_1^2) [ 2 (M^2-s )^2+s_1 (6 M^2-2 s+s_1)]
\times
  \nn  \\ &&
   \left .   \left .
   \ln
\left(\frac{\sqrt{s_1-4 m^2}+\sqrt{s_1}}{2 m}\right) \right ]   \ln
\left[ \frac{[ M^4+\lambda _1 (M^2-s )-M^2 (2 s+s_1 )+s
(s-s_1)]^2}{4 M^2 s s_1^2}\right] \right \}\,.
  \nn
   \ea
We can also write a more complicated analytical expression for $d\sigma/ds_1$ accounting  for
the restriction on the variable $u$. The corresponding result is given in the Appendix.

In the limiting case $s\gg (s_1, M^2)\gg m^2$, which applies for  electron-positron pair production, these expressions are essentially simplified, namely
\ba
\frac{d\sigma_c}{d\,s_1}&=&\frac{\alpha^3}{3\,\pi\,s\,s_1}
\left [\frac{1}{2}+\frac{17\,M^2+2\,s_1}{2\,s}
+\left(1-\frac{3\,M^2+2\,s_1}{s}\right )\ln{\frac{s}{M^2}}\right
]\,,\nn \\
\frac{d \sigma_b}{d\,s_1}&=&\frac{2\,\alpha^3}{\pi\,s_1^2}\left
\{\ln{\frac{s_1}{m^2}}\ln{\frac{s^2}{s_1\,M^2}}-\frac{8}{3}\ln{\frac{s_1}{m^2}}-
\ln{\frac{s^2}{s_1\,M^2}}+\frac{14}{3}+ \right .  \nn \\
&&
\left .
\frac{1}{s}\left
[-s_1\,\ln{\frac{s_1}{m^2}}\ln{\frac{s^2}{s_1M^2}}+(s_1-2M^2)\ln{\frac{s_1}{m^2}}+s_1\,\ln{\frac{s^2}{s_1M^2}}-4
s_1+2 M^2\right ]\right \}\,.\nn
\ea
Eqs.~(\ref{Eq:eq18}) and (\ref{Eq:eq19}) hold for particles with
arbitrary masses. They can be applied to the reactions:
\be
\gamma + e^- \to \mu^+ \mu^- + e^-, \ \gamma + e^- \to \tau^+
\tau^- + e^-, \ \gamma + \mu^- \to \tau^+ \tau^- + \mu^-,\nn
\ee
and the asymptotic formulas, in the limit $s\gg (s_1,\, m^2)\gg
M^2,$ valid for muon pair production, become:
\ba
\frac{d\sigma_c}{d\,s_1}&=&\frac{\alpha^3}{6\pi\,s\,s_1}(2+y)\sqrt{1-y}\left
[\frac{1}{2}+x_1+(1-2 x_1)\ln{\frac{s}{M^2}}\right ]\,, \
x_1=\frac{s_1}{s}\,,
\ y=\frac{4 m^2}{s_1}\,,
\nn \\
\frac{d \sigma_b}{d\,s_1}&=&\frac{2\,\alpha^3}{\pi\,s_1^2}\left
\{ \left[ (2+2 y-y^2)L -(1+y)\sqrt{1-y}\right
](1-x_1)\ln{\frac{s^2}{M^2s_1}}+
\right .
\nn\\
&&\left. \left [\frac{17}{6}y^2-7 y-\frac{16}{3}+x_1\left(2+5
y-\frac{3}{2}y^2\right )\right ]\,L+\sqrt{1-y}\left
[\frac{14}{3}+\frac{17}{6}y-x_1\left(4+\frac{3}{2}y\right
)\right ]\right \},
\nn \\
L&=&\ln{\frac{\sqrt{s_1}+\sqrt{s_1-4 m^2}}{2 m}}\,.
\nn
\ea

%%%%%%%%%%%%%%%%%%%%%%%%%%%%

\section{Analysis of the dark photon signal}

%%%%%%%%%%%%%%%%%%%%%%%%%%%%%%%

We have to estimate first the QED background in order to find the
kinematical conditions where the cross section $d\sigma_c/d s_1$
exceeds $d\sigma_b/d s_1,$ because in our work,  the DP signal is related to a modification of  the Compton-type diagrams. We perform the following  calculations for $e^+ e^-$ pair creation.

It is well known that at
photon energies larger than 10\,MeV, the main contribution to
the cross section of the tripletlike processes arises from the
Borsellino diagrams due to the events
at small values of $t_2$ \cite{Boldyshev:1994bs}.

\begin{figure}
%\captionstyle{flushleft}
\includegraphics[width=0.44\textwidth]{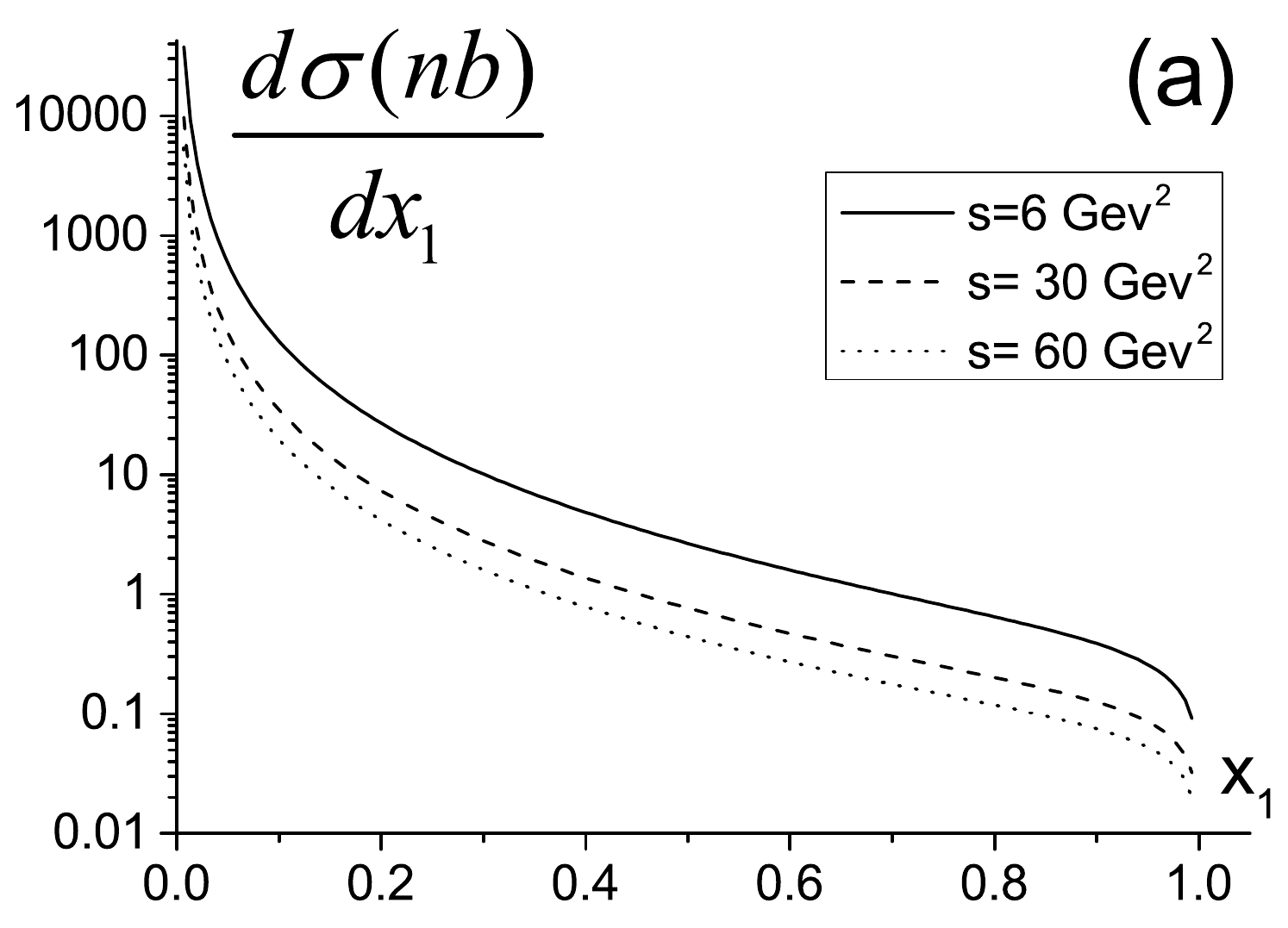}
\hspace{0.4cm}
\includegraphics[width=0.44\textwidth]{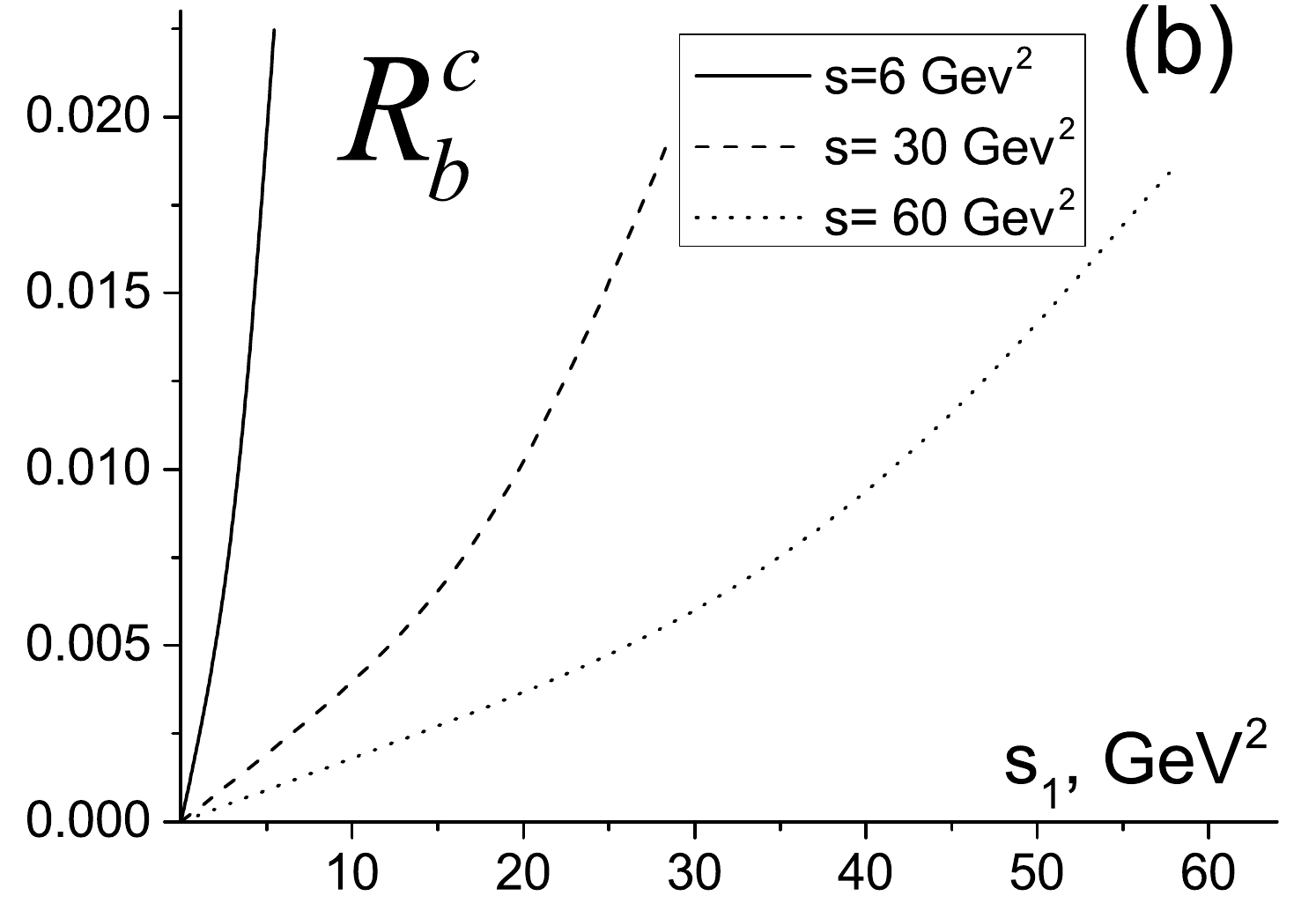}
\caption{(a) Differential cross section of the
process $\gamma+\mu^- \to e^+ e^- +\mu^-,$
as a function of $x_1=s_1/s$ for fixed values of the variable $s$, calculated
with Eqs.~(\ref{Eq:eq18}) and (\ref{Eq:eq19});  (b) Ratio
of the contributions to the cross section of the Compton-type
diagrams to the Borsellino ones as a function of the $e^+e^-$ invariant mass squared,.}
\label{Fig:fig3}
\end{figure}

In Fig.~\ref{Fig:fig3}, we show : (a) the differential cross section, i.e.,  the sum of (\ref{Eq:eq18}) and (\ref{Eq:eq19}) contributions as a function of  the dimensionless variable $x_1=s_1/s$ at fixed $s$,  and (b) the ratio of the Compton-type diagrams
contribution to the Borsellino ones as a function of $s_1$, 
\be
R^c_b=\frac{d\sigma_c}{d\sigma_b}\,, \
\nn
\ee
at different colliding
energies: $s=$6, 30, 60 GeV$^2$, provided that the whole
kinematical region ($s_1,u$) is allowed. We see that in a
wide, physically interesting range of variable $s_1,$ the
quantity $R^c_b(s_1)$ is rather small (does not exceed
2$\cdot$10$^{-2}$) and it is obvious that, in the case of the
nonlimited phase space, the Borsellino contribution leads to a
very
large QED background for searching for a small DP signal which
modifies the Compton-type contribution only.

\begin{figure}
%\captionstyle{flushleft}
\includegraphics[width=0.30\textwidth]{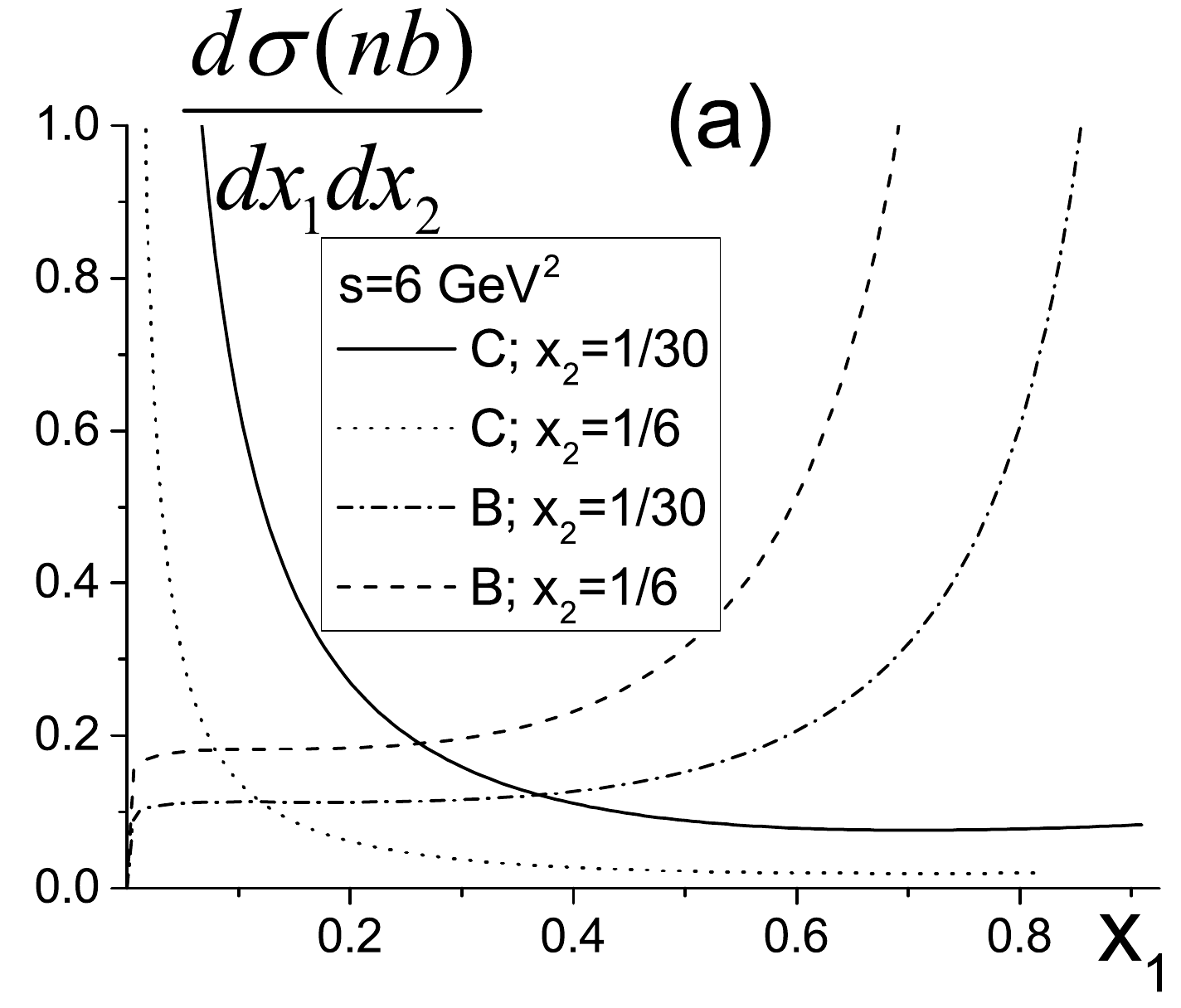}
\includegraphics[width=0.30\textwidth]{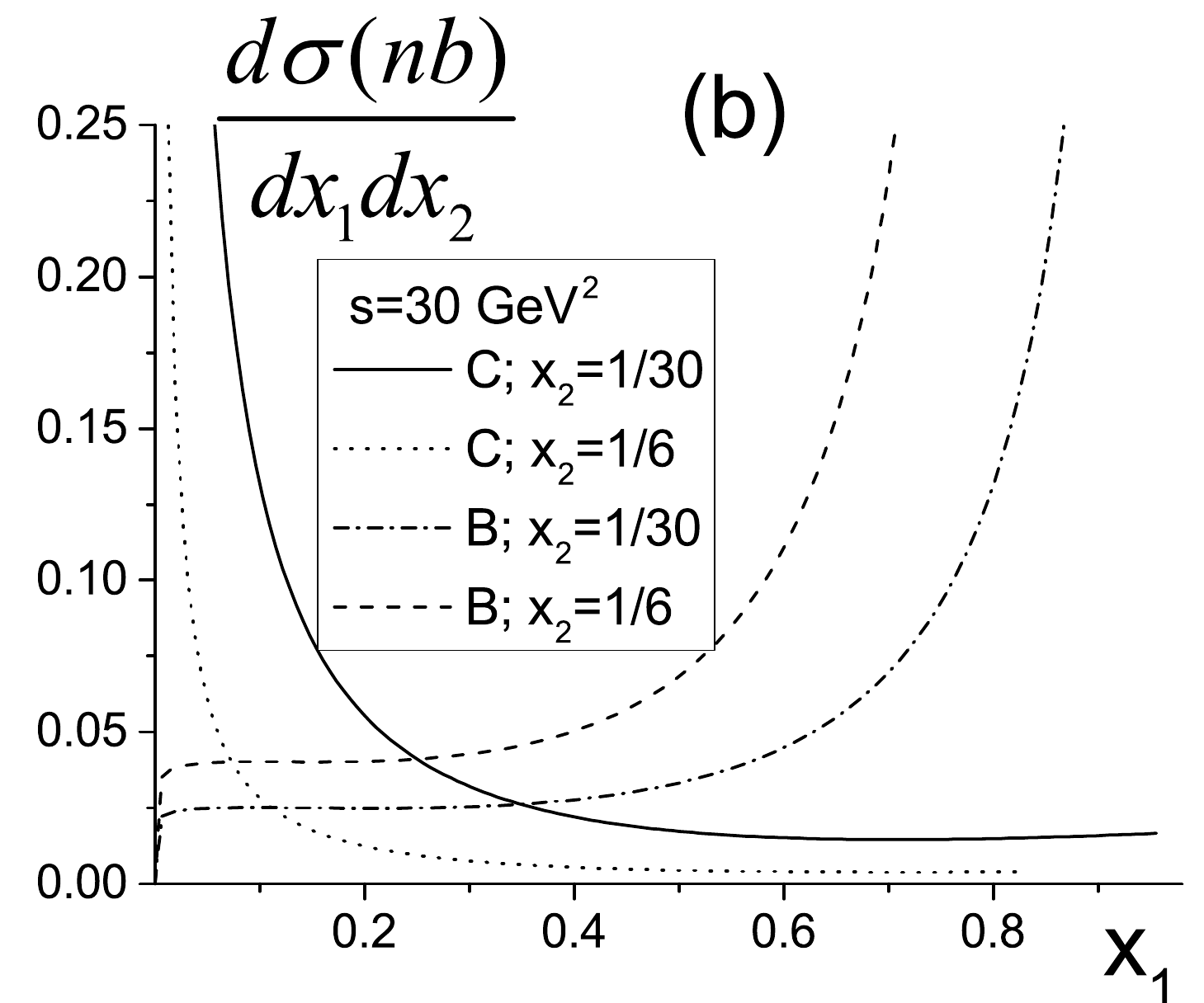}
\includegraphics[width=0.30\textwidth]{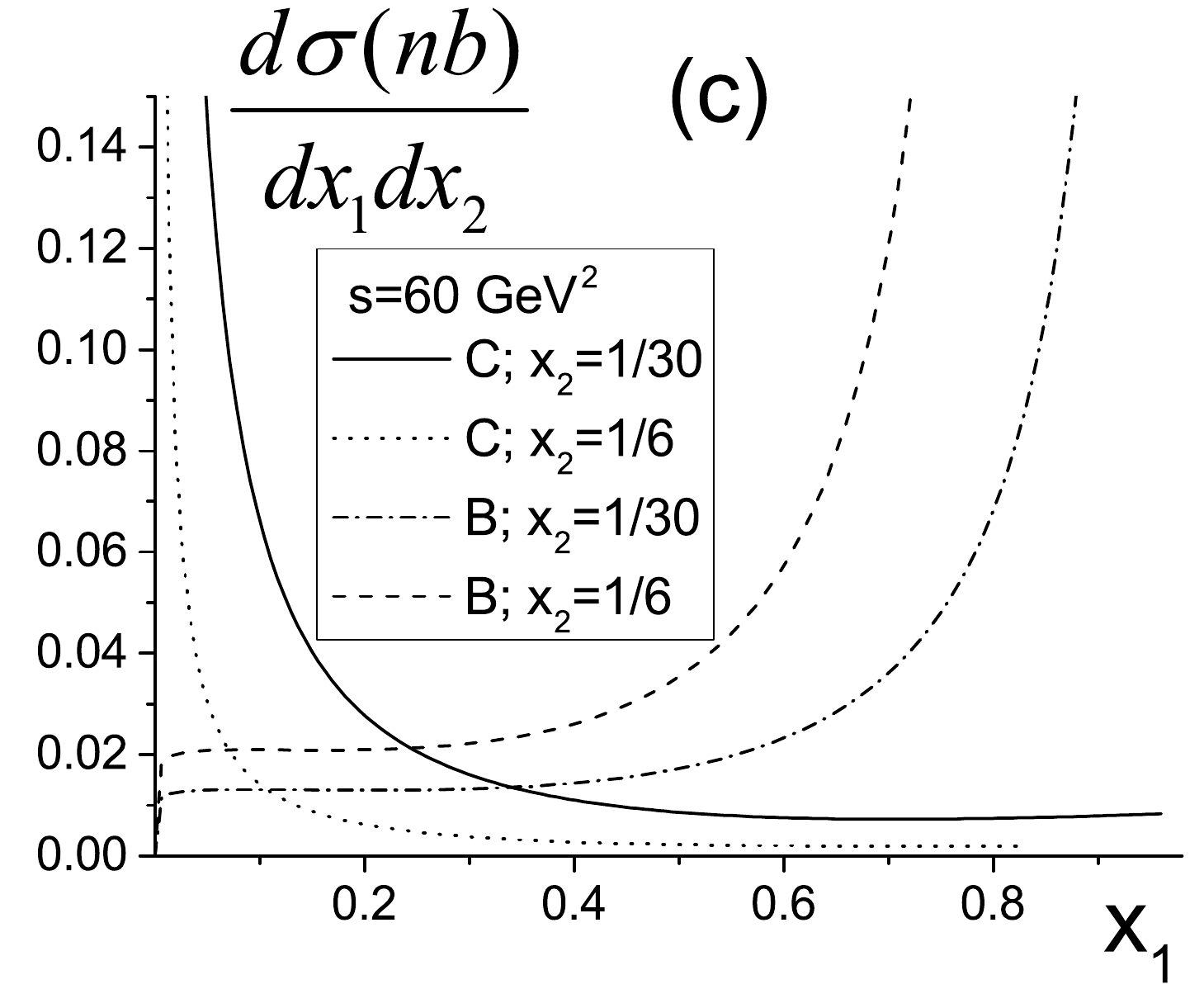}

\vspace{0.4cm}

\includegraphics[width=0.30\textwidth]{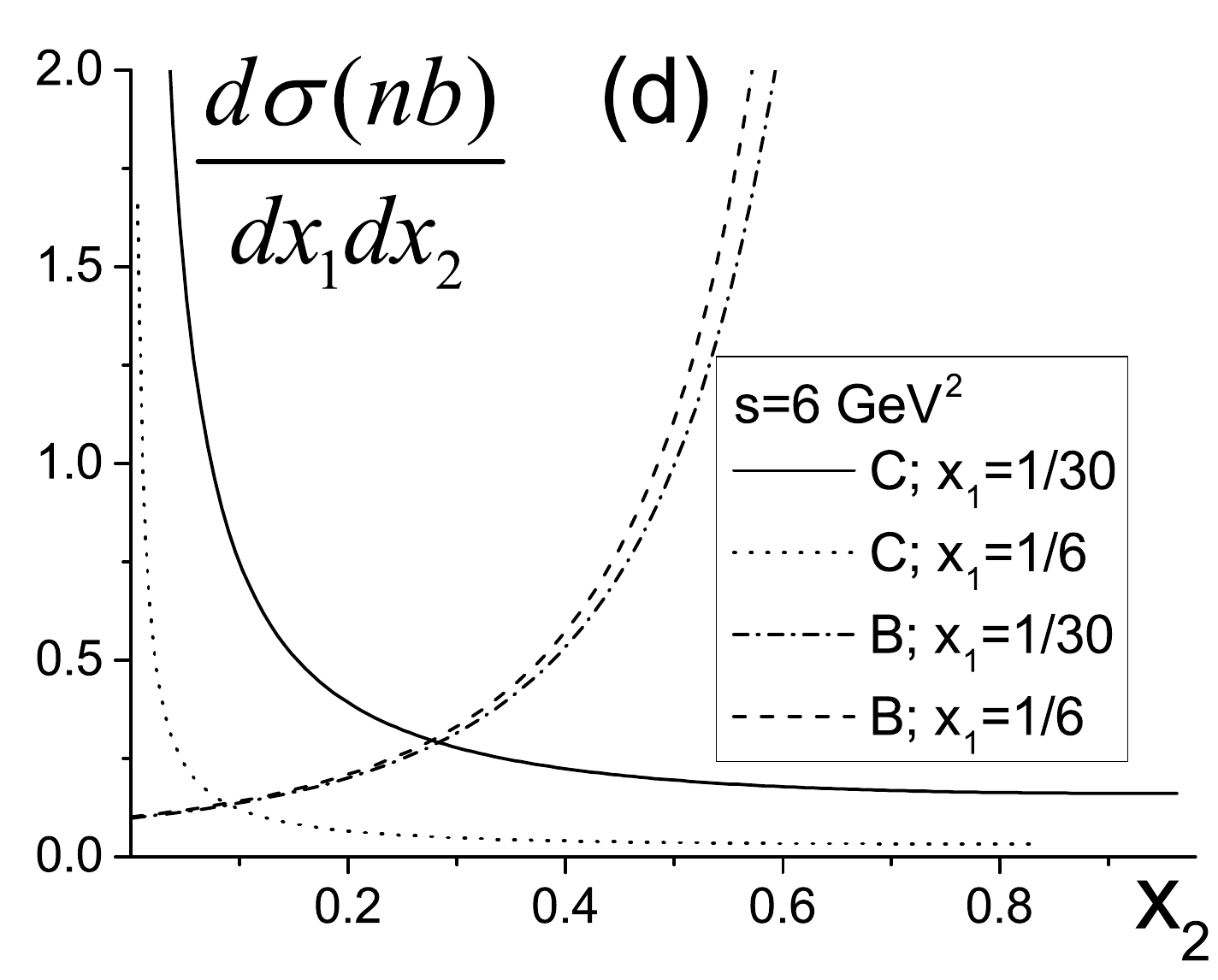}
\includegraphics[width=0.30\textwidth]{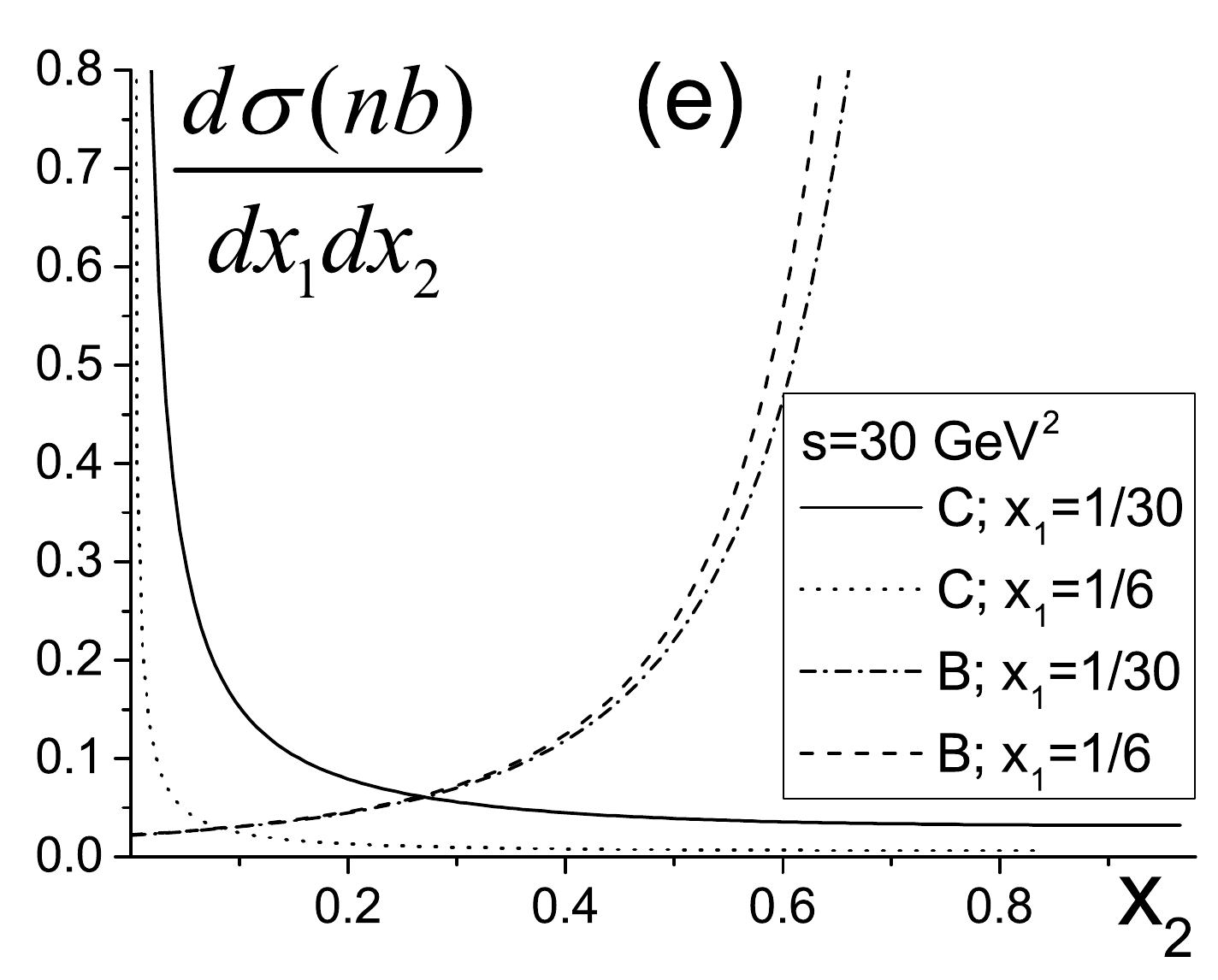}
\includegraphics[width=0.30\textwidth]{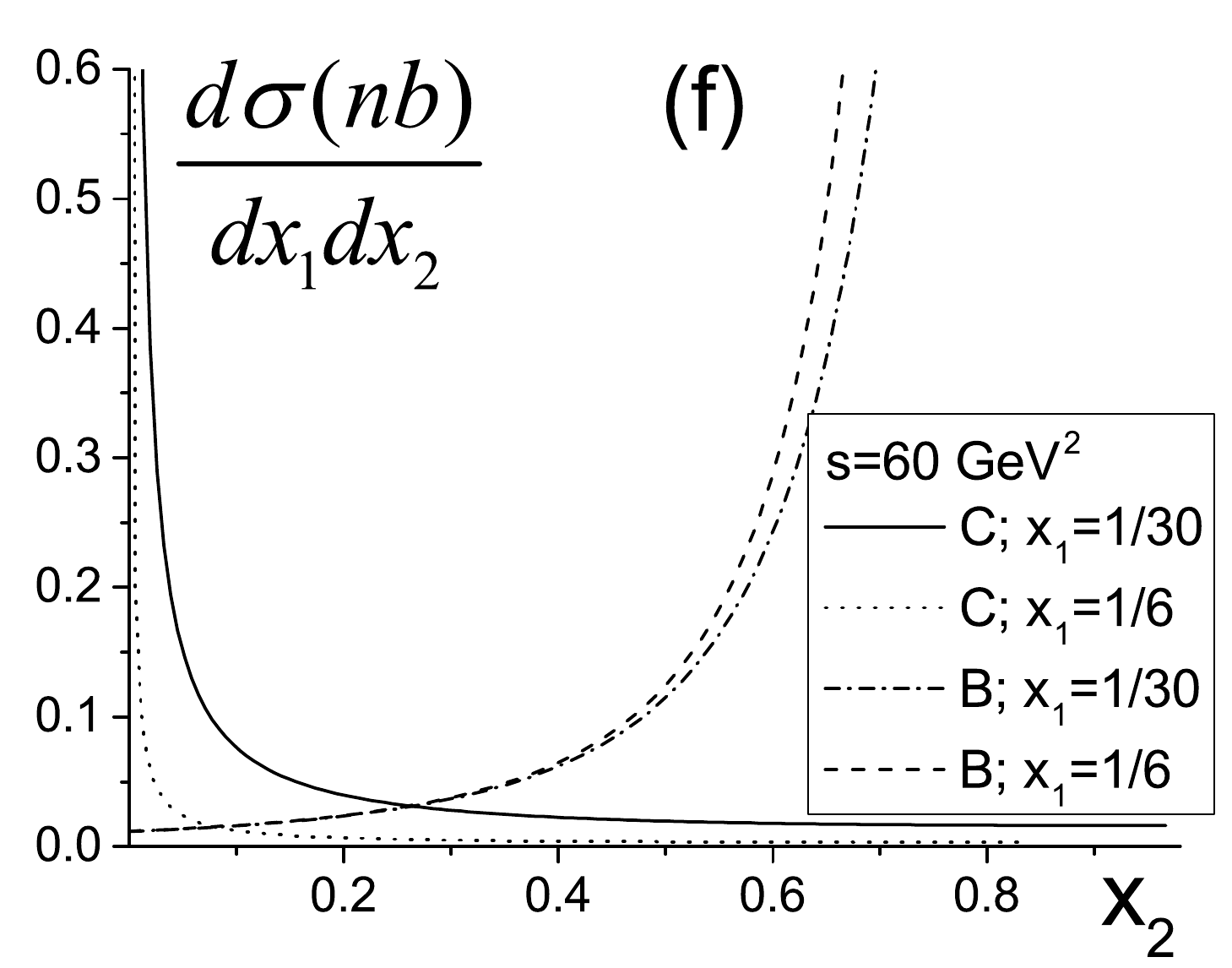}

\caption{Double differential cross section for the reaction
$\gamma+\mu^- \to e^+ e^- +\mu^-.$
In the first row it is given as a function of $x_1=s_1/s$ for fixed values of the variable $x_2=|u|/s$:
for the Compton-type diagrams $x_2=1/30$ (solid line) and $x_2=1/6$ (dotted  line),
for the Borsellino diagrams $x_2=1/30$ (dash-dotted line) and $x_2=1/6$ (dashed line);
and for different values of the total energy squared $s$:
$s=6$ GeV$^2$ (a), $s=30$ GeV$^2$ (b), and  $s=60$ GeV$^2$  (c);
- in the  second row it is given as a function of $x_2$ for fixed  values of the
variable $x_1$:
for the Compton-type diagrams $x_1=1/30$ (solid line) and $x_1=1/6$ (dotted  line),
for the Borsellino diagrams $x_1=1/30$ (dash-dotted line) and $x_1=1/6$ (dashed line),
and for different values of the total energy squared $s:$ \,
$s=6$ GeV$^2$ (a), $s=30$ GeV$^2$ (b), and $s=60$ GeV$^2$  (c).}
 \label{Fig:fig4}
 \end{figure}

To find the kinematical region where the signal over background
ratio is maximized, we analyze the double $(s_1, u)$
distribution separately for the Compton and Borsellino
contributions, using relations (\ref{Eq:eq15}) and
(\ref{Eq:eq16}), and the results are presented in
Fig.~\ref{Fig:fig4}, where we plot the corresponding double
differential cross sections and from which one can easily
determine the regions where the Compton contribution exceeds
the Borsellino one. We see that the contribution due to the
Compton-type diagrams increases with decrease of both variables, $s_1$ and $|u|$,
whereas the contribution of the Borsellino diagrams indicates
just the opposite behaviour. Since we have to scan
$s_1-$distribution, we can restrict the ($s_1,u$) region by
cutting the large values of $|u|$ in order
to reach our goal.

\begin{figure}
\includegraphics[width=0.32\textwidth]{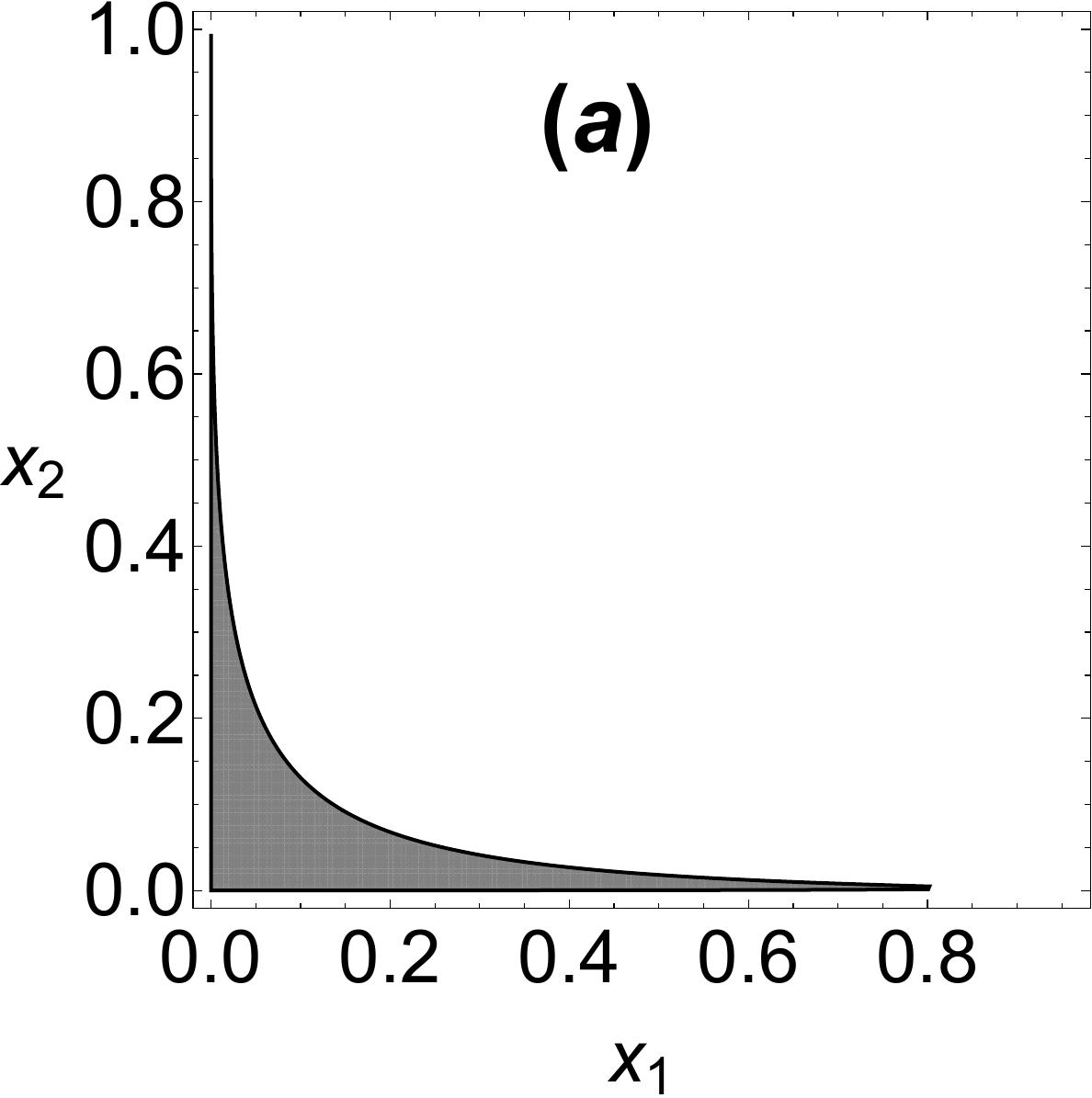}
\includegraphics[width=0.33\textwidth]{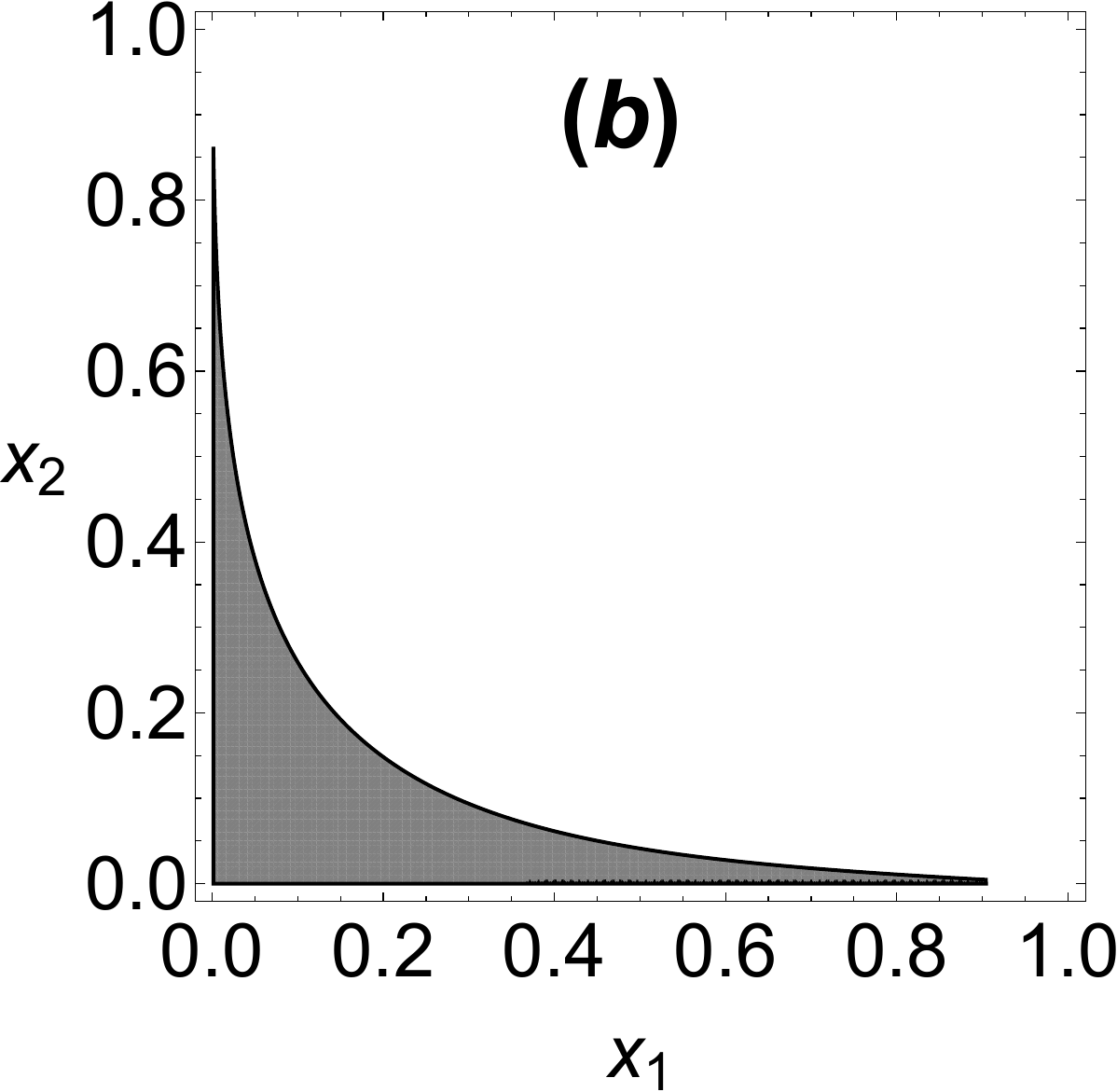}
\includegraphics[width=0.32\textwidth]{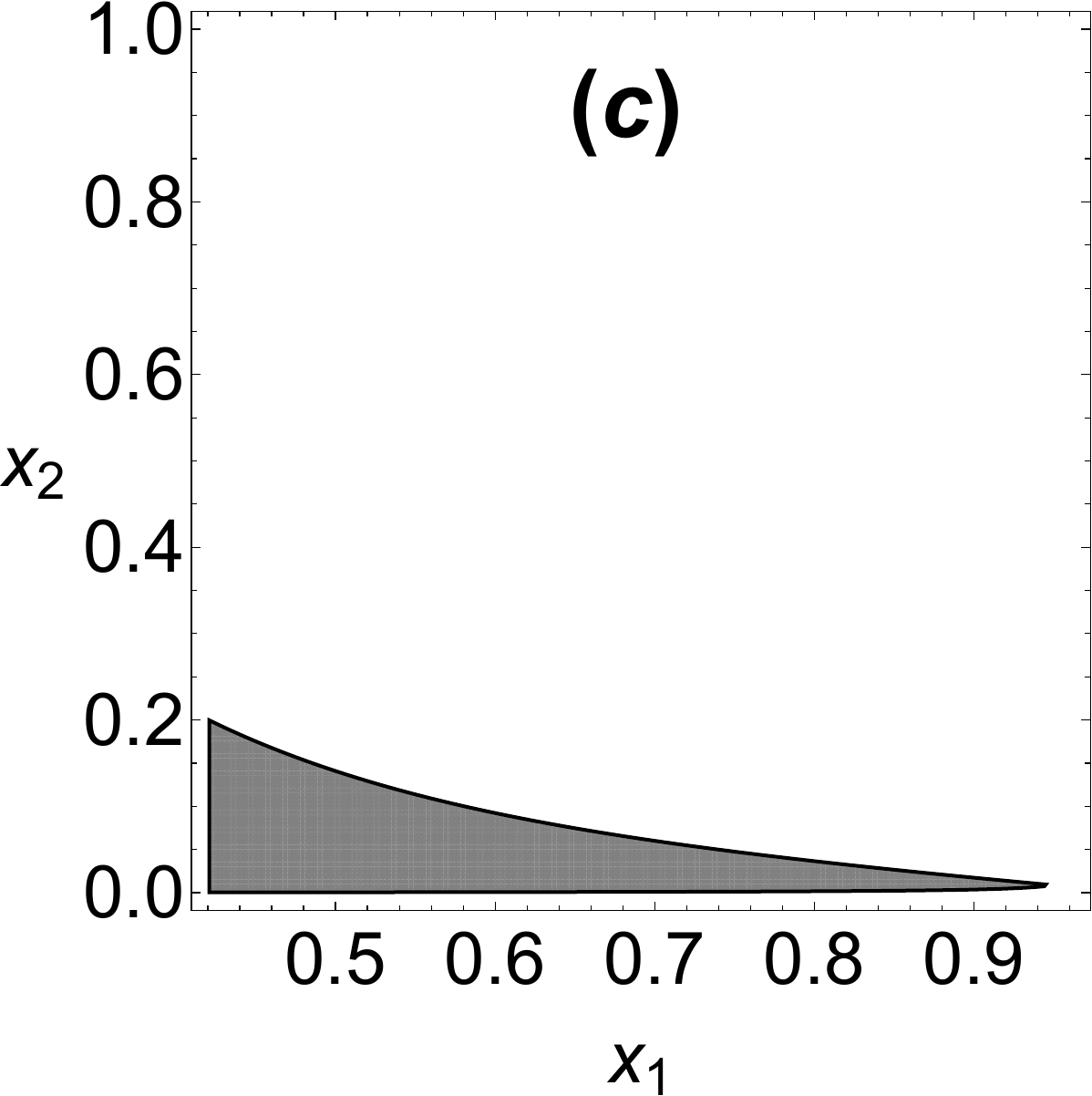}
\caption{Kinematical region where the Compton contribution to
the double differential cross section
exceeds the Borsellino contribution,  at $s$=30~GeV$^2$: (a) for the process $\gamma+\mu^- \to e^+ e^- +\mu^-,$ (b) for
the process $\gamma+e^-\to \mu^+ \mu^- + e^-,$ and (c) for the
processes $\gamma+e^-(\mu^-)\to \tau^+ \tau^- + e^-(\mu^-).$}
 \label{Fig:fig5}
\end{figure}

As one can see from the curves in Fig.~\ref{Fig:fig4}, there is
a large region of variables $u$ and $s_1$ where the Compton
contribution exceeds the Borsellino one, and it indicates that
measurements should be preferentially performed in this region to detect the DP
signal in form of a resonance in the single-photon intermediate
state. The corresponding regions for the different reactions are
shown in Fig.~\ref{Fig:fig5}.

\begin{figure}
%\captionstyle{flushleft}
\includegraphics[width=0.32\textwidth]{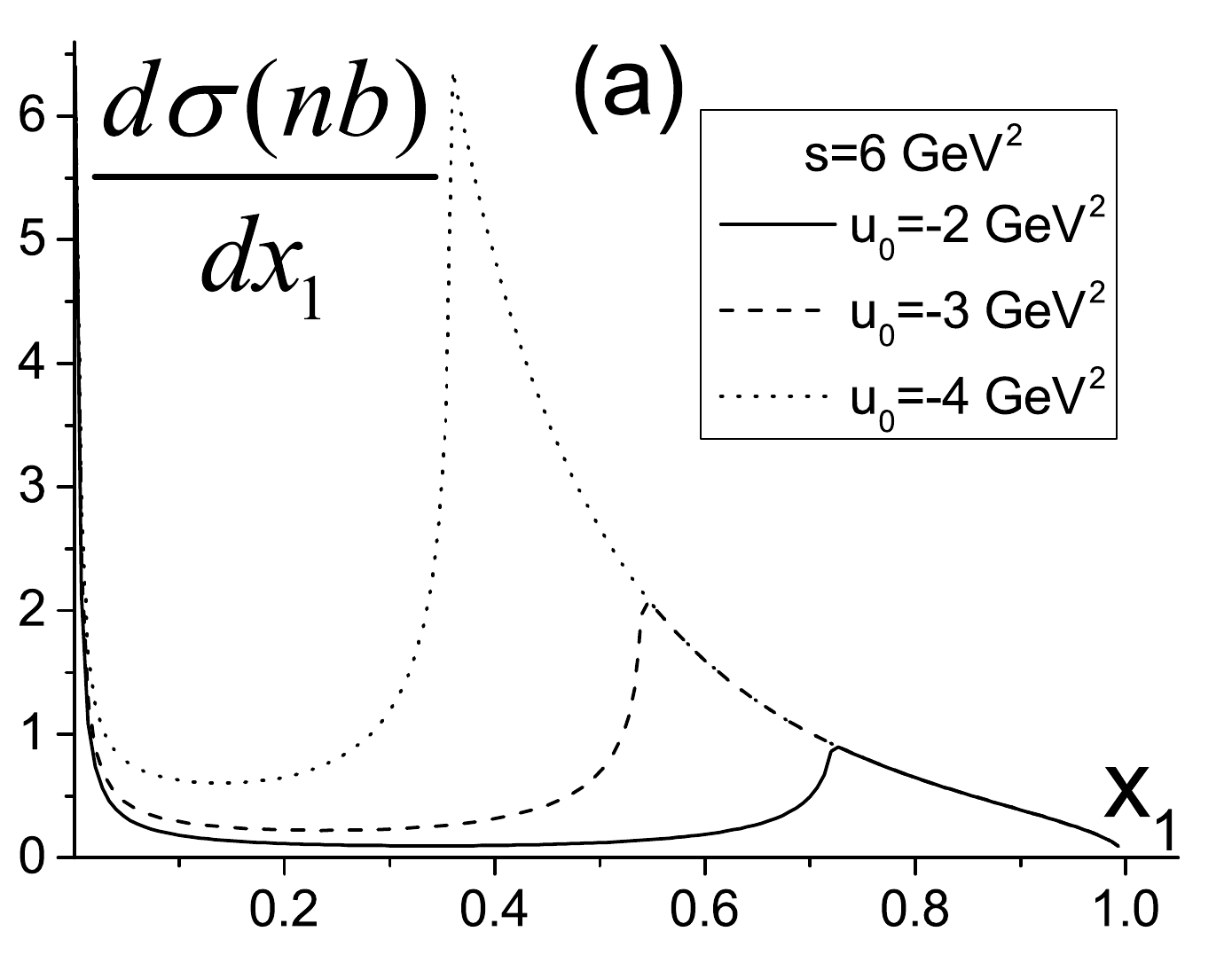}
\includegraphics[width=0.32\textwidth]{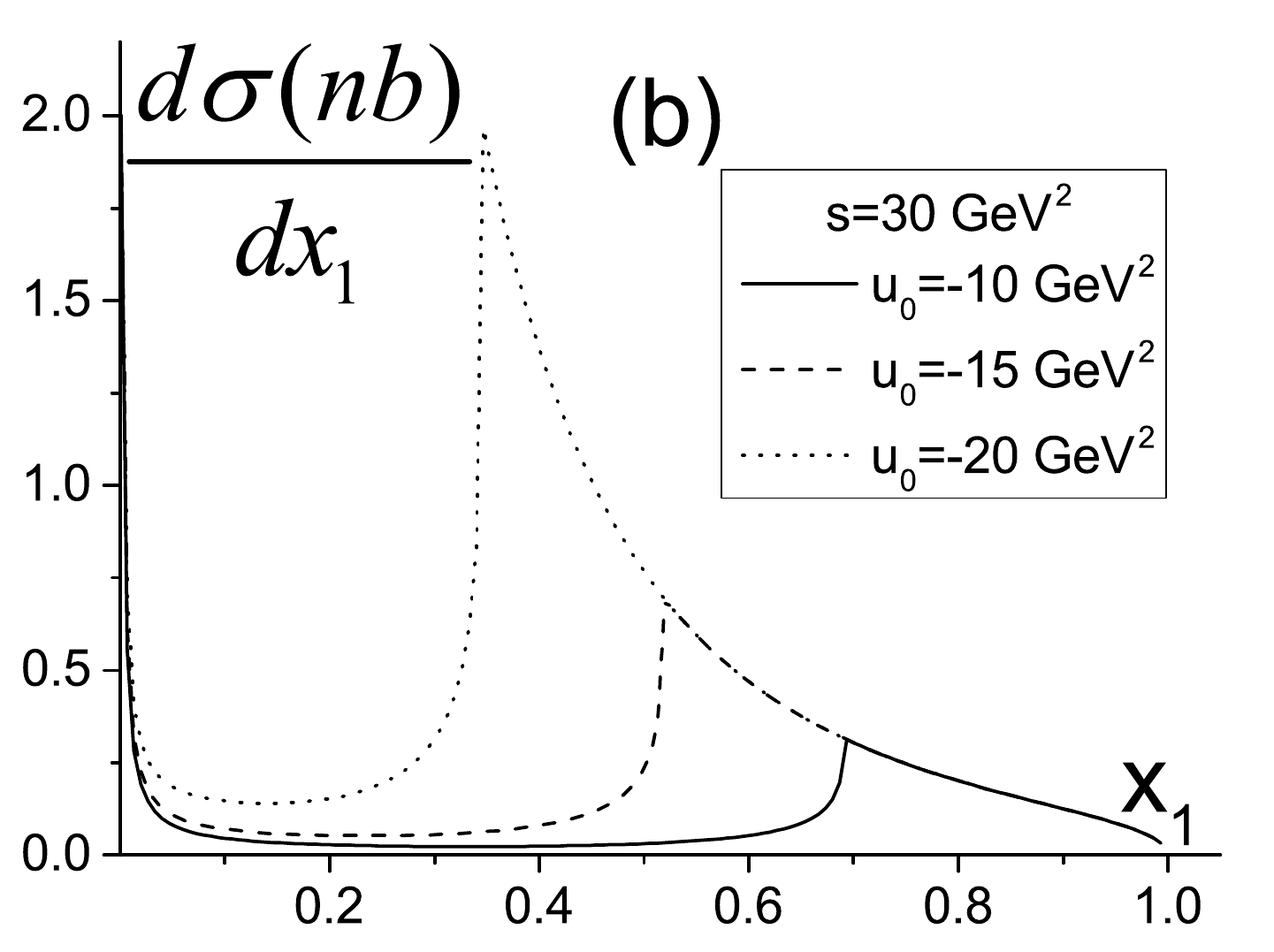}
\includegraphics[width=0.32\textwidth]{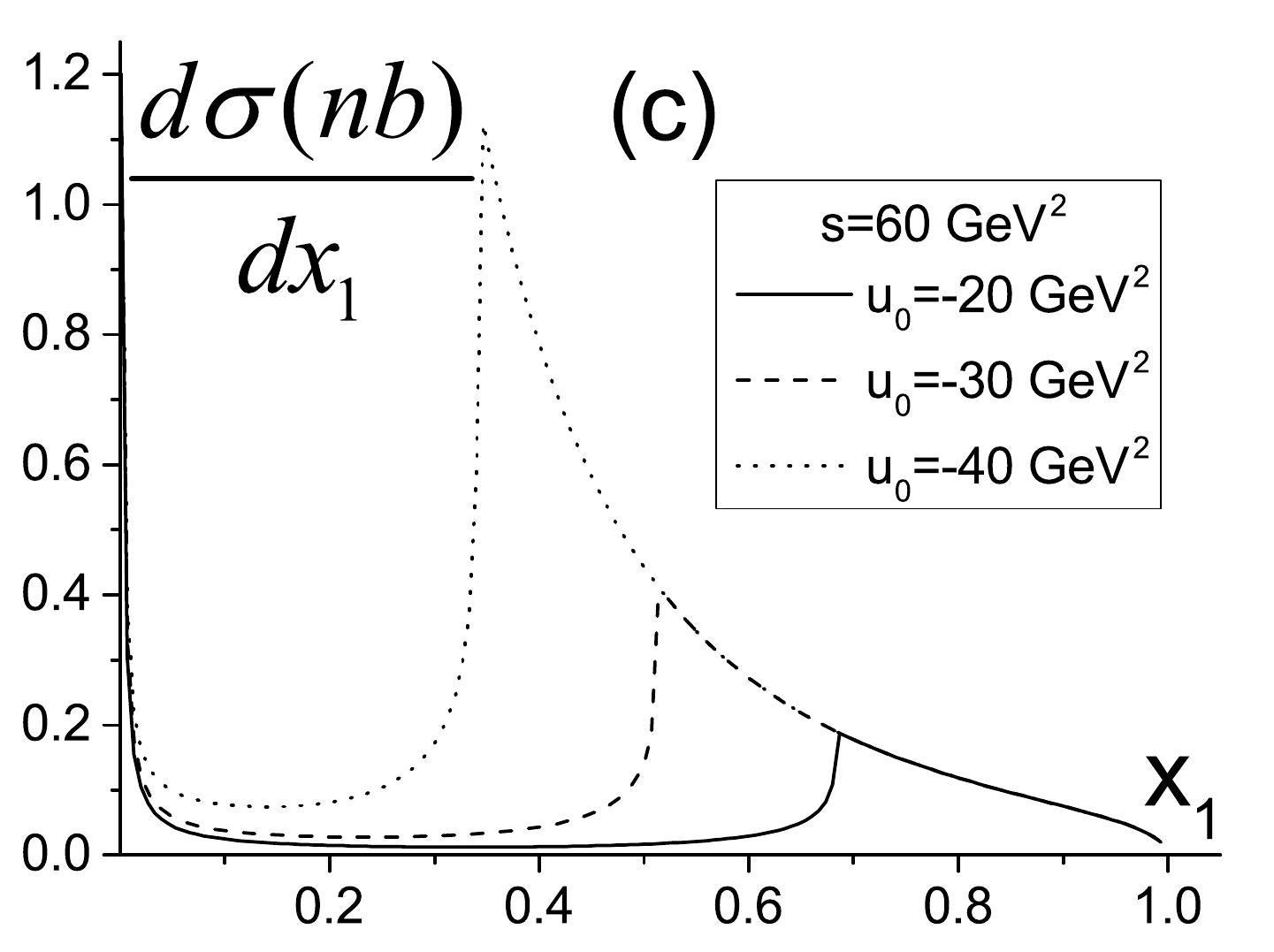}
\vspace{0.5cm}
\includegraphics[width=0.32\textwidth]{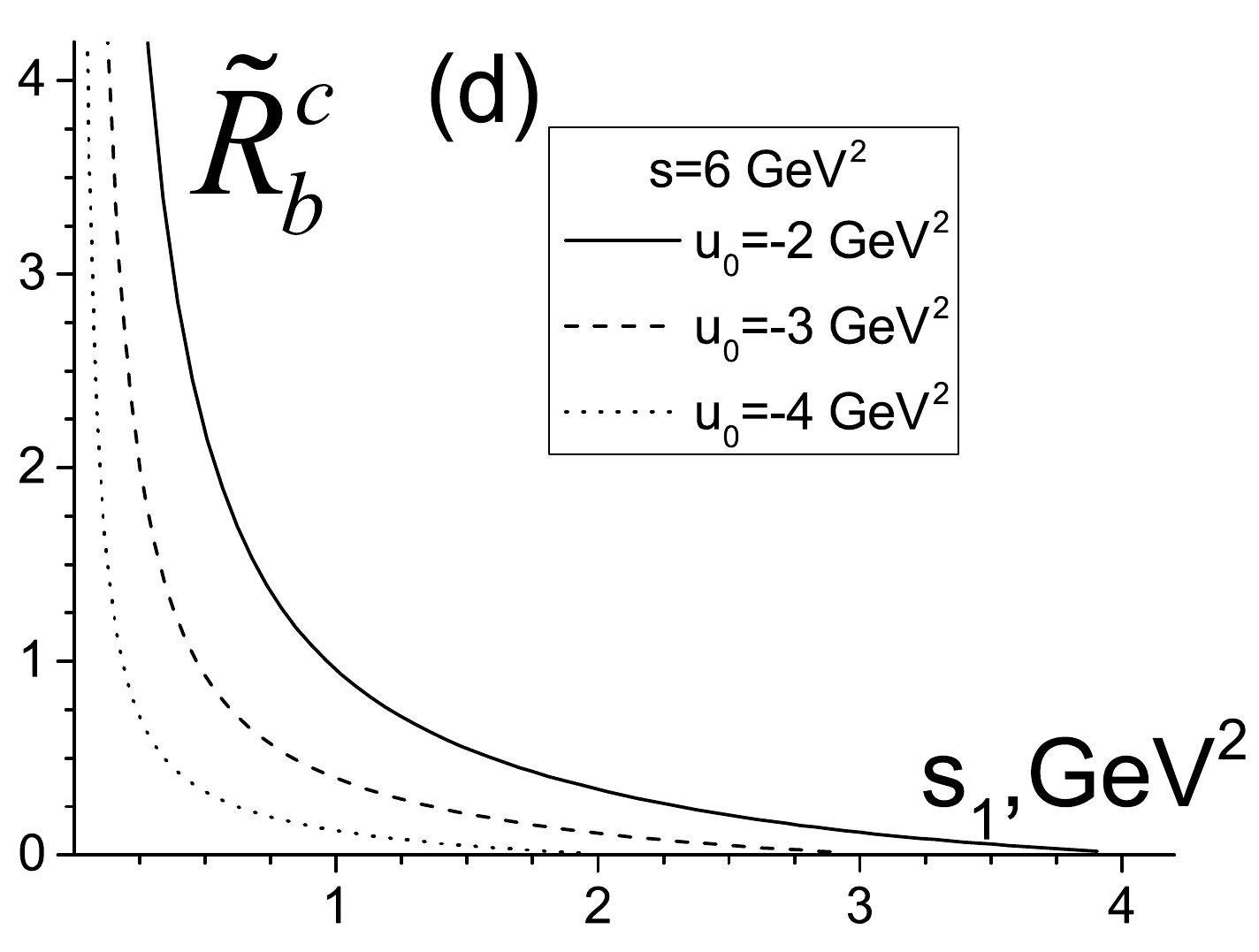}
\includegraphics[width=0.32\textwidth]{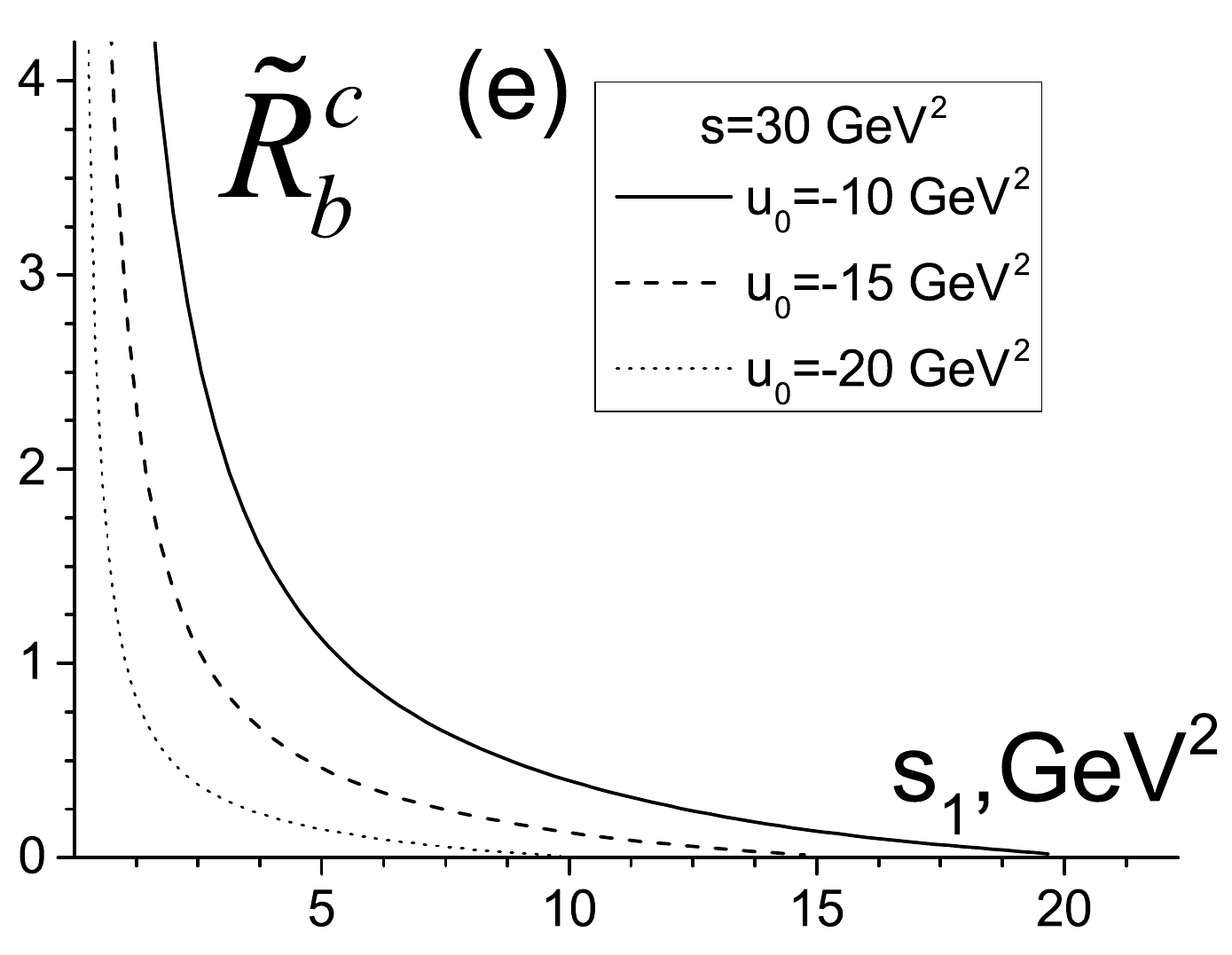}
\includegraphics[width=0.32\textwidth]{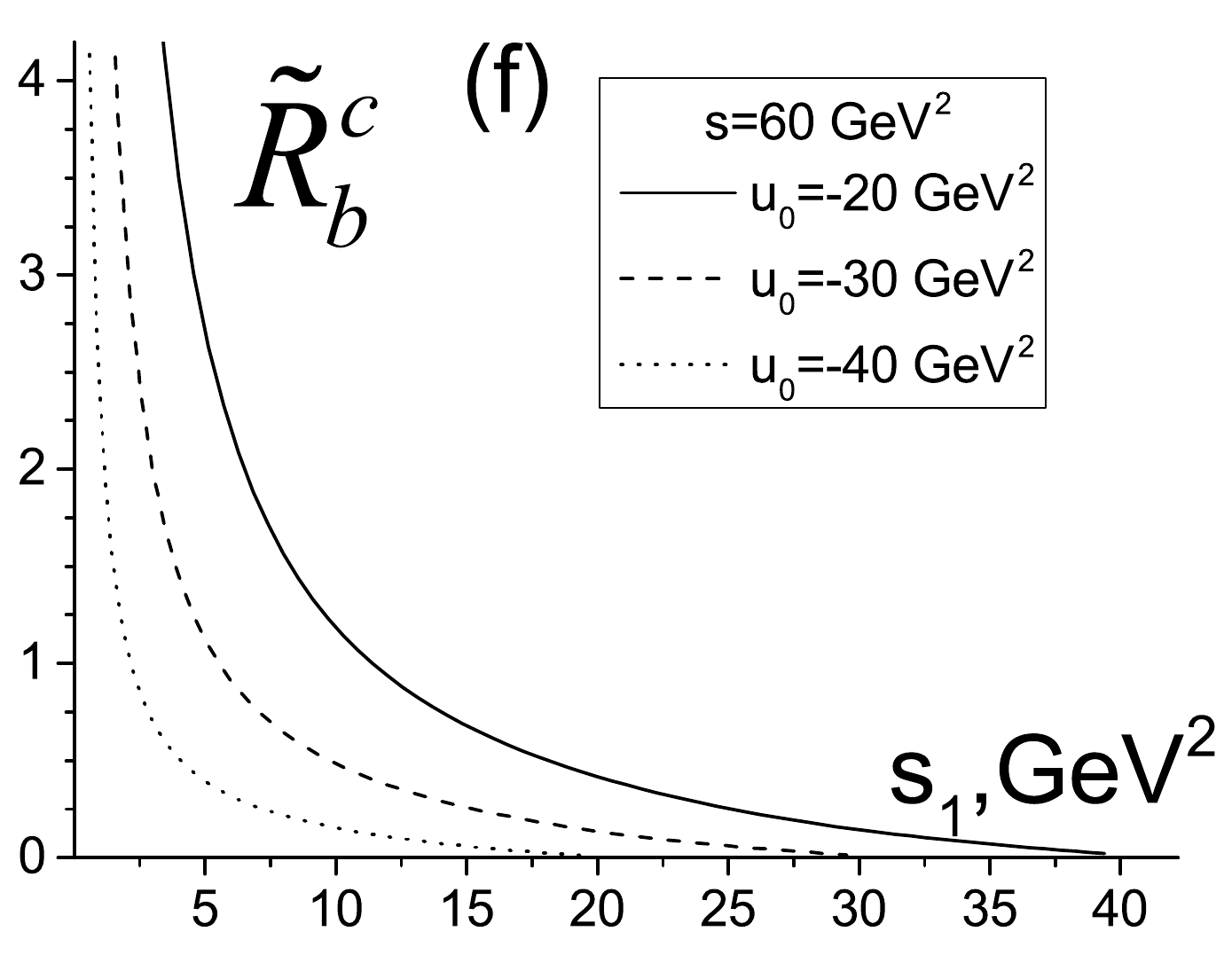}
\caption{Differential cross section of the process
$\gamma+\mu^- \to e^+ e^- +\mu^-$, as a function of the $e^+e^-$
invariant mass squared, for the sum of the Compton-type (\ref{Eq:eq18}) and the Borsellino
(\ref{Eq:eq19}) contributions, with the constraint
(\ref{Eq:eq23}), is shown for $s=6$ GeV$^2$ (a), $s=30$ GeV$^2$ (b), and  $s=60$ GeV$^2$  (c),
and for $u_0$=-2 GeV$^2$ (solid line),  $u_0$=-3 GeV$^2$ (dashed line)  and  $u_0$=-4 GeV$^2$ (dotted line).
The corresponding ratio of the Compton-type (\ref{Eq:eq18}) to the Borsellino contributions is shown in the insets (d-f).}
 \label{Fig:fig6}
\end{figure}

To reduce the Borsellino contribution, we suggest to remove the
events
with small values of the variable $|t_2|$ (or with large values
of $|u|$) by the kinematical
cut:
\be
\label{Eq:eq23}
u\, >  u_0,
\ee
where $u_0$ is a negative parameter that may take different values
in the numerical calculations below. We can perform the integration
over the variable $u$ and derive an analytical result for the
$s_1$ distribution even in the restricted phase space given by
Eq. (\ref{Eq:eq23}).
In this case, the region of the variables $u$ and $s_1$ is
shown in Fig.~\ref{Fig:fig2}, where the quantity $s_{10}$ is the
solution of the equation $u_0=u_-$ and reads
\be
\label{Eq:s10}
s_{10}=\frac{(M^4-s u_0)(2 M^2-s-u_0)}{(M^2-u_0)(s-M^2)}\,, \
u_0 <\tilde{u}\,, \ \tilde{u}=M
\left(\frac{M^2}{\sqrt{s}}+M-\sqrt{s}\right)\,.
\ee
Note that the solution given by Eq. (\ref{Eq:s10}) is the same also
for the equation $u_0=u_+$, with 
$$u_0 > \tilde{u}\,, \ \big[4 m^2 < s_1 < s_{10}\,, \ u_0 < u <
u_+\big]\,.$$
Two subregions can be delimited:
\be
\left[ 4 m^2 < s_1 < s_{10}, \, u_0 < u < u_+\right ]\,;
\ \left[s_{10} <s_1 <(\sqrt{s}-M)^2\,, \ u_- < u < u_+\right
]\,. \nn
\ee
The event selection, under  the constraint
(\ref{Eq:eq23}) (restricted phase space), decreases
essentially the Borsellino contribution, whereas
the Compton-type contribution decreases weakly. Their ratio
\be
\widetilde{R}^c_b=\frac{d\sigma_c(u >  u_0)}{d\sigma_b(u >  u_0)},
\ee
in the limited phase space  is
shown in Fig.\,\ref{Fig:fig6}, to be compared to the ratio $R^c_b$.

 The differential cross section of the
process $\gamma+\mu^- \to e^+ e^- +\mu^-$ is illustrated in Figs.\,\ref{Fig:fig6}(a-c)
taking into account all the contributions in the matrix
element squared (\ref{Eq:eq9}) and the constraint
(\ref{Eq:eq23}).  This cross section is plotted as a function of the
dimensionless variable $x_1$, for different values of $u_0$ .
Figure \ref{Fig:fig6} shows a steep decrease of this cross section, whereas $s$ and $u_0$ increase.
The ratio Compton-type contribution to the Borsellino one for the corresponding kinematics is shown
in Figs.\,\ref{Fig:fig6}(d-f). We apply the approach developed in Ref.~\cite{Gakh:2018ldx} (see also \cite{Bjorken:2009mm, Chiang:2016cyf})
to obtain the constraints on the possible values of DP parameters $\epsilon$ and $M_{A'}$ for a fixed event number
$N$ and standard deviation $\sigma.$
%%%%%%%%%%%%%%%%%%%%%%%%%%%%%%%%%%%%%%%%%%%%%%%%%%%%%%%%%%%%%%%%%%%%%%%%%%%%%%%%%%%%%%%%%%%%%%%%%
The effective QED and DP Lagrangian \cite{Bjorken:2009mm}
\begin{equation}\label{lagr}
\mathcal{L}=\mathcal{L}_{QED} +\displaystyle\frac{1}{2}\epsilon F^{\mu\nu} F'_{\mu\nu}-\displaystyle\frac{1}{4}F^{'\mu\nu} F'_{\mu\nu} +M^2_{A'}A^{'\mu}A'_{\mu}
\end{equation}
ensures the propagation of the fields $A_{\mu}$ and $A'_{\mu}$ and generates the interaction of the DP with the SM leptons in the form
$$e \epsilon \bar{\psi}_j(x)\gamma^{\mu} \psi_j(x) A'_{\mu}, \ \ j=e, \mu, \tau$$
due to the kinetic mixing effect [the second term in the rhs of Eq. (\ref{lagr})].
%%%%%%%%%%%%%%%%%%%%%%%%%%%%%%%%%%%%%%%%%%%%%%%%%%%%%%%%%%%%%%%%%%%%%%%%%%%%%%%%%%%%%%%%%%%%%%%%
We assume that selected values of the $\ell^+_j~\ell^-_j$ invariant mass fall in
the energy region
\be\label{bin}
 M_{A'}-\delta m/2 < \sqrt{s_1}  < M_{A'}+\delta m/2\,, \nn
\ee
where $\delta m$ is the experimental resolution for the invariant mass,
$i.e.$, the bin width containing the events of the possible signal.

The DP modifies the Compton-type matrix element by the spin-one particle Breit-Wigner propagator
in such a way that its contribution into the cross section reads
\be\label{sdp}
d\sigma_{A'}=\frac{\epsilon^2 s_1 [2(s_1-M^2_{A'})+\epsilon^2 s_1] d\sigma_c}{D(s_1)}\,, \ D(s_1)=(s_1-M^2_{A'})^2 + M^2_{A'} \Gamma^2\,,
\ee
where the DP width $\Gamma$ is defined by its decays to SM lepton pairs
\be
\label{Eq:eq20}
\Gamma (\gamma\,' \to
{\ell}^+{\ell}^-)=\epsilon^2\,\frac{\alpha}{3M_{A'}^2}(M_{A'}^2+2m_{\ell}^2)\sqrt{M_{A'}^2-4m_{\ell}^2}\,\,\Theta(M_{A'}-2
m_l)= \epsilon^2\,\Gamma_0,
\ee
where $m_{\ell}$ is the SM lepton mass and $\Theta(x)$ is the
Heaviside $\Theta$ function.
%%%%%%%%%%%%%%%%%%%%%%%%%%%%%%%%%%%%%%%%%%%%%%%%%%%%%%%%%%%%%%%%%%%%%%%%
The term $2(s_1-M^2_{A'})$ in the numerator of $d\sigma_{A'}$ in Eq. (\ref{sdp}) describes the QED-DP interference, whereas the term $\epsilon^2\,s_1$ corresponds to the pure DP contribution.
%%%%%%%%%%%%%%%%%%%%%%%%%%%%%%%%%%%%%%%%%%%%%%%%%%%%%%%%%%%%%%%%%%%%%%%%%%%%%%

In our numerical calculations, where 
$e^+ e^-$ or $\mu^+ \mu^-$ pairs are created, we restrict
ourselves to the analysis of a light DP signal with mass $M_{A'}<1$ GeV.
Then, the decay $A'\to \tau^+ \tau^-$ is forbidden
and, therefore, $m_{\ell}$ in (\ref{Eq:eq20}) is the electron or the
muon mass.

Bearing in mind that
$$\Gamma \ll \delta m \ll M_{A'}\,,$$
we can apply the narrow resonance approximation and write:
\be
\left [D(s_1)\right
]^{-1}=\frac{\pi}{M_{A'}\,\epsilon^2\Gamma_0}\delta(s_1-M_{A'}^2)\,.
\nn
\ee
Applying \cite{Chiang:2016cyf, Gakh:2018ldx}:
\be
\label{Eq:eq26}
\sigma d\,\sigma_Q=\sqrt{N} d\,\sigma_{A'}\,,
\ee
one obtains, by integrating both sides over the bin range 
\be
\label{Eq:eq27}
\epsilon^2 =\frac{2\,\sigma}{\pi\sqrt{N}}\frac{\delta
m\Gamma_0}{M_{A'}^2}\frac{d\sigma_Q (M_{A'}^2)}{d\sigma_c
(M_{A'}^2)}\,,
\ee
where $d\sigma_Q$ is the total QED contribution.

Equation (\ref{Eq:eq27})  defines the constraints on the parameters
$\epsilon^2, M_{A'}^2$ and the number of the detected events
$N$ for a given
standard deviation $\sigma.$
%%%%%%%%%%%%%%%%%%%%%%%%%%%%%%%%%%%%%%%%%%%%%%%%%%%%%%%%%%%%%%%%%%%%%%%%
The effect due to the  QED-DP interference relative to the pure DP contribution can be easily estimated
by integrating  Eq.  (\ref{sdp}) at $d\sigma(s_1=M^2_{A'})$ over the above described bin range,
giving 
$$\frac{4 \delta m \Gamma_0}{\pi M^2_{A'}} \ll 1 .$$
%%%%%%%%%%%%%%%%%%%%%%%%%%%%%%%%%%%%%%%%%%%%%%%%%%%%%%%%%%%%%%%%%%%

\begin{figure}
%\captionstyle{flushleft}
\includegraphics[width=0.45\textwidth]{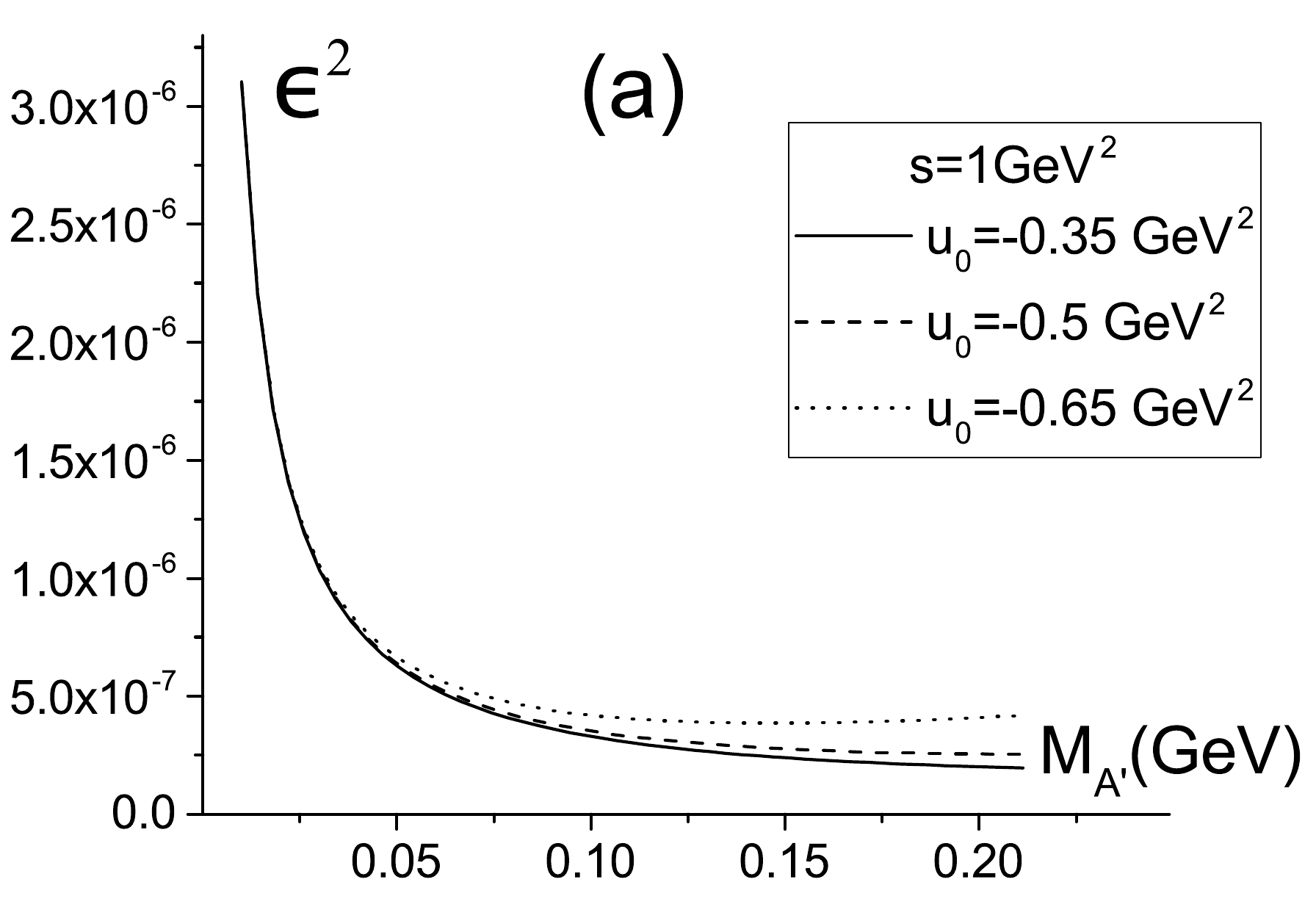}
\includegraphics[width=0.45\textwidth]{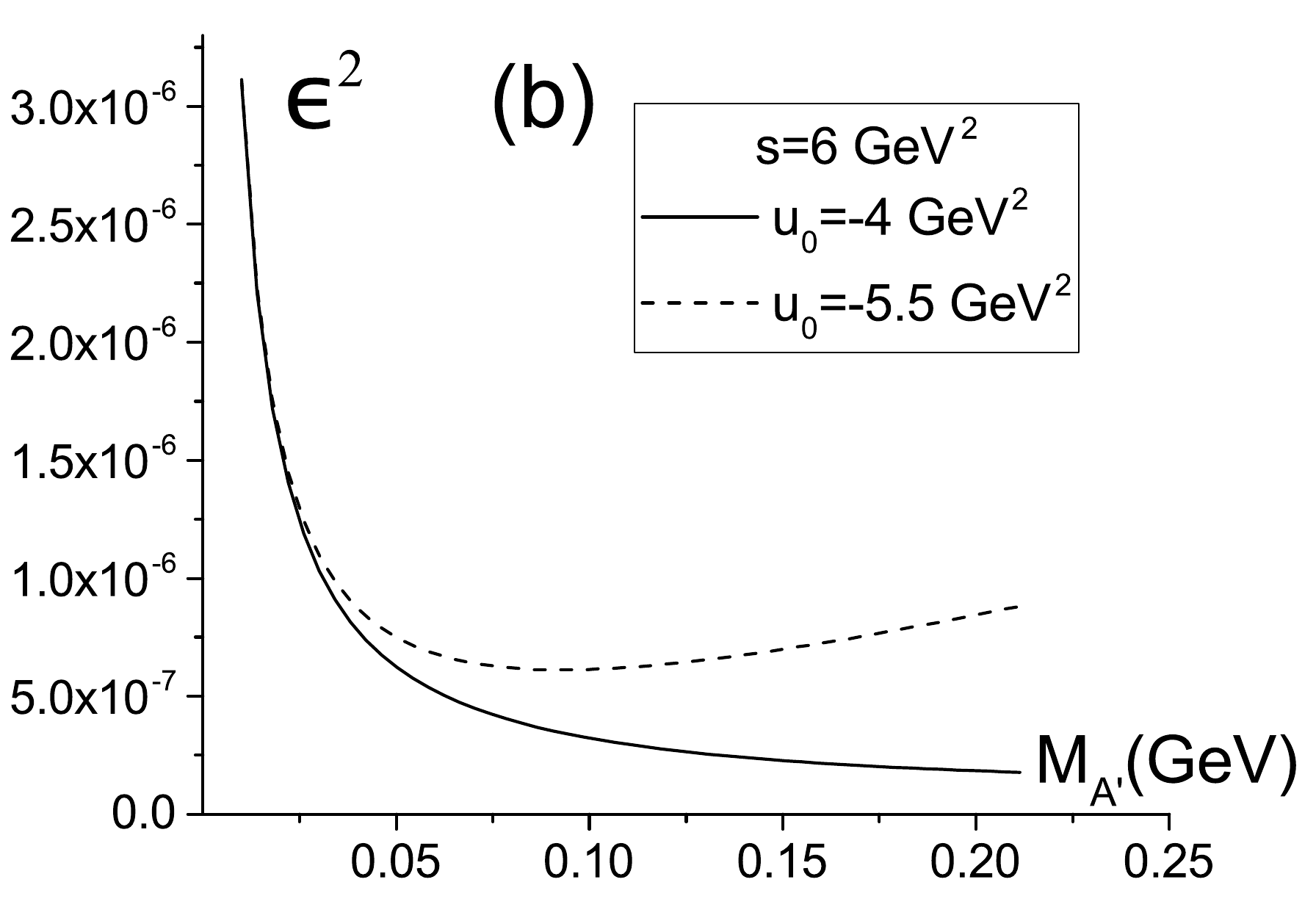}

\vspace{0.5cm}

\includegraphics[width=0.45\textwidth]{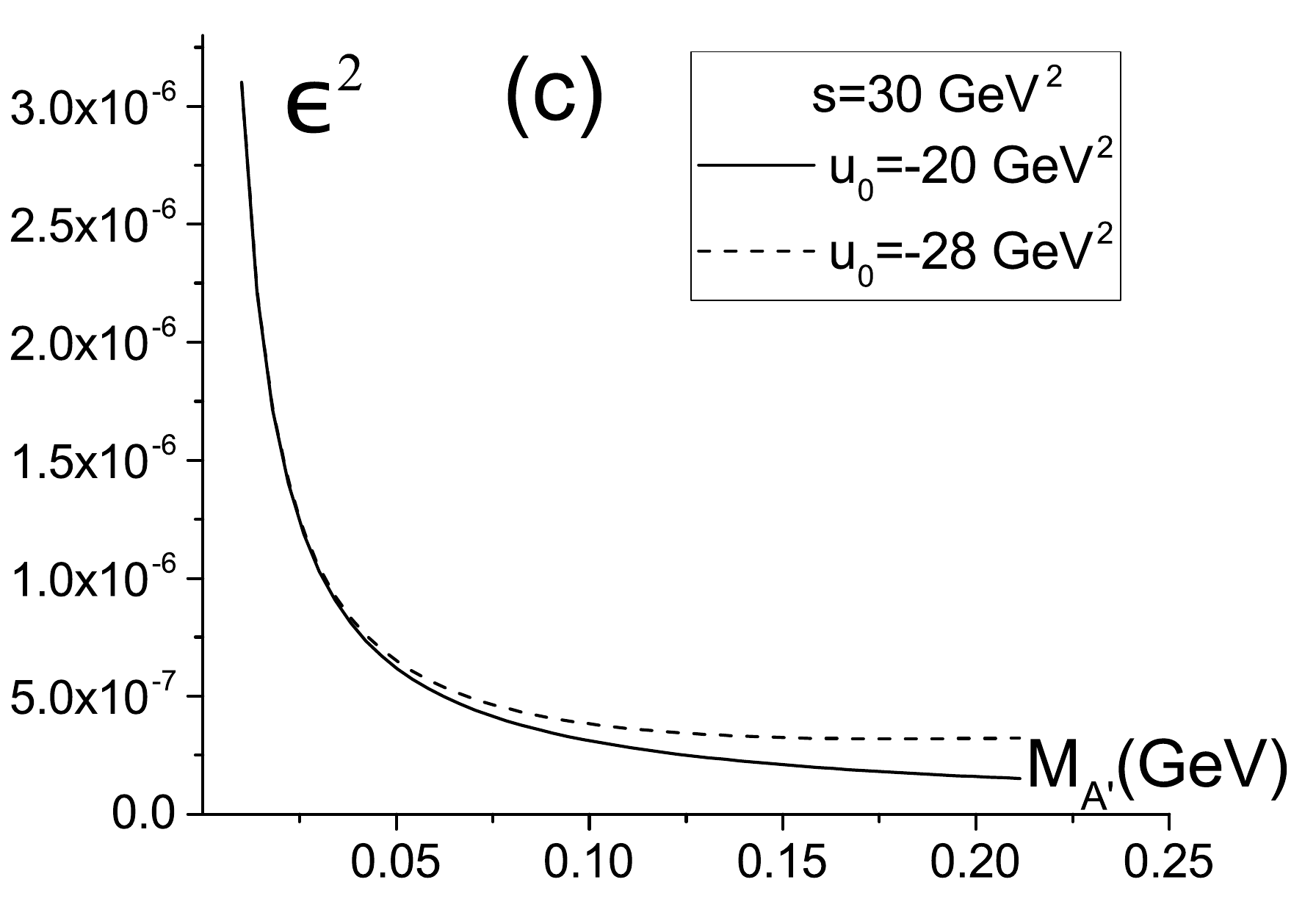}
\includegraphics[width=0.45\textwidth]{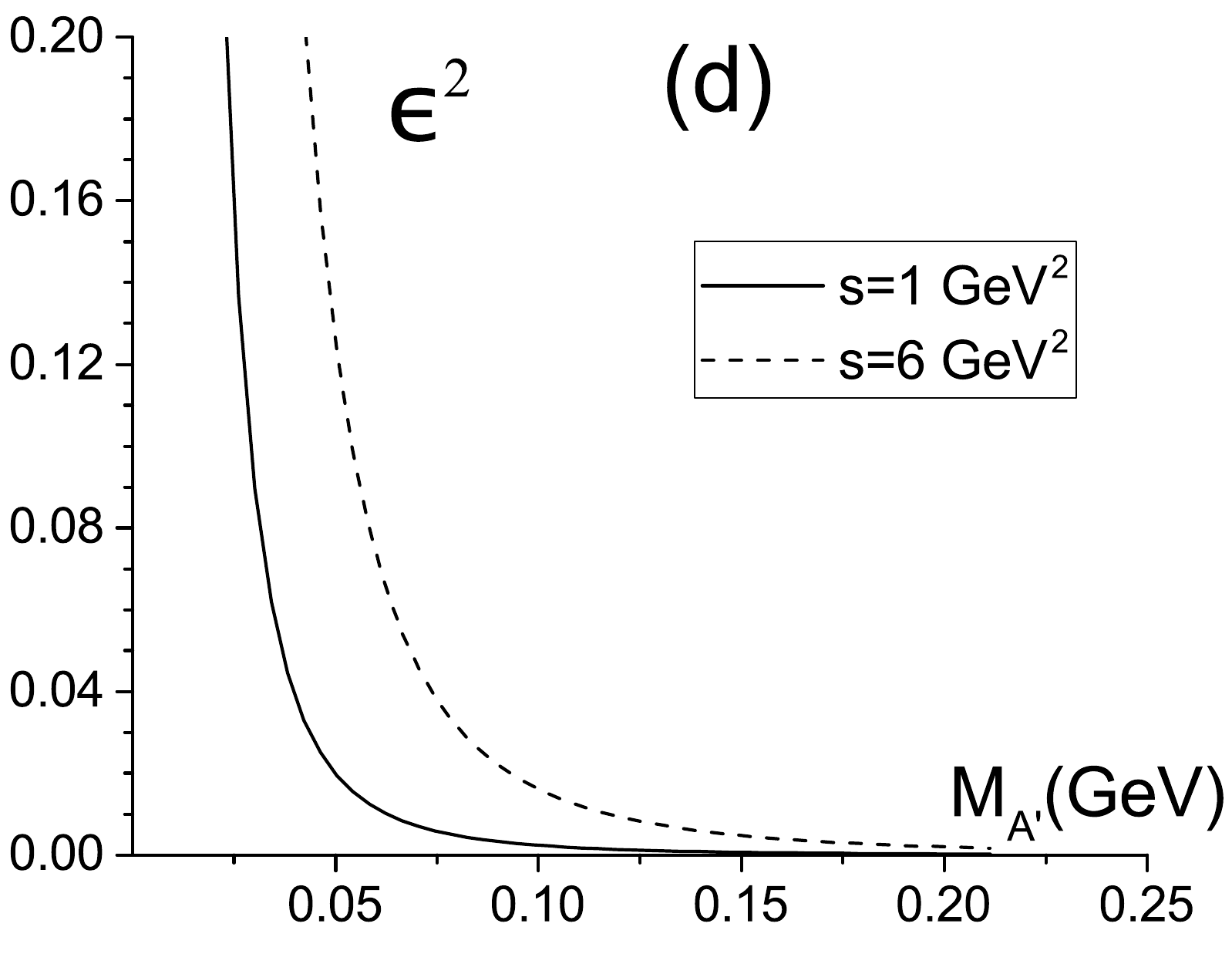}
\caption{ Constraints on the DP parameters
in terms of $\epsilon^2$ as a function of $M_{A'}$ for the conditions:
$N=10^{4}$, $\sigma=2$, and $\delta m = 1 $ MeV, in  the process
$\gamma +\mu^-\to e^+ e^- +\mu^- $,   for $M_{A'}<2 M$,  where only
one channel of DP decay is allowed,  for $s=1$ GeV$^2$ (a), $s=6$ GeV$^2$ (b), and $s=30$ GeV$^2$  (c).
Different curves correspond to different values of $u_0$, i.e., different kinematical cuts, and
represent the lower limit on the DP parameters in the case of no DP event detected.
Inset (d) shows the  ($\epsilon^2$,\,$M_{A'})$ dependence for $s=1$ GeV$^2$ (solid line) and $s=6$ GeV$^2$ (dashed line)
 without kinematical cuts.}
 \label{Fig:fig7}
\end{figure}

\begin{figure}
%\captionstyle{flushleft}
\includegraphics[width=0.45\textwidth]{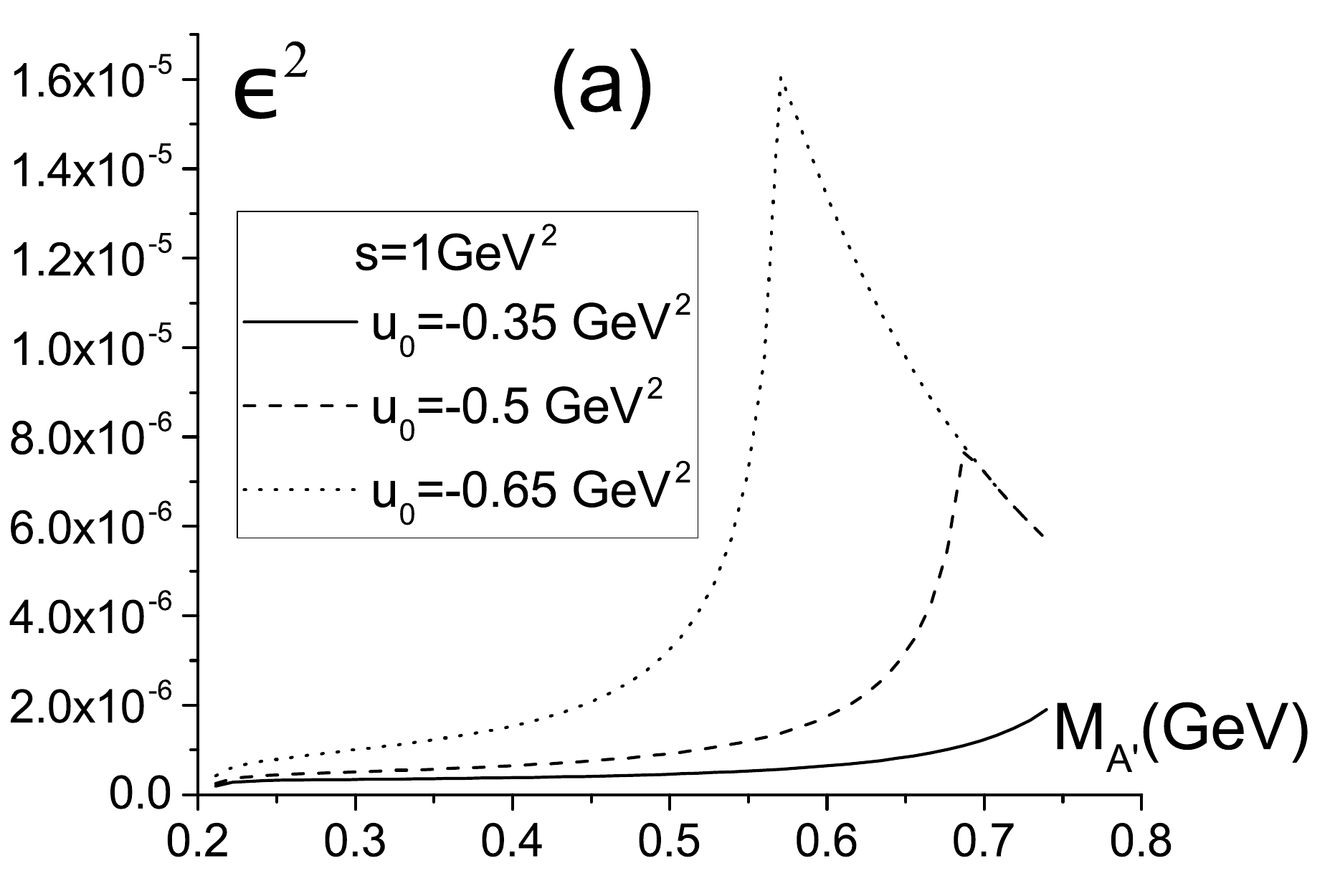}
\includegraphics[width=0.45\textwidth]{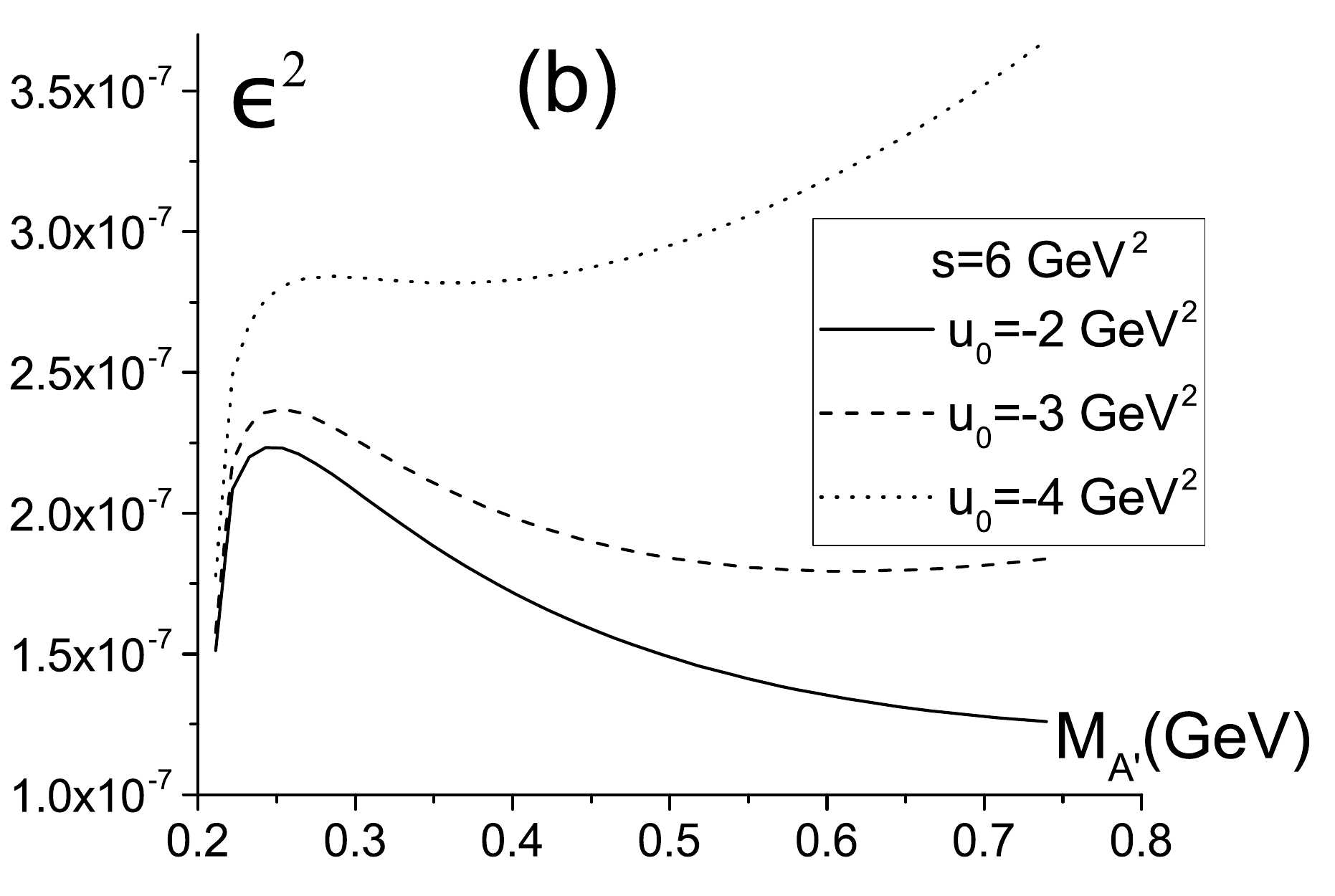}

\vspace{0.5cm}

\includegraphics[width=0.45\textwidth]{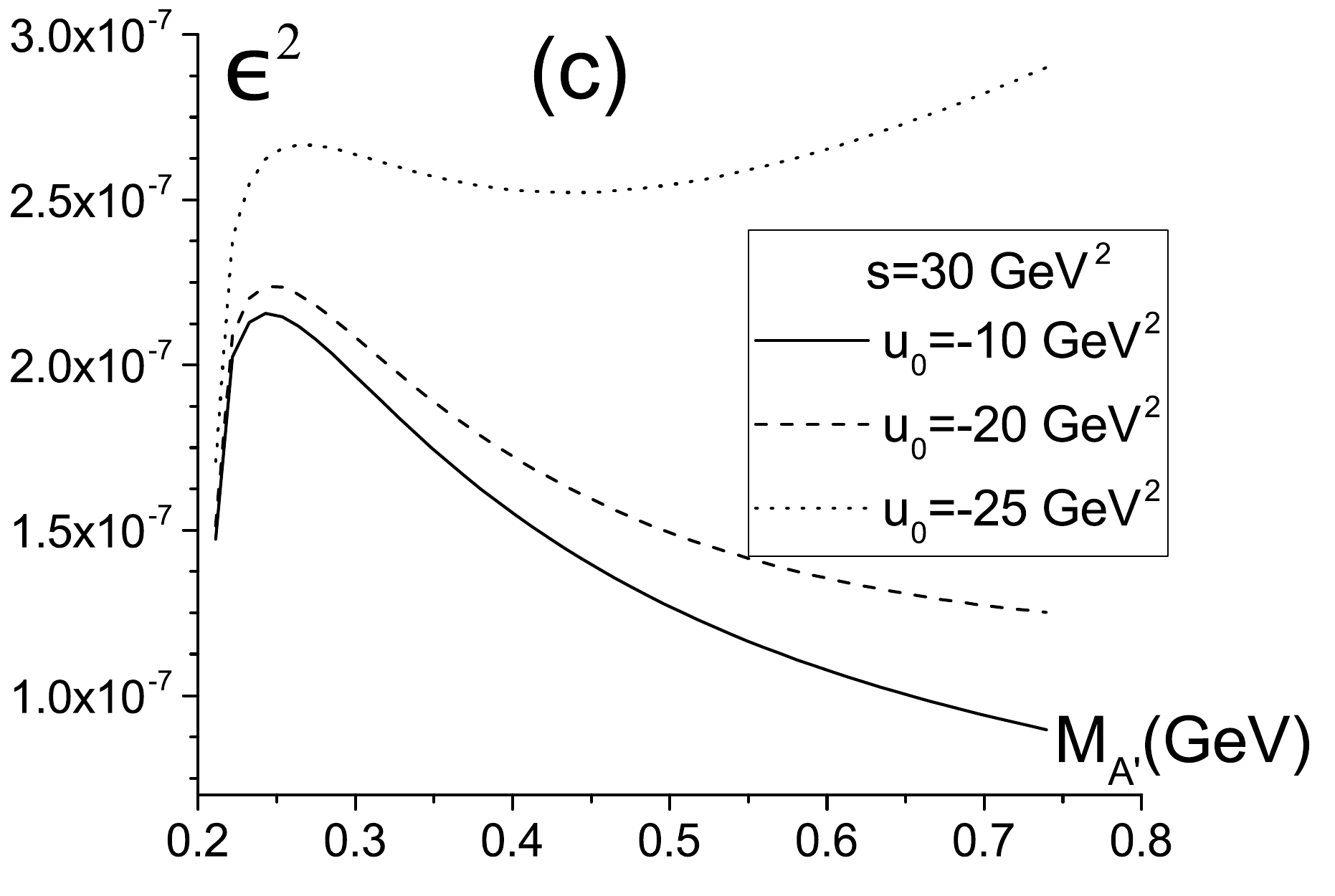}
\includegraphics[width=0.45\textwidth]{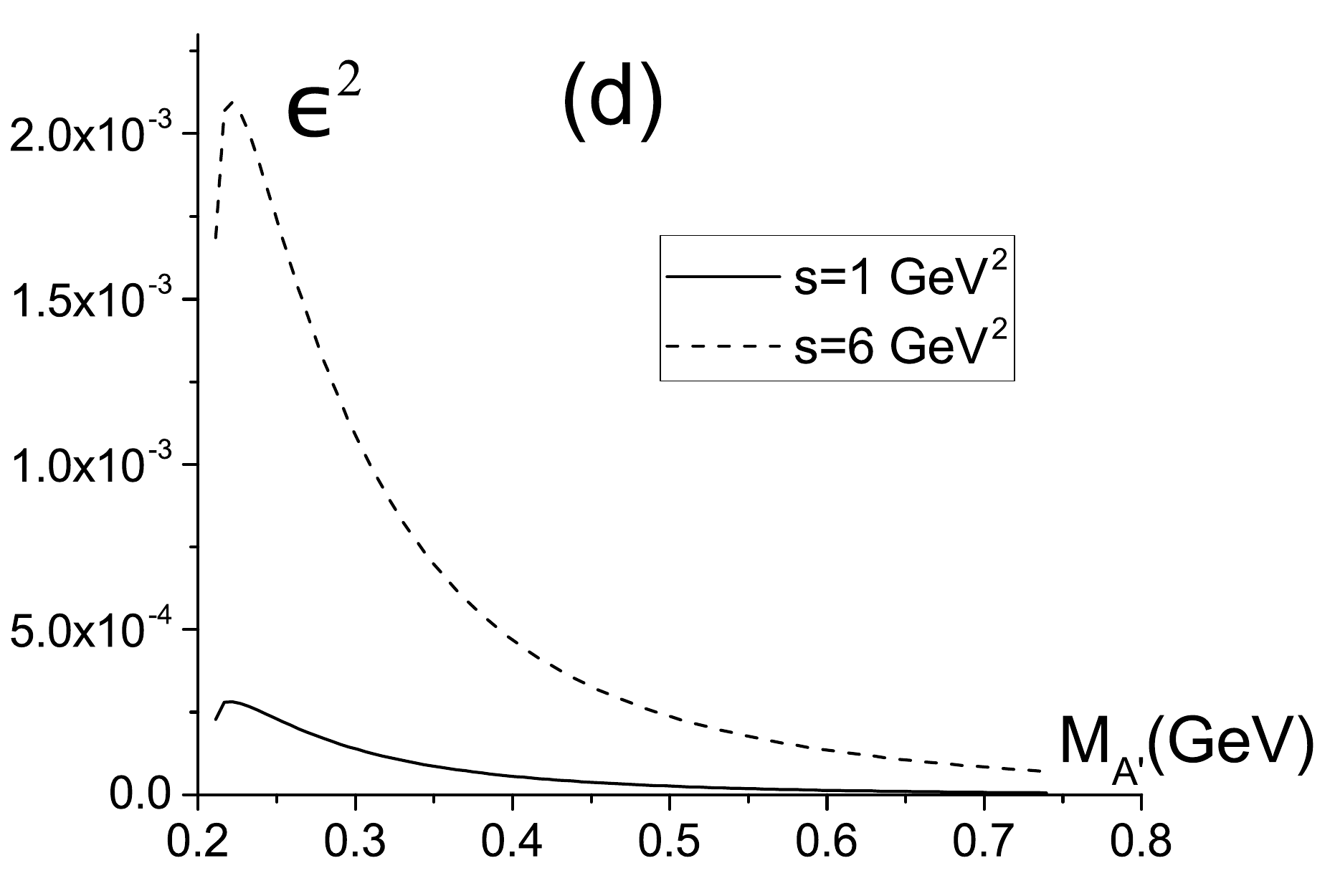}
\caption{The same as in Fig.\ref{Fig:fig7} but in the region of
greater DP masses, $M_{A'} > 2 M$ where two channels are open for DP decays.}
 \label{Fig:fig8}
\end{figure}

\begin{figure}
%\captionstyle{flushleft}
\includegraphics[width=0.45\textwidth]{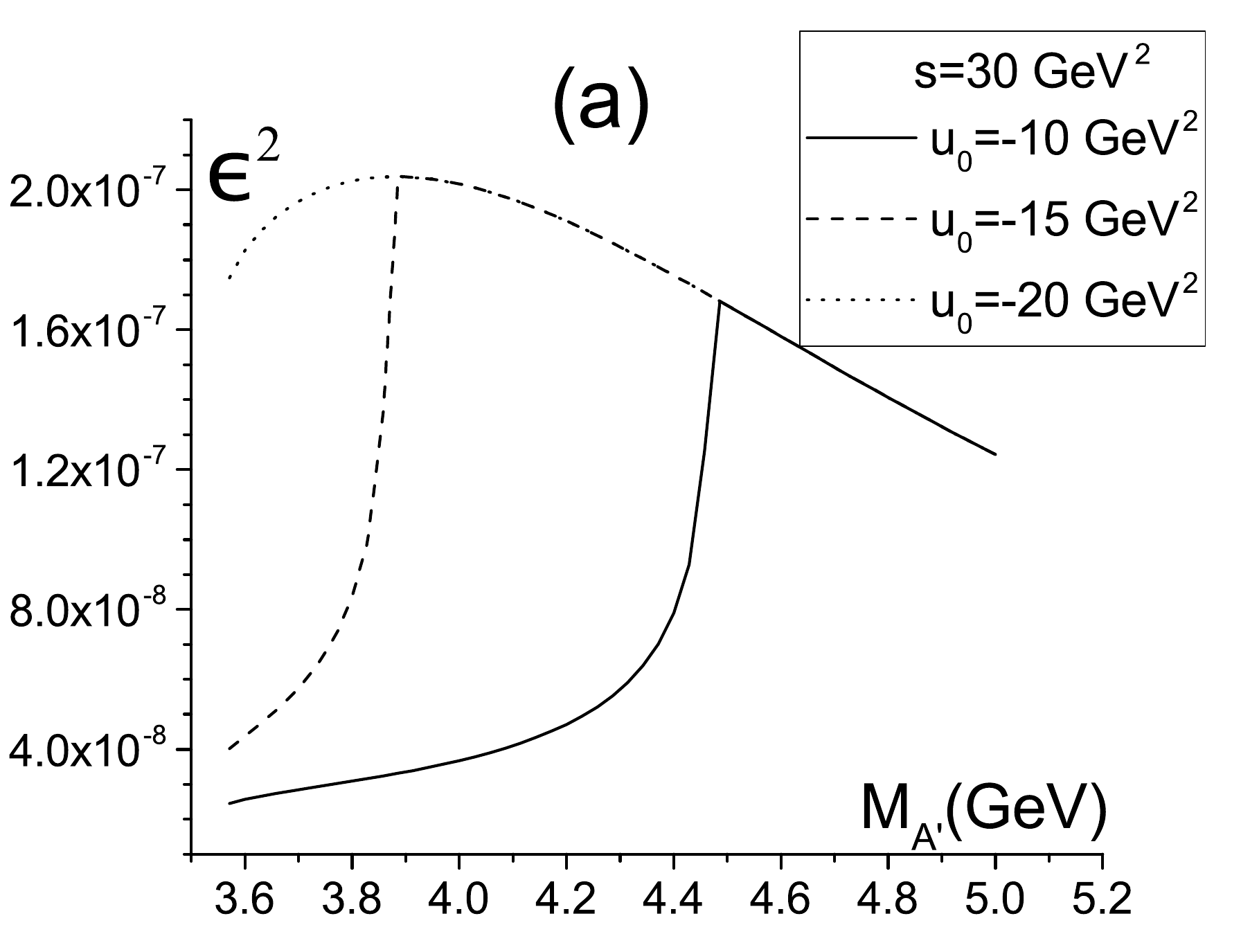}
\includegraphics[width=0.45\textwidth]{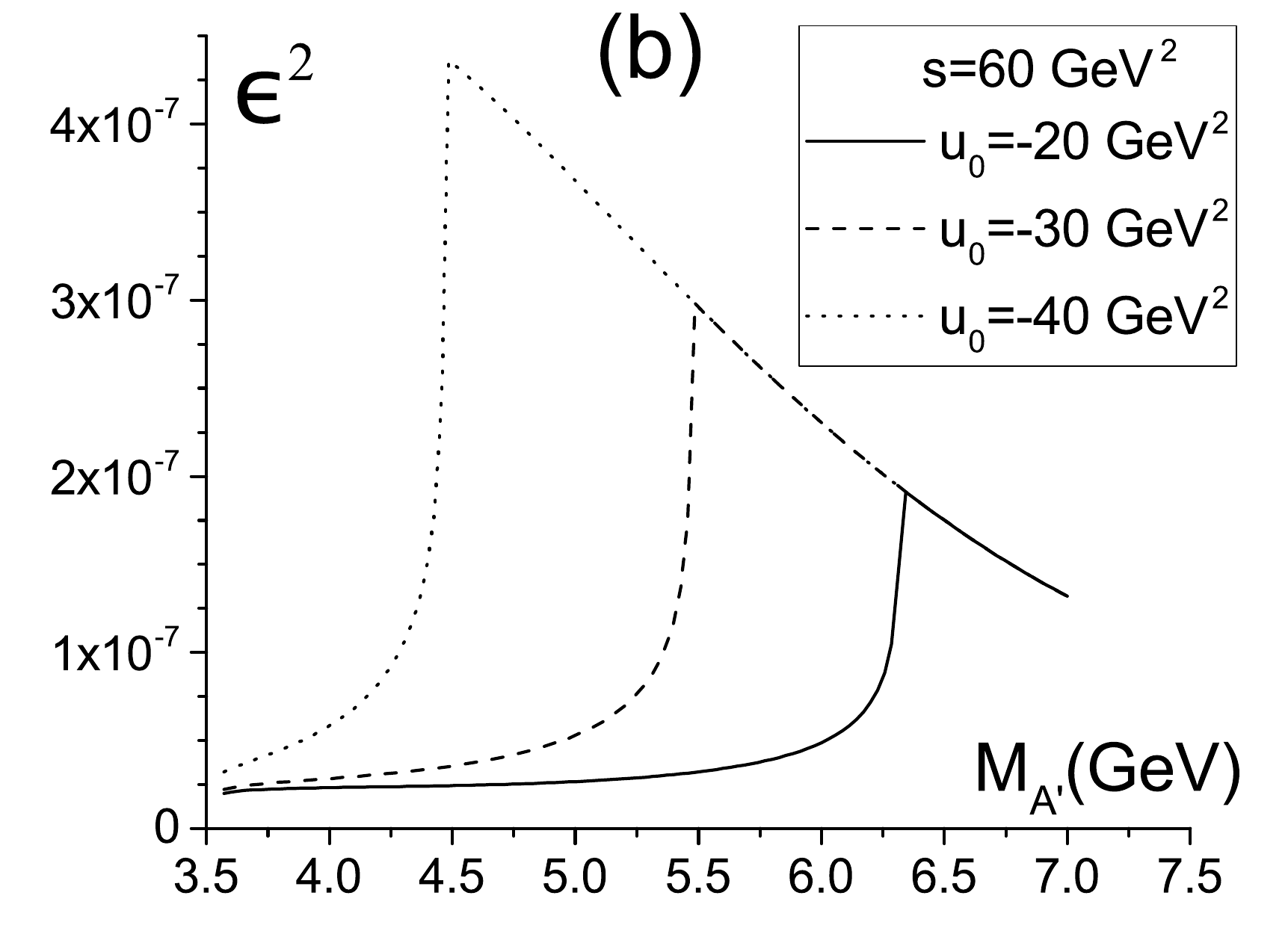}

\caption{$\epsilon^2$ versus $M_{A'}$ for the reaction $\gamma +\mu^-\to
\tau^+ \tau^- +\mu^-$, for $s=30$ GeV$^2$ (a) and $s=60$ GeV$^2$ (b).
Different curves correspond to different values of $u_0$, i.e., different kinematical cuts.}
 \label{Fig:fig9}
\end{figure}

These constraints  are illustrated below (Figs.~\ref{Fig:fig7}-\ref{Fig:fig9}) 
at $\sigma=2$ for the energy bin value $\delta m= 1$~MeV. In our
calculations we assume that there are three channels of the
DP decays into SM leptons: $A'\to e^+e^-$, 
$\mu^+\mu^-$, $\tau^+\tau^-$. The plots of $\epsilon^2$ versus $M_{A'}$  for the reaction $\gamma + \mu^- \to e^+ e^- + \mu^-$
are shown in Figs.~\ref{Fig:fig7}, and \ref{Fig:fig8} for $M_{A'} <
2 M$ and $M_{A'} > 2 M$, correspondingly. In the case of
$e^+ e^-$ pair production we restrict ourselves with the DP
 mass values $M_{A'}< 2 m_{\tau} \ (m_\tau$ is the $\tau$
lepton mass) and, therefore, take into account DP decays into
$e^+e^-$ and $\mu^+\mu^-$ when calculating the 
quantity $\Gamma_0.$ The plots for the 
$\gamma + \mu^- \to \tau^+\,\tau^- + \mu^-$ process are shown in
Fig.~\ref{Fig:fig9}. In this case, all three channels of DP
decays contribute into $\Gamma_0.$

We can conclude that the kinematical limits to reduce the
contribution of the Borsellino diagrams in the cross section increase essentially the
sensitivity to the DP signal and should be implemented in the experimental event selection.

\begin{figure}
%\captionstyle{flushleft}
\includegraphics[width=0.32\textwidth]{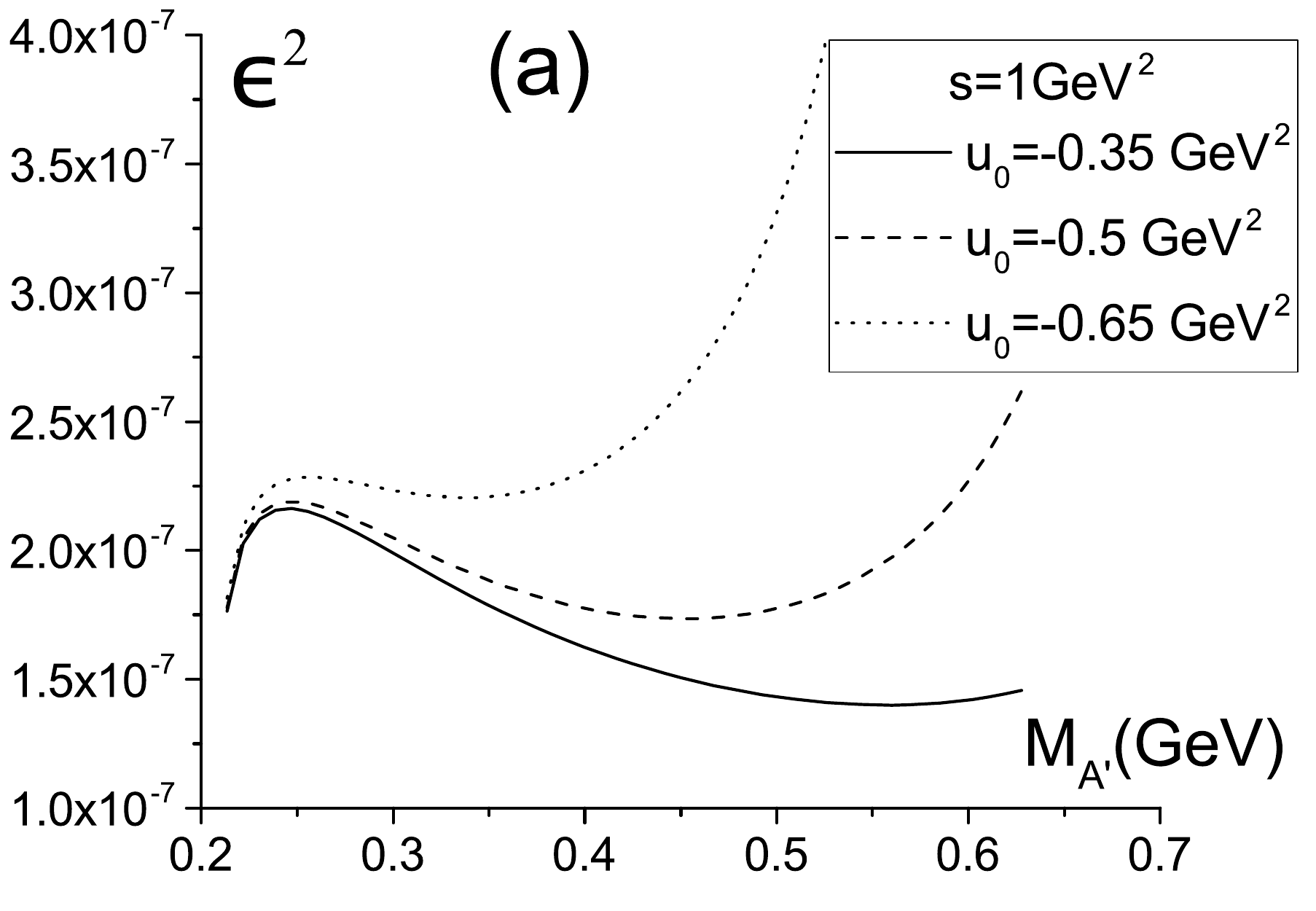}
\includegraphics[width=0.32\textwidth]{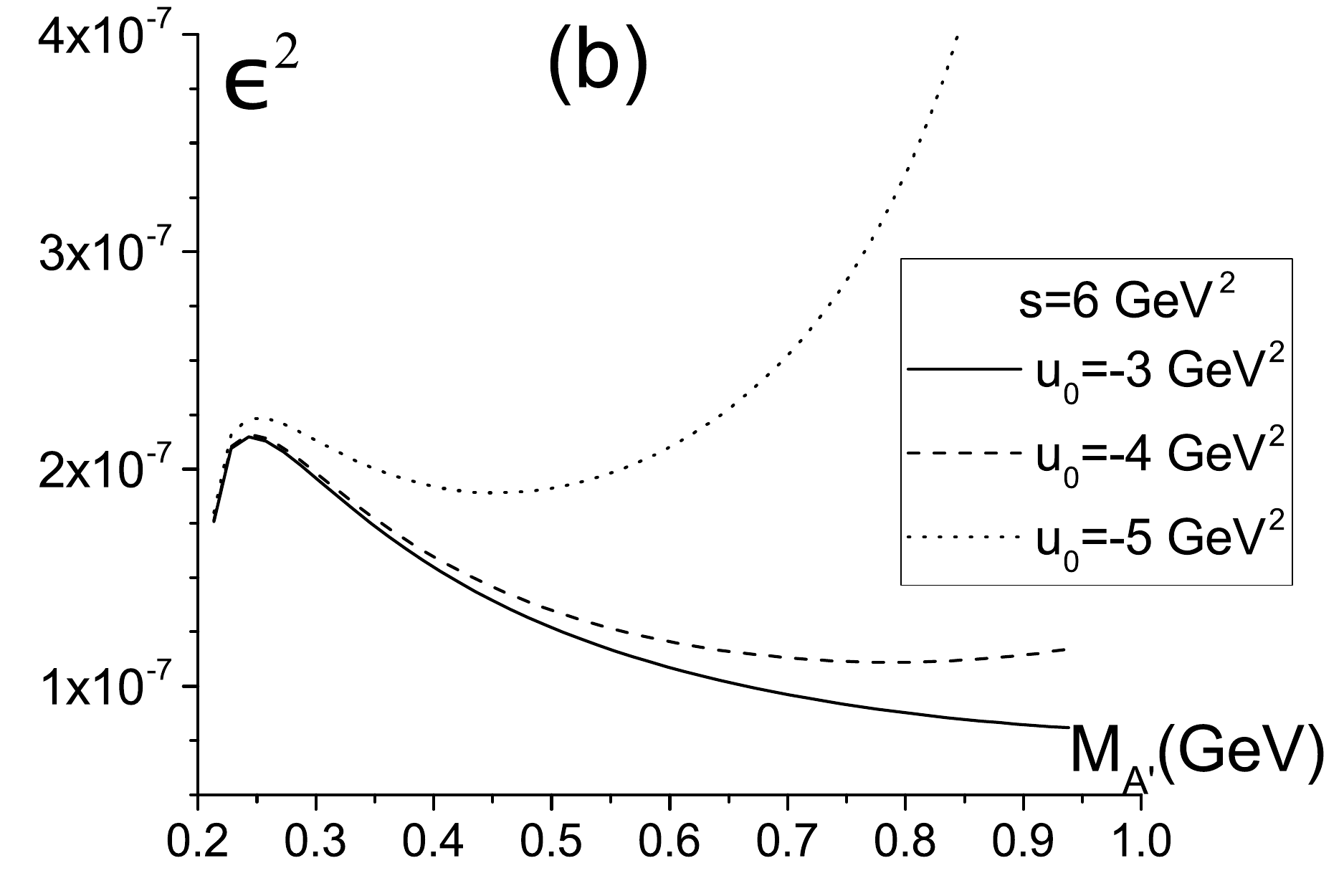}
\includegraphics[width=0.32\textwidth]{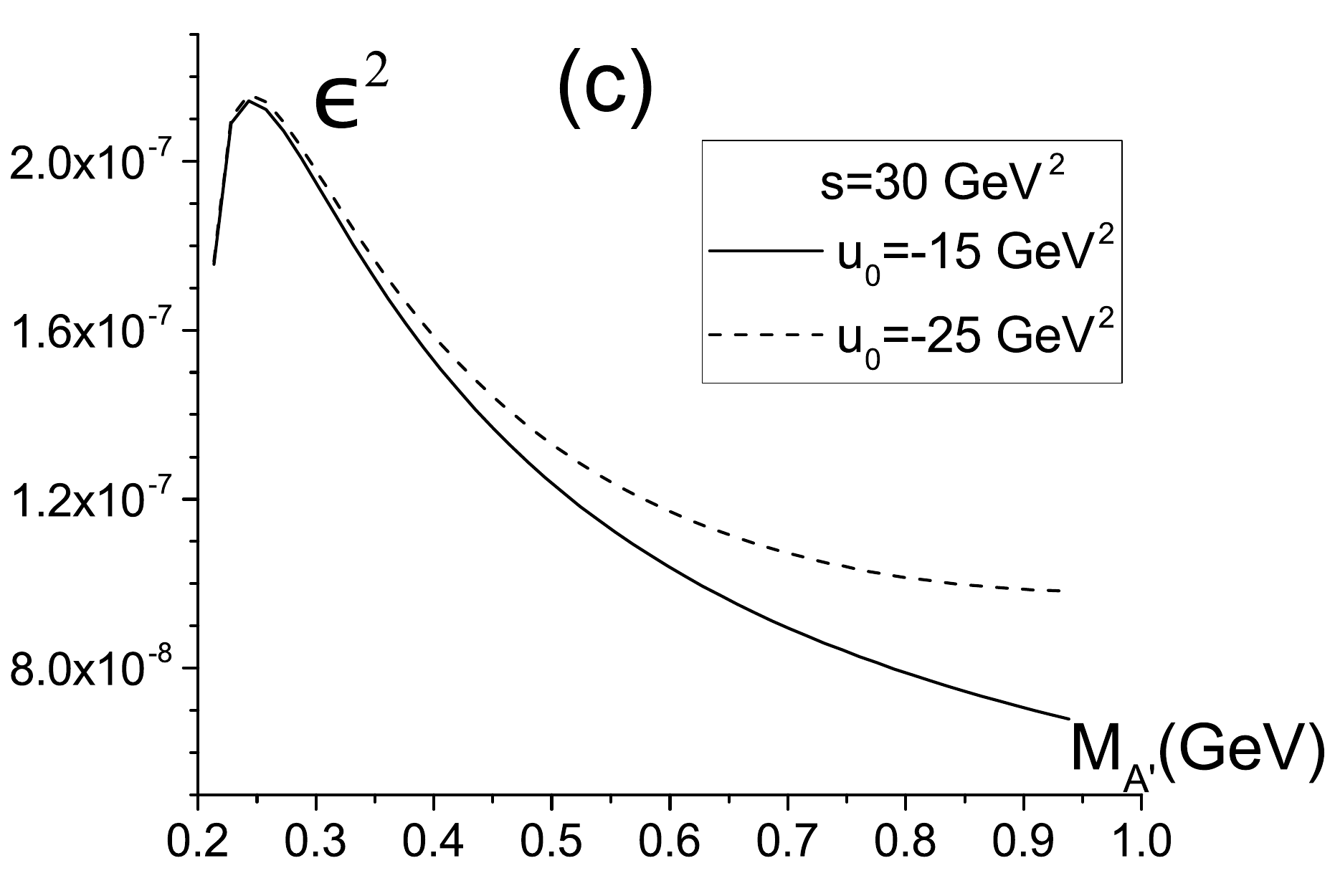}

\caption{The same as in Figs.~\ref{Fig:fig8}(a-c) but for the process
$\gamma +e^- \to \mu^+ \mu^- +e^- $.}
 \label{Fig:fig10}
\end{figure}

\begin{figure}
%\captionstyle{flushleft}
\includegraphics[width=0.45\textwidth]{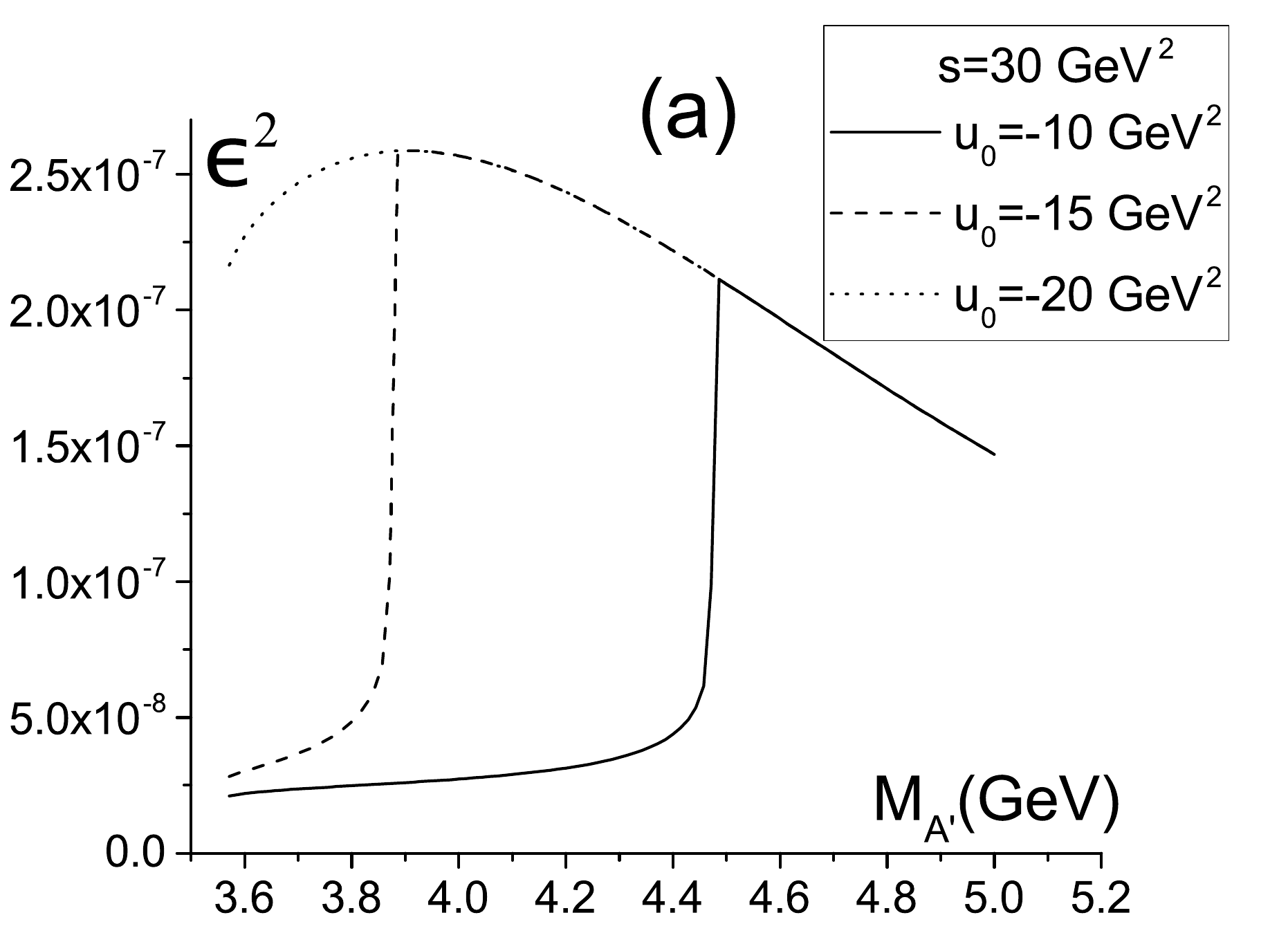}
\includegraphics[width=0.45\textwidth]{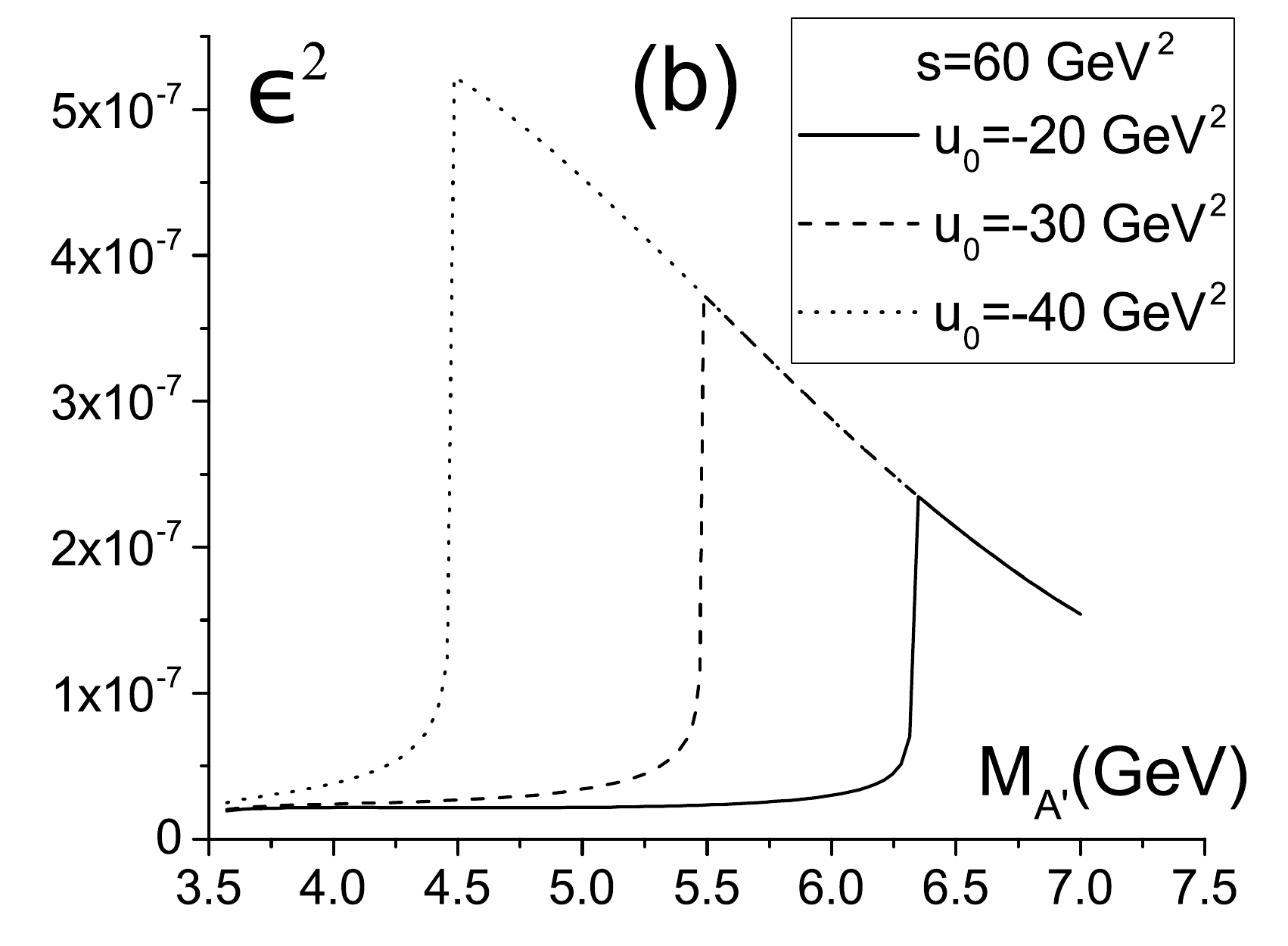}

\caption{The same as in Fig.~\ref{Fig:fig9} but for the process
 $\gamma +e^- \to \tau^+ \tau^- + e^- $.}
 \label{Fig:fig11}
\end{figure}

The corresponding results of calculations for the reactions
$\gamma + e^- \to \mu^+\,\mu^- + e^-$ and $\gamma + e^- \to
\tau^+\,\tau^- + e^-$ are shown in
Figs.~\ref{Fig:fig10} and \ref{Fig:fig11}. Again, in the case of
$\mu^+\mu^-$ production, the limit  $M_{A'} < 2
m_{\tau}$ is applied.

\begin{figure}
%\captionstyle{flushleft}
\includegraphics[width=0.32\textwidth]{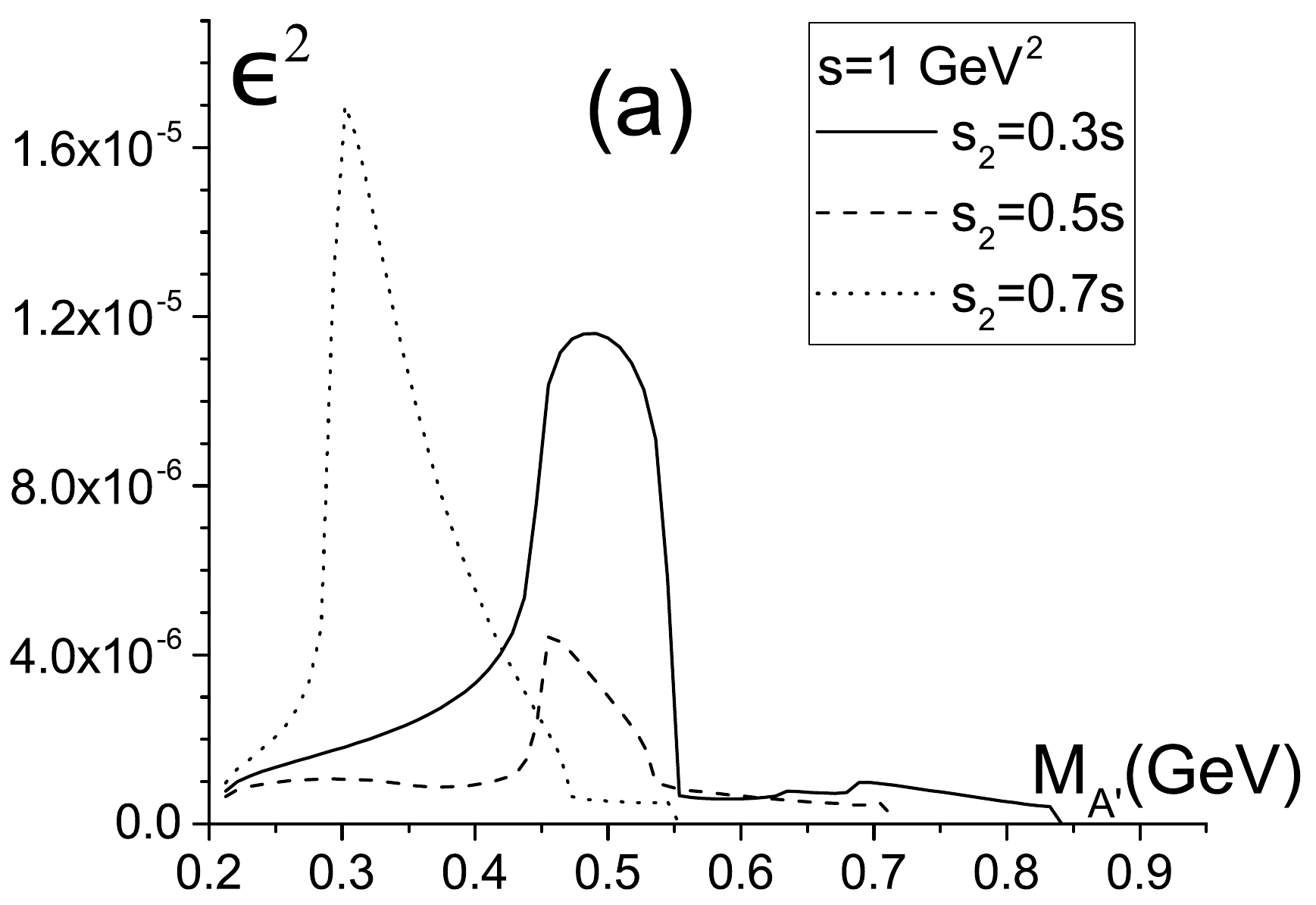}
\includegraphics[width=0.32\textwidth]{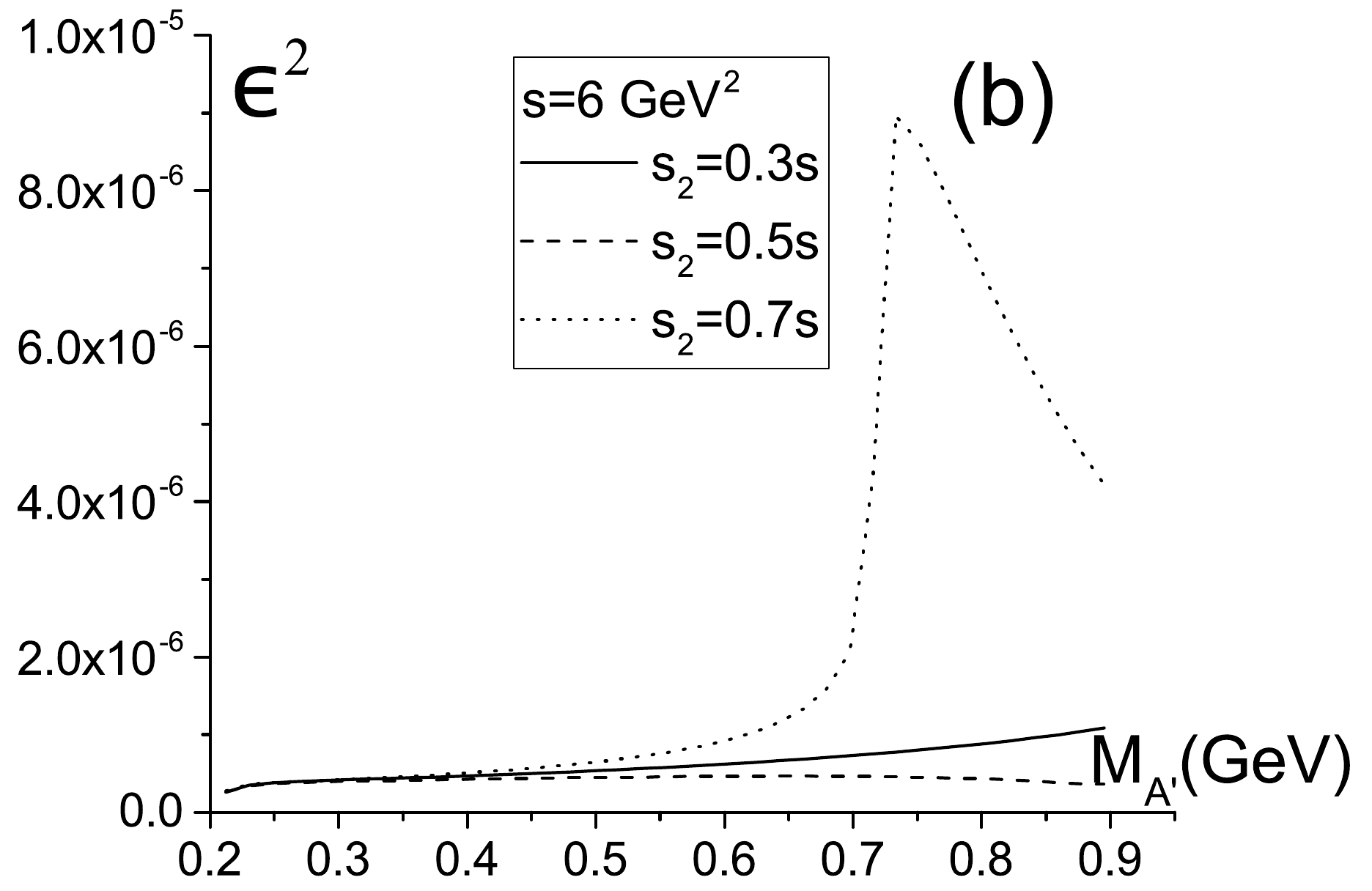}
\includegraphics[width=0.32\textwidth]{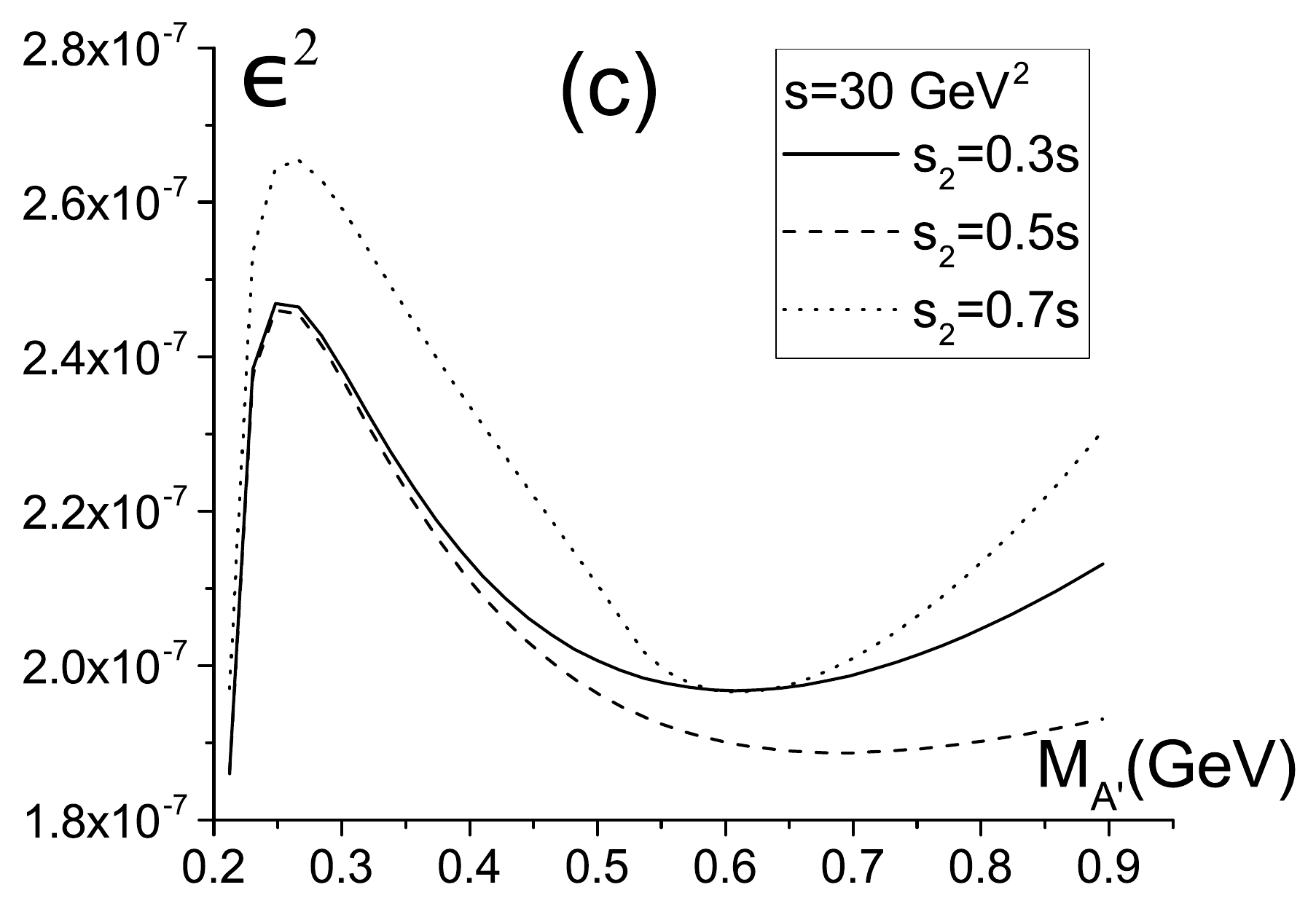}
\caption{The $(\epsilon^2, M_{A'})$ bounds for the DP manifestation in the
process $\gamma + e^- \to e^+ e^- + e^-$ calculated by formulas of Ref. \cite {Gakh:2018ldx} but with corrected
value of $\Gamma$ that takes into account decay $A'\to \mu^+\mu^-$.}
 \label{Fig:fig12}
\end{figure}

For every point in the $(\epsilon^2, M_{A'})$ region (at a
given values of $s$ ) below curves, $\sigma<2$ and above curves,
$\sigma>2$. If the real $A'$ signal corresponds, at least, to
three (or more) standard deviations, then the quantities
$\epsilon^2$ (at fixed $M_{A'}$), when this signal can be
recorded, increase by 1.5 times (or more) as compared with the
corresponding points on the curves in
Figs. \ref{Fig:fig7}-\ref{Fig:fig11}.
%%%%%%%%%%%%%%%%%%%%%%%%%%%%%%%%%%%%%%%%%%%%%%%%%%%%%%%%%%%%%
To be complete, we plot also in Fig.\ref{Fig:fig12} the possible  $\epsilon^2$ and $M_{A'}$ bounds which can be obtained in the triplet production
process $\gamma + e^- \to e^+ e^- + e^-$ for the DP masses enlarged up to 1 GeV values as compared with Ref.\cite{Gakh:2018ldx}.

%%%%%%%%%%%%%%%%%%%%%%%%%%%%%%%%%%%%%%%%%%%%%%%%%%%%%%%%%%%%%%%%%%%%%%%%%%%%%%%%%%%%%%%%%%%%%%%
%%%%%%%%%%%%%%%%%%%%%%%%%%%%%%%%%%%%%%%%%%%%%%%%%%%%%%%%%%%%%%%%%%%%%%%%%%%%%%%%%%%%%%%%%%%%%%%%%

\begin{figure}
%\captionstyle{flushleft}
\includegraphics[width=0.60\textwidth]{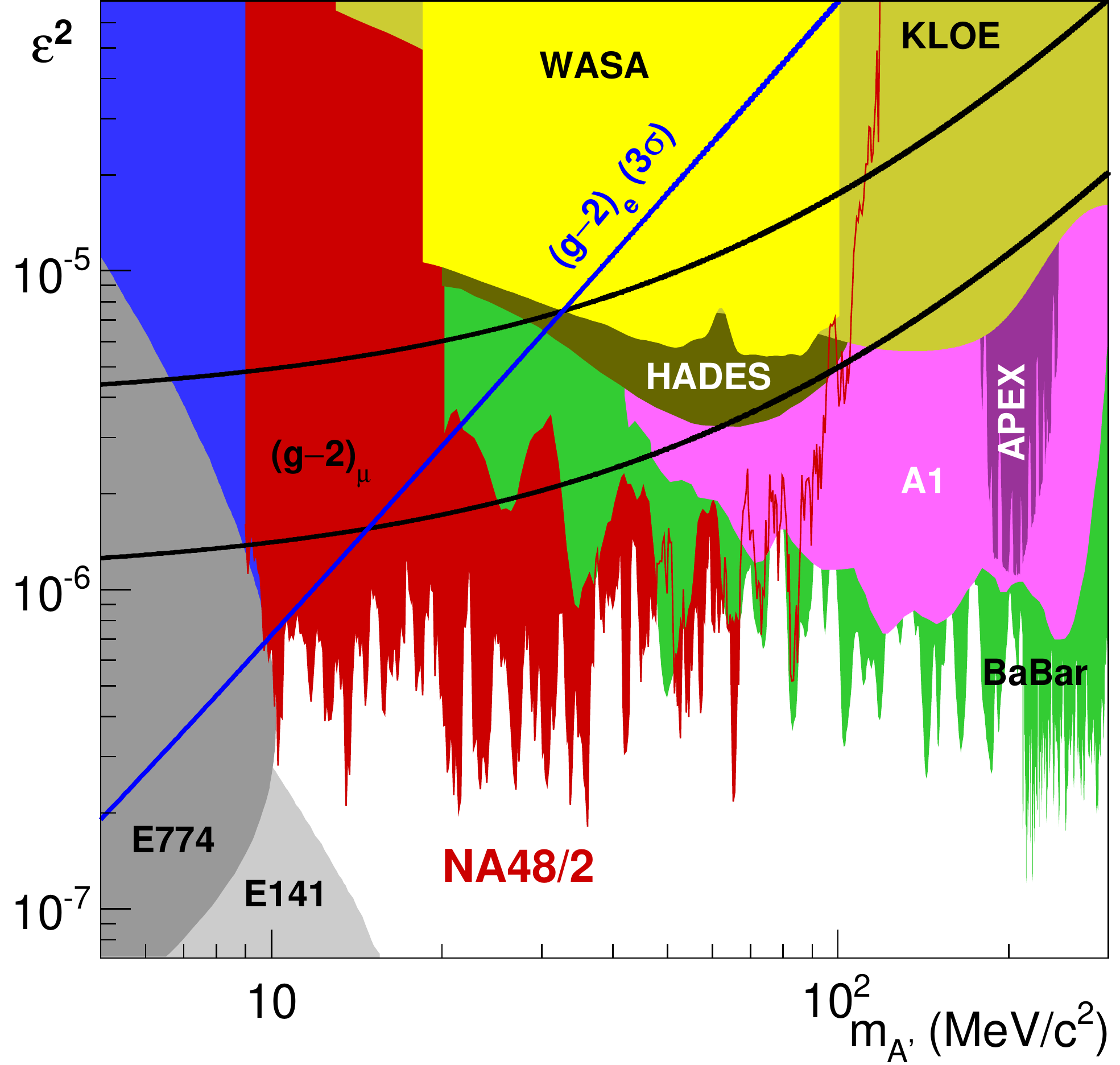}
\caption{The NA48/2 preliminary upper limits at 90 $\%$ CL on the mixing parameter $\epsilon^2$
versus the DP mass $m_{A'}$ \cite {Goudzovski:2014rwa}, compared to the other published exclusion limits from meson decay,
beam dump and $e^+e^-$ collider experiments \cite{Let14}. The results obtained in Refs. \cite{Davoudiasl:2014kua}
and \cite{Pospelov:2008zw} from the muon and electron $g-2$ factors are also shown.}
 \label{Fig:fig13}
\end{figure}
%%%%%%%%%%%%%%%%%%%%%%%%%%%%%%%%%%%%%%%%%%%%%%%%%%%%%%%%%%%%%%%%%%%%%%%%%%%%%%%%%%%%%%%%%%%%%%%%%

It is easy to see from Eq.~(\ref{Eq:eq27}) that an increase of the
energy bin value $\delta m$ decreases the sensitivity of the
detection of the $A'$ signal in the process (\ref{Eq:eq1}). The
reason is evident, because such an experimental device increases
the QED background which is, in fact, proportional to $\delta m$
and leaves unchanged the number of events due to the DP signal in the 
narrow $A'$ resonance.
The dependence of the sensitivity on the DP mass $M_{A'}$ is
determined by the interplay of the $M_{A'}$-dependences of
$\Gamma_0$,  $d\sigma_Q$ and $d\sigma_c$ entering in 
Eq.~(\ref{Eq:eq27}). This statement takes place also if we use the exact form of $D(s_1)$ when integrating both sides of
Eq.~(\ref{Eq:eq26}). Accounting for the kinematical restriction
(\ref{Eq:eq23}) increases essentially the sensitivity, due to the 
suppression of the QED background. To illustrate the
corresponding effect, $\epsilon^2(M_{A'})$ for the reaction
$\gamma+\mu^- \to e^+e^- + \mu^-$ is also plotted without the kinematical cuts.

The number of the events $N$ in the denominator of the rhs of Eq.~(\ref{Eq:eq27}), in the frame of the described event
selection, can be written with a good approximation as
\be
\label{Eq:eq28}
N=\frac{2\delta
mM_{A'}}{s}\frac{d\sigma_Q}{dx_1}\Big(x_1=\frac{M_{A'}^2}{s}\Big)L\cdot T
\,.
%\nn
\ee
that allows one to estimate the necessary integral
luminosity $(L \cdot T)$ for collecting 10$^4$ events.
Taking  values for $d \sigma_Q$ in the range
$10^{-31}-10^{-34}$ cm$^{-2}$ (as it follows from the curves in
Fig. \ref{Fig:fig6} ) and values for $2\delta m M_{A'}/s$ in the range 
10$^{-3}$-10$^{-1}$, one finds the interval
\be 
L\cdot T\approx (10^{36} - 10 ^{41})\  {\rm cm}^{-2}
\nn
\ee
at the considered values of $s$ between 1 and
60 GeV$^2$.
The largest energies require the largest integral luminosity and
vice versa.
As concerns the radiative corrections to the QED contribution,
they do not essentially change the ratio $\sigma_Q/\sigma_c$ entering
Eq.~(\ref{Eq:eq27}) and cannot essentially change the
curves in Figs.~\ref{Fig:fig7}-\ref{Fig:fig11}.

Let us compare the constraints on the parameter $\epsilon$ estimated in our paper
with the ones obtained in various experiments (see Fig. \ref{Fig:fig13}). The A1 Collaboration at MAMI  \cite{Merkel:2011ze} measured the mass distribution of the reconstructed $e^+e^-$ pair in the
reaction $e^-+^{181}Ta \to e^-+e^-+e^++^{181}Ta $. The measured limit is 
$\epsilon^2 < 10^{-6}$ in the following range of DP masses: 210 MeV/$c^2$ $< m_{e^+e^-} < 300$ MeV/$c^2$ to be compared to our estimation for the $\epsilon^2$ value in this
DP mass range :  $(1-2)\times 10^{-7}$. We suggest, instead,  to  measure the
distribution over the invariant mass of the produced lepton pair in the reactions
$\gamma+\mu^-\to e^++e^-+\mu^-$ and $\gamma+e^-\to \mu^++\mu^-+e^-$ (see Figs. 8-10). In the DP search with the KLOE detector
at the DA$\Phi$NE $e^+e^-$ collider \cite{Babusci:2014sta},  the authors looked for
a dimuon mass peak in the reaction $e^+e^-\rightarrow \mu^+\mu^-\gamma$,
corresponding to the decay $A'\rightarrow \mu^+\mu^-$.  With an
integrated luminosity of 239.3 pb$^{-1}$ (5.35$\times$10$^5$ events), they set a 90$\%$ C.L. upper
limit for the kinetic mixing parameter $\epsilon^2$ of 1.6$\times10^{-5}$ to
8.5$\times10^{-7}$ in the mass region 520 $< m_{A'} < $980 MeV. Our estimation for
this DP mass range is $\epsilon^2\leq 10^{-7}$ ($\gamma+\mu^-\to
e^++e^-+\mu^-$) and $\epsilon^2\leq (10^{-7}- 6\cdot 10^{-8})$ ($\gamma+e^-\to
\mu^++\mu^-+e^-$). DP was searched as a maximum in the
invariant mass distribution of $e^+e^-$ pairs originating from the radiative decays
as e.g. $\phi\rightarrow\eta e^+e^-$ in the experiment \cite{Moskal:2013asd}. Only an upper
limit was set with 90$\%$ C.L. on $\epsilon^2 < 1.7\times 10^{-5}$ for 30 $< M_{A'} < $400 MeV
and $\epsilon^2 < 8\times 10^{-6}$ for the subregion 50 $< M_{A'} <$ 210 MeV. Our
results are competitive also in these conditions:  $\epsilon^2\leq 10^{-7}$ ($\gamma+\mu^-\to
e^++e^-+\mu^-$).
%%%%%%%%%%%%%%%%%%%%%%%%%%%%%%%%%%%%%%%%%%%%%%%%%%%%%%%%%%%%%%%%

Some published exclusion limits on the DP parameter space from meson decay, beam dump
and $e^+ e^-$ collider experiments \cite{Let14}
 are presented in Fig.~\ref{Fig:fig13}, readapted from Fig. 6 in Ref. 
\cite {Goudzovski:2014rwa}.
%%%%%%%%%%%%%%%%%%%%%%%%%%%%%%%%%%%%%%%%%%%%%%%%%%%%%%%%%%%%%%%%%%%%%%%%%

Let us summarize the advantages of the photoproduction reactions with different lepton
flavors in the final state
\begin{enumerate}
\item In order to obtain the master bidimensional  distribution  $(s_1,~u)$,  it is sufficient to measure
only the $\ell_i$ particle 4-momentum. The master distribution does not contain the  interference between the Compton-type and the Borsellino
diagrams, which simplifies essentially the theoretical analysis. 
\item In order to suppress the background due to the Borsellino contribution it is sufficient  to set a cut only in one invariant variable (if all final flavors are the same, one needs a cut in two variables).
\item The DP physical events are accumulated in a wide region of invariant variables, allowing one to reduce the necessary integral luminosity. In the case of triplet photoproduction, besides the two cuts on
invariant variables mentioned above, the events must be selected at a fixed value of one (from two final) $e^+~e^-$ pair.
\end{enumerate}
%%%%%%%%%%%%%%%%%
\section{Conclusion}
%%%%%%%%%%%%%%%%%

We investigated  the direct detection of dark photon, one of the new particles that may
possibly shed light on the nature and on the interaction of dark
matter. The DP is mixed with the ordinary photon due to the effect of
the kinetic mixing and it can, therefore, interact with the SM
leptons. It is characterized by a mass $M_{A'}$ and a small parameter
$\epsilon$ describing the coupling strength relative the
electric charge $e$.

We analyzed a possible way to detect a DP signal when its mass
lies in the range between few MeV and few GeV. This mass region  is presently
accessible at the existing accelerators. The idea is to scan the distribution of the
invariant mass squared of the $\ell^+_j
\ell^-_j$ system, $s_1$,  in the reactions
$\gamma + \ell_i \to \ell^+_j \ell^-_j + \ell_i$ with $i\neq j$
and $i=e,\mu$; $j=e,\mu,\tau,$ where  few tens MeV photons
collide with high-energy electron or muon beams. Because of the
interaction with SM leptons, DP appears as a narrow resonance in the $\ell^+_j
\ell^-_j-$system over a background, and modifies the
Compton-type diagrams by the Breit-Wigner term. Choosing processes with $i\neq j$ one avoids
the ambiguity of the measurements arising from the final particle identity.

First, we calculated the double differential cross section with
respect to the variables $s_1$, and $u$ and then derived  the
$s_1$ distribution after  integration over the variable $u$.
Note that to measure such double differential cross section it is sufficient to
measure the four-momentum of the final lepton $\ell_i$.

The advantage of this reaction is that the
background is of a pure QED origin  and
can be calculated exactly with the necessary precision.

In the case when all kinematically possible values of the variable $u$ are
taken into account, a large QED background arises due to the contribution of the
Borsellino diagrams, as illustrated in
Fig.~\ref{Fig:fig3}. To suppress this background, we analyzed the
$(s_1,u)$ distribution and identified the kinematical regions where the
contribution of the Compton-type diagrams exceed the
Borsellino ones (see Fig.~\ref{Fig:fig4}.).
The background contribution increases when  the  variables $s_1$ and $u$ decrease, whereas the
DP signal has just the opposite behavior. Therefore we applied
different cuts $(u>u_0)$ to exclude the range of large values of
$|u|$ (see Figs.~\ref{Fig:fig5} and \ref{Fig:fig6}).

Selecting  the restricted $(s_1,u)$ region,  we estimated the constraints
on the possible values of the parameters $\epsilon^2$ and $M_{A'}$
for a given number of the detected events, $N=10^4$, and standard deviation
$\sigma=2$ for all considered reactions (see
Figs.~\ref{Fig:fig7} -\ref{Fig:fig11}). Our results suggest that a convenient bin width containing
all the events of the possible DP signal near $s_1=M_{A'}$ could be 
$\delta m$= 1 MeV. Equation (\ref{Eq:eq27}) determines the relation between
$\epsilon^2$ and $M_{A'}$  as a function of the parameters
 $N$, $\sigma$ and $\delta m$.
%Estimates of the
%integral luminosity necessary to obtain $10^4$ events in the
%considered experimental conditions, show that such experiment is
%indeed presently feasible.

\section{Acknowledgments}

Two of us (N.P.M. and M.I.K) acknowledge the hospitality of IRFU/DPhN, under the Paris-Saclay grant AAP-P2I-DARKPHI. This work was partially supported by the Ministry of Education and Science of Ukraine (project n. 0118U002031). The 
research was conducted in the scope of the IDEATE International 
Research Project (IRP). This work was partially supported by the 
ERASMUS+ program (code F PARIS 011).

\section{Appendix}

To obtain the $s_1$ distribution it is convenient to define the quantity
%\ba
$$\int \frac{d\sigma_i}{ds_1 du} du \equiv F_i(s,s_1,u)\,,  \ i=b, \,c.$$
Then we have; 
$$ F_b(s,s_1,u)=
\frac{\alpha ^3}{\pi
(s-M^2)} \Biggl\{ \frac{\sqrt{s_1-4 m^2}}{\sqrt{s_1}}
 \Biggr[ \frac{2 (4 m^2+s_1 ) M^2}{ (M^2-s ) (-2 M^2+s+u-s_1 )
s_1}+$$
$$ \frac{8 (M^2-s) (m^2+2 s_1)}{3 (-2 M^2+s+u )^3} +
   \frac{4 [  (M^2-s+s_1) m^2+2 s_1^2 ]}{(-2 M^2+s+u)^2 s_1}+$$
$$\frac{2}{(M^2-s) (-2 M^2+s+u ) s_1^2} \big[ 2 \left ( 2 M^4-4
(s-s_1 ) M^2+ 2 s^2+s_1^2-2 s s_1\right ) m^2+ $$
$$   s_1 \big[ M^4-2 (s-2 s_1) M^2+s^2+2 s_1^2] \big] \Biggr ]
-\ln \left(\frac {\sqrt{s_1}+\sqrt{s_1-4 m^2}}
{2m}\right)\times $$
$$\Biggl [ \frac{4 (-8 m^4+4 s_1 m^2+s_1^2 ) M^2}{(M^2-s) (-2
M^2+s+u-s_1) s_1^2}+\frac{8 (M^2-s)
(-4 m^4+6 s_1 m^2+s_1^2 )}{3 (-2 M^2+s+u)^3 s_1}+ $$
 $$  + \frac{4 \left[ -4 (M^2-s+s_1) m^4+2 s_1 (M^2-s+3 s_1 )
   m^2+s_1^3\right]}{(-2 M^2+s+u)^2 s_1^2}+$$
$$\frac{4}{(M^2-s) (-2 M^2+s+u ) s_1^3} \big[ s_1^2 [(M^2-s
)^2+s_1 ( 2 M^2+s_1) ]+$$
$$2 m^2 (2 m^2-s_1) [ 2 M^4-4 (s-s_1) M^2+2 s^2+s_1^2- 2 s s_1]
\big ] \Biggr ] +$$
$$\frac{2 \ln (-2 M^2+s+u )}{(M^2-s) s_1^4} \Biggl [ -\sqrt{s_1}
\sqrt{s_1-4 m^2}\times $$
 $$  \big [ 2 [2 M^4+(6 s_1-4 s ) M^2+2 s^2
   +s_1^2-  2 s  s_1] m^2+
    s_1 [ M^4+(5 s_1-2 s ) M^2+   s (s-s_1) ] \big] \Biggr] -$$
  $$
2 \ln \left(\frac{\sqrt{s_1}+\sqrt{s_1-4 m^2}}{2 m}\right)
\bigl [ 2 m^2 (2 m^2-s_1) [ 2 M^4+(6 s_1-4 s) M^2+2 s^2+s_1^2-2
s s_1] +
 $$
   $$s_1^2 [ M^4+(3 s_1-2 s) M^2+ s(s-s_1)] \big]+$$
 $$
   \frac{\ln (-2 M^2+s+u-s_1)}{(M^2-s ) s_1^4}
\Biggl [ 2 \ln \left(\frac{\sqrt{s_1}+\sqrt{s_1-4 m^2}}{2
m}\right) (8 m^4-4 s_1 m^2-s_1^2)\times$$
$$[2 M^4+(6 s_1-4 s) M^2+2 s^2+s_1^2-2 s s_1]+\sqrt{s_1}
\sqrt{s_1-4 m^2}\times$$
$$      \big [ 4 [2 M^4+(6 s_1-4 s) M^2+2 s^2+s_1^2-2 s s_1] m^2+
      $$
$$s_1[2 M^4+(10 s_1-4 s) M^2+2 s^2+s_1^2-2 s s_1]\big ] \Biggr]\  \Biggr\} \,,$$

$$F_c(s, s_1, u)= -\frac{\alpha ^3 \sqrt{s_1-4 m^2}
(2 m^2+s_1 )}{3 \pi s_1^{5/2} (M^2-s)^4}
\Biggl[ -\frac{1}{2} u^2 (M^2-s)-\frac{2 M^2 (2 M^2+s_1)
   (M^2-s)^2}{M^2-u}+$$
$$[ 3 M^4+M^2 (6 s-2 s_1)-s^2-2 s_1^2+2 s s_1] (M^2-s ) \log
(u-M^2)+u
   (M^4-5 M^2 s-2 s s_1 )\Biggr] \,.$$
   %\ea
The single $s_1$ distribution [Eqs. (\ref{Eq:eq18} and \ref{Eq:eq19})]
is the difference
$$F_i(s,s_1,u_+)-F_i(s,s_1,u_-), $$
where the analytical form of the $s_1$-distribution for the restricted
phase space defined by Eq.~(\ref{Eq:eq23}), is given as
\be
F_i(s,\,s_1,\,u=u_+)-F_i(s,\,s_1,\,u=u_0),
\ \ i=b,~c.
\nn
\ee
This expression is valid if $s_1<s_{10},$ where $s_{10}$ is the
solution of the equation $u_-=u_0$ (see Fig.~\ref{Fig:fig2}).

%\bibliography{Biblion}

\end{document}